\documentclass[oneside, hidelinks, 12pt]{article}
\usepackage[mathscr]{eucal}
\usepackage{amssymb,latexsym,amsthm,pifont,graphicx,anysize,multicol}
\usepackage{verbatim}
\usepackage{graphicx}
\usepackage{amsmath}
\usepackage{amsthm}
\usepackage{enumerate}
\usepackage{framed}
\usepackage{authblk}
\usepackage{cancel}
\usepackage{color}
\usepackage{transparent}
\usepackage{bm}
\usepackage{tocvsec2}
\usepackage{comment}
\usepackage{setspace}
\usepackage{hyperref}
\usepackage{epigraph}
\usepackage[margin=1.5cm]{caption}
\usepackage[title]{appendix}



\marginsize{1.25 in}{1.25 in}{0.7 in}{1.3 in}
\DeclareFontFamily{OT1}{rsfs}{}
\DeclareFontShape{OT1}{rsfs}{m}{n}{ <-7> rsfs5 <7-10> rsfs7 <10->
rsfs10}{} \DeclareMathAlphabet{\mycal}{OT1}{rsfs}{m}{n}

\definecolor{dgray}{RGB}{90,90,90}
\definecolor{gray}{RGB}{120,120,120}
\definecolor{lgray}{RGB}{150,150,150}
\definecolor{purple}{RGB}{150,0,150}

\theoremstyle{plain}
\newtheorem{thm}{Theorem}[section]
\newtheorem{lem}[thm]{Lemma}
\newtheorem{prop}[thm]{Proposition}
\newtheorem{cor}[thm]{Corollary}

\newtheorem*{conj*}{Conjecture}
\newtheorem*{BC*}{Bartnik Splitting Conjecture}

\theoremstyle{definition}
\newtheorem{Def}[thm]{Definition}

\newtheorem{exm}[thm]{Example}

\newtheorem{rmk}[thm]{Remark}

\theoremstyle{remark}

\newcommand{\field}[1]{\mathbb{#1}}

\renewcommand\l{\lambda}

\newcommand\s{\sigma}
\renewcommand\d{\partial}

\newcommand\e{\epsilon}
\renewcommand\b{\beta}

\renewcommand\l{\lambda}
\newcommand\g{\gamma}
\newcommand\8{\infty}
\renewcommand\a{\alpha}

\newcommand\beq{\begin{equation}}
\newcommand\eeq{\end{equation}}
\newcommand\ben{\begin{enumerate}}
\newcommand\een{\end{enumerate}}
\newcommand\bit{\begin{itemize}}
\newcommand\eit{\end{itemize}}

\makeatletter

\newcommand{\Rmnum}[1]{\expandafter\@slowromancap\romannumeral #1@}

\makeatother

\newcommand{\Lim}[1]{\raisebox{0.0ex}{\scalebox{1}{$\displaystyle \lim_{#1}\;$}}}
\newcommand{\Sup}[1]{\raisebox{0.0ex}{\scalebox{1}{$\displaystyle \sup_{#1}\;$}}}

\newcounter{mnotecount}[section]

\begin{document}

\title{Spacetime distances: an exploration}
\date{}

\author{Carlos Vega\thanks{The author originally began this research as a postdoc at MSRI funded by NSF Grant No. 0932078 000, and also continued this research as a postdoc at CUNY funded by Professor Christina Sormani's NSF grants DMS - 1006059 and DMS - 1309360.} }

\maketitle

\begin{abstract} What is the distance between two points in spacetime? This is a basic geometric question, which so far has no single, definitive answer. Unlike their Riemannian cousins, Lorentzian manifolds are not known to carry a canonical distance function. There is, however, a well-known way to `Riemannianize' a Lorentzian metric tensor, under suitable conditions, (e.g., `Wick rotation'). In a fairly different vein, a new `null distance function' was also introduced in \cite{nulldist}. Our goal here is to begin an exploration of the broad question of distance functions on spacetimes in general, including a concrete comparison of the aforementioned constructions. We devote special attention to the model `generalized Robertson-Walker' (GRW) setting, and also study how the `classical FLRW big bangs' manifest in terms of such distance functions, and associated metric completions.
\end{abstract}

\newpage
\renewcommand\contentsname{}
\setcounter{tocdepth}{2}
\tableofcontents

\pagebreak
\section{Introduction}

\vspace{1pc}
The simplest description of general relativity is, perhaps, as follows:
$$gravity \; \; = \; \; geometry$$
Indeed, general relativity is a highly sophisticated \emph{geometric} theory of the universe, in which spacetime is modeled by a Lorentzian manifold, $(M^{n+1},g)$, within which, the dynamics of matter and the curvature of spacetime are intertwined in an inseparable cosmic relationship. Yet some extremely basic \emph{geometric} questions, like (Q1) below, remain unanswered. The higher-order question (Q2) arises when asking, for example, how well a certain idealized model spacetime, $M_{model}$, `\emph{approximates}' our actual physical universe, $M_{actual}$. More mathematically, setting the mass $m = 0$ in the Schwarzschild model gives Minkowski space, $\field{M}$. If a Schwarzschild model has \emph{small mass}, how `\emph{close}' is it to Minkowski space? If we consider a \emph{sequence} of Schwarzchild spacetimes, $M_k$, with masses $m_k \to 0$, does $\{M_k\}$ `\emph{converge}', in some sense, to $\field{M}$? Note that these are also in the spirit of the `stability question' for the Positive Mass Theorem. See, for example, \cite{LeeSormani} for a discussion and results in such directions. Another question, of particular interest here, is (Q3). Naively, we might describe a universe as having emanated from a `\emph{big bang}' if the universe `sprang from a single point', i.e., if it was initially arbitrarily `small'. But `small' \emph{with respect to what measure?}

\vspace{1.5pc}
\hspace{2pc} \emph{(Q1): What is the distance between two points in spacetime?}

\vspace{1pc}
\hspace{2pc} \emph{(Q2): What is the distance between two spacetimes?}

\vspace{1pc}
\hspace{2pc} \emph{(Q3): What is a big bang?}

\vspace{2pc}
Using distances to understand geometric objects may be the most fundamental idea in geometry. It is, in the first place, in this broad and fundamental spirit that we here explore the question of how to meaningfully measure distances in spacetimes, as in (Q1). At the same, this investigation also serves as another step in a broader effort to understand (Q2), including and especially the question of spacetime convergence theories, and how to meaningfully define limits of sequences of spacetimes. Every Riemannian manifold carries a canonical, intrinsic `Riemannian distance function', defined by minimizing lengths of connecting curves. In the Lorentzian setting, this approach fails, catastrophically; given any two points $p$ and $q$ in a spacetime, there is always a curve from $p$ to $q$ of (Lorentzian) length \emph{zero}. While the so-called `Lorentzian distance function' is a canonical construction, and is in some ways analogous to its Riemannian cousin, it is not a true distance function in the sense of metric spaces. In fact, it fails every required property. Indeed, there is no known `canonical' distance function on a spacetime. Of course, being a manifold, any spacetime is certainly metrizable; we can throw a distance function on it, somewhat at random. The real question is that of identifying distance functions which reflect the original spacetime structure in meaningful and useful ways. For example, distance functions which interact nicely with the causal structure of the spacetime would be of particular interest. 

\vspace{1pc}
The question of spacetime convergence theories, and more specifically, the development of a potential Lorentzian version of `intrinsic flat' convergence, was first suggested to Christina Sormani by Shing-Tung Yau, and some preliminary discussions and investigations were then pursued with Lars Andersson, and  Ralph Howard. The idea involved using the so-called `cosmological time function', studied together with Greg Galloway in \cite{AGHcosmo}, to produce a `canonical Riemannianization' of (an appropriate class of) spacetime, and to then apply the established Riemannian tools. This basic Riemannianization trick, (e.g., `Wick rotation'), already well-known to Lorentzian geometers, when applied using a time function, $\tau$, requires $\tau$ to be \emph{smooth}. As cosmological time $\tau_c$ is nonsmooth in general, the idea of adapting this process was then discussed, and some progress was made in this direction by R. Howard. Completing the resulting metric space, and extending the cosmological time function could then also be used to quantify the `initial singularity' of the spacetime geometrically.

\vspace{1pc}
In \cite{nulldist}, a new `null distance function' was introduced, which is a kind of `time-traveling taxicab distance'. This also involves a choice of time function, $\tau$, but notably does not require $\tau$ to be smooth, or even continuous, though a certain `anti-Lipschitz' condition is required to ensure definiteness. This construction was, in fact, motivated by explorations of question (Q3) above. While this may not seem quite as natural as the `Riemannianization' approach, at first glance, the null distance has some important advantages. In addition to readily accepting time functions $\tau$ of lower regularity, the induced null distance $\hat{d}_\tau$ satisfies an important causality property. For all points $p, q \in M$, we have:
$$q \in J^+(p) \; \; \Longrightarrow \; \; \hat{d}_\tau(p,q) = \tau(q) - \tau(p)$$
Moreover, it was shown in \cite{nulldist} that the converse also holds in a broad class of model spaces, and thus the causal structure of the spacetime is essentially encoded in the resulting metric space structure. That is, this process takes as input, a spacetime $(M,g)$ and (appropriate) time function $\tau$, and outputs a metric space $(M, \hat{d}_\tau)$, and, at least in model settings, the causal relation on the original spacetime can be read entirely from the metric space $(M, \hat{d}_\tau)$. And indeed, while it remains an open question, this kind of equivalence, (roughly) the converse of the above, should generalize much more broadly. 

\vspace{1pc}
Below, we continue the study of both the Riemannianized and null distance functions. In the `generalized Robertson-Walker' (GRW) setting, we also study an `elevator distance' function, which is a kind of warped taxicab distance. Several important similarities emerge. We show, for example, broad agreement in the model GRW setting, (for appropriate choices of time function). Indeed, we abstract some underlying behavior in this setting, encoding these in a list of `GRW faithfulness' properties. On the other hand, some notable differences between these distance functions also appear. For example, the null distance function is observed to exhibit a high degree of sensitivity to the choice of time function, whereas this is less so for the Riemannianized distance. Whether this is interpreted as a feature, or bug, for either distance, depends perhaps on the context. As for causality, we noted above that the null distance function is very closely connected to the causal structure, and indeed encodes it in model settings. It turns out that there is no hope for a simple such relationship, in general, for the Riemannianized distance. 

\vspace{1pc}
The GRW setting serves as a useful preliminary testing ground for comparison and illustration. As an application, we study GRW `singularities' and `big bangs', though we use these terms in a somewhat naive sense. (See, for example, \cite{Ling}, \cite{Sbierski}, \cite{Ringstrom_wave_bang}, for studies of a different nature.) In the spirit of the discussions described above, we quantify such `singularities' here in terms of distance functions and their associated metric completions. Both the spatially open and spatially closed cases are addressed. This also serves as a preliminary and complimentary study to separate and upcoming work discussed together with C. Sormani on more general `big bangs', as described in \cite{SormaniOberwolfach}.

\vspace{2.5pc}
\emph{Style, Goals, and Outline:}

\vspace{1pc}
The presentation below is fairly detailed and explicit. Certain generalities are reviewed in some detail. Some basic examples are carried out in full. The emphasis here is not on efficiency nor compactness, but rather on clarity, ease of use, and illustration.

\vspace{1pc}
Our most basic goal is perhaps to begin a comparison and understanding of the similarities and differences of the more conventional Riemannianized distance and the newer null distance. For one, this study confirms qualitative agreement in the model GRW setting, (for appropriate choice of time function). However, some notable distinctions are also observed, including some important advantages to the null distance, especially in terms of causality. But we also study spacetime distance functions more generally. Many related questions arise, some of which we address here, while others are left for future investigation.

\vspace{1pc}
In Section \ref{sec_Prelim}, we begin with some preliminaries, especially on metric spaces and metric completions, and basic Riemannian and spacetime geometry, including the Riemannian distance function, the Lorentzian `distance' function, generalized Robertson-Walker (GRW) spacetimes, and time functions.

\vspace{1pc}
In Section \ref{sec_Distances}, we recall the well-known, though perhaps underutilized `Riemannianization' process, as well as the newer null distance function. We also make some general remarks about length metrics, and metric completions.

\vspace{1pc}
In Section \ref{sec_GRW}, we focus on the GRW setting. Here, a fairly natural `elevator distance' is also available, which is a kind of a warped taxicab metric. Making use also of \cite{nulldistprops}, we make several new observations about the null distance, including some comparisons with the other distance constructions, notably with respect to the interplay with the causal structure in the final subsection, and Example \ref{Riem_does_not_encode}.

\vspace{1pc}
In Section \ref{sec_GRW_faith_sing}, continuing in the GRW setting, we abstract some of the basic properties exhibited by the standard Riemannianized, null, and elevator distances in a list of `GRW faithfulness' properties, from which various results are derived. The concrete distance functions studied here are all defined using an appropriate choice of time function on the spacetime. The dependence on this choice is investigated, with a special focus on `faux' vs `genuine' `big bang points'. Finally, we turn attention to the `classical big bangs', and in Theorem \ref{GRW_faith_bang_thm}, show that these manifest in the same qualitative way for all GRW faithful distance functions, including the standard Riemannianized, null, and elevator distances.

\vspace{2.5pc}
\emph{Acknowledgements}:

\vspace{1pc}
We express our deep and broad gratitude to Prof. Christina Sormani, for her support, and for the many discussions during our joint work together, which have inspired and shaped the present study in several fundamental ways. This collaboration originally started at MSRI, to which we are grateful for having hosted us both during the semester program on Mathematical General Relativity during Fall 2013. It then continued with a postdoctoral appointment of the author at Lehman College and The Graduate Center CUNY over Spring 2014. We are grateful to Prof. Sormani, and to these institutions, for that opportunity as well.

\vspace{1pc}
We were also very much inspired by and have relied upon the work done by B. Allen and A. Burtscher in \cite{nulldistprops}. We thank S. Harris for a helpful clarification. Finally, we thank G. Galloway, for his continued support, and for several helpful discussions.

\pagebreak
\section{Preliminaries} \label{sec_Prelim}

\vspace{0pc}
We begin with a brief review of some basic material on metric spaces, and Riemannian and Lorentzian geometry. See, for example, \cite{burago} for further background on metric spaces. For further Lorentzian background, we note the standard references \cite{BEE}, \cite{HE}, \cite{Penrose}, \cite{Wald}, \cite{ON}, the latter also covering much of the basic Riemannian theory. Throughout this entire work, we assume that all manifolds are smooth, and that all Riemannian and Lorentzian metrics are smooth.

\vspace{1pc}
We also note the following conventions. The end of a proof will, as usual, be marked with a square $\square$ symbol. But similarly, longer examples will sometimes be closed with a diamond $\diamond$ symbol. When interpreting algebraic expressions involving, for example, infima or suprema, we interpret $1/\infty = 0$ and $1/(0^+) = \infty$, and $- \infty \le a \le \infty$, for all $a \in \field{R}$. We also assume, for example, that $a \ge \infty$ means $a = \infty$. Some proofs are omitted below, when they are straightforward enough to verify.

\vspace{1pc}
\subsection{Metric Spaces}

\vspace{1pc}
Consider a set $X$. By a \emph{metric} or \emph{distance function} on $X$, we mean a function $d : X \times X \to [0,\infty)$ which satisfies the following properties for all $x, y, z \in X$,
\ben
\item [(1)] \emph{Symmetry}: $d(x,y) = d(y,x)$
\item [(2)] \emph{Triangle Inequality}: $d(x,z) \le d(x,y) + d(y,z)$
\item [(3)] \emph{Definiteness}: $d(x,y) = 0 \; \Longleftrightarrow \; x = y$
\een
In this case, we call the pair $(X,d)$ a \emph{metric space}. Less stringently, by a \emph{pseudometric} or \emph{semi-metric} on $X$, we mean a function $d : X \times X \to [0,\infty)$ which satisfies (1) and (2) above, and also $d(x,x) = 0$, for all $x \in X$, but with the possibility that $d(x,y) = 0$ for some $x \ne y$. 

\vspace{1pc}
\begin{exm} [Euclidean Plane] \label{Euclidean_plane} The prototypical example of a metric space is perhaps the Euclidean plane, $X = \field{R}^2 = \{(x,y) : x,y \in \field{R}\}$, together with its standard Euclidean distance function:
$$d_{\field{E}^2}((x_1,y_1), (x_2,y_2)) = \sqrt{(x_2-x_1)^2 + (y_2-y_1)^2}$$
\end{exm}

\vspace{1pc}
\begin{exm} [Standard Euclidean Norm and Distance]  \label{Euclidean_norm_distance} More generally, consider Euclidean $n$-space, $\field{R}^n = \{ {\bf x} = (x^1, ..., x^n) : x^i \in \field{R}\}$. We recall that the standard Euclidean norm of a vector ${\bf x} = (x^1, ..., x^n) \in \field{R}^n$ is given by:
$$\|{\bf x}\|_{\field{E}^n} = \|(x^1, ..., x^n)\|_{\field{E}^n} = \sqrt{(x^1)^2 + (x^2)^2 + \cdots + (x^n)^2}$$
The standard Euclidean distance between ${\bf x}, {\bf y} \in \field{R}^n$ is then given by:
$$d_{\field{E}^n}({\bf x}, {\bf y}) = \sqrt{(y^1-x^1)^2 + (y^2-x^2)^2+ \cdots + (y^n-x^n)^2} = \|{\bf y} - {\bf x}\|_{\field{E}^n}$$
Note that for $n = 1$, the Euclidean norm on $\field{R}^1 = \field{R}$ is just absolute value, and the Euclidean distance between $x, y \in \field{R}$ is $d_{\field{E}}(x,y) = |y-x|$. We sometimes abuse notation and refer to the metric space $(\field{R}^n, \| \cdot \|_{\field{E}^n})$, especially for $n =1$.
\end{exm}

\vspace{2pc}
Consider two metric spaces $(X,d_X)$ and $(Y,d_Y)$. A map $F : X \to Y$ is \emph{distance-preserving} if, for all $x_1, x_2 \in X$, we have $d_Y(F(x_1), F(x_2)) \; = \; d_X(x_1, x_2)$. Two metric spaces $(X,d_X)$ and $(Y,d_Y)$ are \emph{isometric} if there is a bijection $F : X \to Y$ which is distance-preserving, in which case we call $F$ a \emph{(metric space) isometry}. When two metric spaces are isometric, we will also write $(X,d_X) \; \approx \; (Y,d_Y)$.

\vspace{1pc}
Let $(X,d)$ be a metric space. The corresponding \emph{metric (space) topology}, or the \emph{topology induced by $d$}, is the topology on $X$ generated by the open metric balls, $\{B_r(p) : p \in X, r > 0\}$. Note that any metric space isometry $F : (X,d_X) \to (Y,d_Y)$ is then also a homeomorphism $F : X \to Y$, with respect to the metric space topologies. 

\vspace{1pc}
Fixing a set $X$, we will say that two metrics $d_1$ and $d_2$ on $X$ are \emph{topologically equivalent} if they induce the same topology on $X$. More stringently, two metrics $d_1$ and $d_2$ on $X$ are called \emph{Lipschitz equivalent} if there are positive constants $C_1, C_2 > 0$ such that, for all $x, y \in X$,
$$C_1 \, d_1(x,y) \le d_2(x,y) \le C_2 \, d_1(x,y)$$
(Note that, if $C_1d_1 \le d_2 \le C_2d_1$ as above, then also $(1/C_2)d_2 \le d_1 \le (1/C_1)d_2$, so that the condition is indeed symmetric in $d_1$ and $d_2$.) Lipschitz equivalent metrics are also topologically equivalent, but the converse is not true in general.

\vspace{1pc}
Now fix a metric space $(X,d)$. We recall that a sequence $\{x_n\}$ in $X$ is called \emph{Cauchy} if for all $\e > 0$, there is an $N \in \field{N}$, such that $d(x_j,x_k) < \e$ for all $j, k \ge N$. A metric space $(X,d)$ is \emph{complete} if all Cauchy sequences in $X$ converge in $X$. Not all metric spaces are complete, but given any metric space $X = (X,d)$, there exists a unique (up to isometry) metric space $\overline{X} = (\, \overline{X}, \overline{d} \,)$ such that $X$ is a dense subspace of $\overline{X}$, and $\overline{X}$ is complete. The metric space $\overline{X}$ is called the \emph{metric completion} of $X$, and is constructed as a set of equivalence classes of Cauchy sequences in $X$, as follows. For two Cauchy sequences $\{x_n\}, \{y_n\}$ in $(X,d)$, define
$$d_{\mathscr{C}}(\{x_n\}, \{y_n\}) := \lim_{n \to \infty} d(x_n,y_n)$$
Let $\mathscr{C}(X)$ denote the set of all Cauchy sequences in $(X,d)$. Taking as granted that the real numbers are complete, then $d_{\mathscr{C}}$ is well-defined, and gives a pseudometric on the set $\mathscr{C}(X)$. Defining $\{x_n\} \sim \{y_n\}$ iff $d_{\mathscr{C}}(\{x_n\}, \{y_n\}) = 0$ gives an equivalence relation on $\mathscr{C}(X)$. Letting $\overline{X} = \mathscr{C}(X)/\sim$ be the set of all such equivalence classes of Cauchy sequences in $(X,d)$, and setting $\overline{d}([\{x_n\}], [\{y_n\}]) := d_{\mathscr{C}}(\{x_n\}, \{y_n\})$, then $(\overline{X}, \overline{d})$ is a metric space, with the properties listed above. 

\vspace{2pc}
We package the following for easy reference:

\begin{prop} [Metric Completion] \label{completion_prop} Consider a metric space $(X,d)$. Let $(\overline{X}, \overline{d})$ be the metric completion, as defined above. Then we have the following:
\ben
\item [(1)] $(\overline{X}, \overline{d})$ is a complete metric space, with distance function given $\overline{d}$ by
$$\overline{d}([\{x_n\}], [\{y_n\}]) = \lim_{n \to \infty} d(x_n,y_n)$$
\item [(2)] For any point $x \in X$, let $\{x_n = x, \forall n\}$ denote the corresponding constant sequence in $X$, which is thus a Cauchy sequence in $(X,d)$. Let $[x]$ denote the corresponding equivalence class in $\overline{X}$, that is, define:
$$[x] := [\{x_n = x, \forall n\}] \in \overline{X}$$
Then the map $\mathcal{I} : X \to \overline{X}$, $\mathcal{I}(x) = [x]$ is an isometry onto its image, with 
$$(X, d) \; \approx \; (\mathcal{I}(X), \overline{d} \,) \; \subset \; (\,\overline{X}, \overline{d}\,)$$
The image $\mathcal{I}(X) \subset \overline{X}$ is precisely the subset of equivalence classes in $\overline{X}$ for which  any representative sequence in $X$ converges to a point in $X$. Furthermore, the set $\mathcal{I}(X) \approx X$ is dense in $\overline{X}$. In practice, for each point $x \in X$, we essentially identify $x = \mathcal{I}(x) = [x]$, and thus also $X = \mathcal{I}(X)$.
\item [(3)] Fix an element $[\{x_n\}] \in \overline{X}$. Note that each entry, $x_j$, is a point in $X$, and thus corresponds to a point $x_j \approx [x_j] = \mathcal{I}(x_j)$ in $\overline{X}$. Putting these together then gives the sequence $\{[x_j]\}$ in $\overline{X}$. This sequence converges to the original element $[\{x_n\}]$ in $(\overline{X}, \overline{d})$. In other words, identifying $x_j = [x_j]$, then in the metric completion $(\overline{X}, \overline{d})$, we have $\lim_{n \, \to \, \infty} (x_n) \;  = \; \lim_{n \, \to \, \infty} \, [x_n] \; = \; [\{x_n\}]$.
\een
\end{prop}

\vspace{2pc}
We note the following:

\begin{prop} [Lipschitz Equivalent Completions]\label{Lipequivcompletionprop} Let $X$ be a set. Consider two metrics $d_1$ and $d_2$ on $X$, and let $(\overline{X}_1, \overline{d}_1)$ and $(\overline{X}_2, \overline{d}_2)$ denote the two corresponding metric completions. If $d_1$ and $d_2$ are Lipschitz equivalent, then, as sets, $\overline{X}_2 = \overline{X}_1 =: \overline{X}$, and the two metrics $\overline{d}_1$ and $\overline{d}_2$ on $\overline{X}$ are Lipschitz equivalent (with the same constants). In particular, they induce the same topology on $\overline{X}$.
\end{prop}

\vspace{1pc}
Let $(X,d_X)$ and $(Y,d_Y)$ be two metric spaces. A function $f : X \to Y$ is called \emph{uniformly continuous} if for any $\e > 0$, there is a $\delta > 0$, such that, for all $x_1, x_2 \in X$, 
$$d_X(x_1, x_2) < \delta \; \; \Longrightarrow \; \; d_Y(f(x_1), f(x_2)) < \e$$
We will say that a map $f : (X,d_X)  \to (Y,d_Y)$ is \emph{Lipschitz}, or more specifically, \emph{Lipschitz, with Lipschitz constant $\lambda$}, or simply \emph{$\lambda$-Lipschitz}, if there is a positive constant $\lambda > 0$, such that, for all $x_1, x_2 \in X$, we have:
$$d_Y(f(x_1), f(x_2)) \; \le \; \lambda \, d_X(x_1, x_2)$$

\vspace{1pc}
\begin{lem} \label{metric_space_maps} Consider the following conditions on a map $f : (X,d_X) \to (Y,d_Y)$ between two metric spaces.
\ben
\item [(1)] $f$ is Lipschitz.
\item [(2)] $f$ is uniformly continuous.
\item [(3)] $f$ takes Cauchy sequences to Cauchy sequences.
\een
We have (1) $\implies$ (2), and (2) $\implies$ (3).
\end{lem}

\vspace{2pc}
Note that the function $f : (0, 1] \to \field{R}$, $f(x) = 1/x$ is continuous, but does not extend continuously to the metric completion $[0,1]$ of its domain. We note:

\begin{prop} [Extending Functions to Completions] \label{extendfunctionprop} Let $(X,d_X)$ and $(Y,d_Y)$ be any two metric spaces, and let $(\overline{X}, \overline{d}_X)$ and $(\overline{Y}, \overline{d}_Y)$ be their metric completions. Fix any continuous function $f : X \to Y$. Then any continuous function $\overline{f} : \overline{X} \to \overline{Y}$, which extends $f$, in the sense that $\overline{f}([x]) = [f(x)]$, for all $x \in X$, must satisfy:
$$\overline{f}([\{x_n\}]) = \Lim{n \, \to \, \infty} [f(x_n)]$$
If the target space $(Y,d_Y)$ is complete, and $f : (X,d_X) \to (Y,d_Y)$ takes Cauchy sequences to Cauchy sequences, (for example, if it is Lipschitz, or uniformly continuous), then there is a unique continuous function $\overline{f} : \overline{X} \to Y$, with $\overline{f} \, |_X = f$, which is well-defined by:
$$\overline{f}([\{x_n\}]) := \Lim{n \, \to \, \infty} f(x_n)$$
\end{prop}

\vspace{1pc}
\subsection{Riemannian Manifolds}

\vspace{1pc}
Now let $M$ be a smooth manifold. We recall that a \emph{Riemannian metric (tensor)} $g$ on $M$ is a $(0,2)$-tensor on $M$ which is symmetric and positive definite. Then $M = (M,g)$ is called a \emph{Riemannian manifold}.

\vspace{.5pc}
\begin{exm} [Euclidean Space as Riemannian Manifold] \label{Euclidean_as_Riemannian} The prototypical Riemannian manifold is Euclidean space, $\field{R}^n = (\field{R}^n, g_{\field{E}^n})$, where the standard Euclidean metric tensor $g_{\field{E}^n}$ is given at each point by the standard dot product. For example, in dimension $n = 3$, this is given by $\field{R}^3 = \{(x,y, z) : x, y, z \in \field{R}\}$ together with the metric tensor $g_{\field{E}^3} = dx^2 + dy^2 + dz^2$. In arbitrary dimension, we may write $\field{R}^n = \{{\bf x} = (x^1,..., x^n) : x^i \in \field{R}\}$ and
$$g_{\field{E}^n} = d(x^1)^2 + \cdots + d(x^n)^2 =: d{\bf x}^2$$ 
\end{exm}

\vspace{1pc}
Fixing a Riemannian metric tensor $g$ on $M$, we recall that the arc length with respect to $g$ of a piecewise smooth curve $\a: [a,b] \to M$ is defined by $L_g(\a) := \int_a^b \sqrt{g(\a'(s), \a'(s))}ds$. For $p, q \in M$, let $\Omega(p,q)$ be the set of all piecewise smooth curves from $p$ to $q$. The \emph{Riemannian distance function} induced by the Riemannian metric tensor $g$ is then defined by
$$d_g(p,q) := \inf \{ \, L_g(\a) : \a \in \Omega(p,q) \, \}$$

\vspace{1pc}
\begin{exm} [Riemannian vs Euclidean Distance on $\field{R}^n$] \label{Riem_vs_Euclid_dist} Consider a Euclidean space $\field{R}^n$. This has a standard Riemannian metric tensor, $g_{\field{E}^n}$, as in Example \ref{Euclidean_as_Riemannian}. The corresponding Riemannian distance $d_{g_{\field{E}^n}}$ between two points $p$ and $q$ in $\field{R}^n$ is effectively the length of the shortest curve connecting them, which is the straight line segment. This is precisely the Euclidean distance $d_{\field{E}^n}$ between $p$ and $q$, as in Example \ref{Euclidean_norm_distance}. That is, we have $d_{g_{\field{E}^n}} = d_{\field{E}^n}$.
\end{exm}

\vspace{1pc}
More generally, we have the following:

\begin{prop} [Riemannian Manifold as Metric Space] \label{Riem_distance_function} Let $(M,g)$ be a Riemannian manifold. The Riemannian distance function $d_g$ defines a (definite) distance function on $M$, making $(M,d_g)$ a metric space. The topology induced by $d_g$ coincides with the manifold topology. 
\end{prop}

\vspace{1pc}
\subsection{Lorentzian Manifolds}

\vspace{1pc}
Let $M^{n+1}$ be a smooth manifold of dimension $n + 1$. A \emph{Lorentzian metric} $g$ on $M$ is a $(0,2)$-tensor on $M$, which is symmetric and nondegenerate, with constant index $\nu = 1$, for which we shall take the signature to be $(-, +, +, \cdots, +)$. Then $M = (M,g)$ is called a \emph{Lorentzian manifold}. 

\vspace{1pc}
\begin{exm} [Minkowski Space as Lorentzian Manifold] The prototypical Lorentzian manifold is \emph{Minkowski space}, $\field{M}^{n+1} = (\field{R}^{n+1}, g_{\field{M}^{n+1}})$. In total dimension $n+1 = 3+1$, for example, this is realized by $\field{R}^{3+1} = \{(t,x,y,z) : t,x,y,z \in \field{R}\}$ together with standard Minkowski metric tensor
$$g_{\field{M}^{3+1}} = -dt^2 + dx^2 + dy^2 + dz^2 = -dt^2 + g_{\field{E}^3}$$
More generally, we may write $\field{R}^{n+1} = \field{R} \times \field{R}^n = \{(t = x^0,x^1,..., x^n) : x^i \in \field{R}\} = \{(t,{\bf x}) : t \in \field{R}, {\bf x} \in \field{R}^n\}$, where $t$ is thought of as a moment or location in time, and ${\bf x} = (x^1, ..., x^n)$ is thought of as a spatial location in $\field{R}^n$. The standard Minkowski metric tensor is then given by $g_{\field{M}^{n+1}} \; = \; -dt^2 \, + \, g_{\field{E}^n}$. Setting $d {\bf x}^2 := d(x^1)^2 + \cdots + d(x^n)^2  = g_{\field{E}^{n}}$ as above, we may write either:
$$\field{M}^{n+1} = (\, \field{R} \times \field{R}^n \; , \; -dt^2 + g_{\field{E}^n} \, ) = (\, \field{R} \times \field{R}^n \; , \; -dt^2 + d{\bf x}^2 \, )$$
\end{exm}

\vspace{1pc}
A Lorentzian metric $g$ on $M$ classifies a vector $X \in TM$ as \emph{timelike}, \emph{null}, or \emph{spacelike} according as $g(X,X)$ is negative, zero, or positive. Timelike and null vectors are referred to collectively as \emph{causal}, and form a double cone in each tangent space. A \emph{spacetime} is a Lorentzian manifold $(M,g)$ which is \emph{time-oriented}, i.e., has a continuous assignment of `future cone' at each point. Hence, every nontrivial causal vector in a spacetime is either \emph{future-pointing} or \emph{past-pointing}. A Lorentzian manifold $M$ is time-orientable iff $M$ admits a smooth everywhere-timelike vector field $T$. Throughout the following, we will assume that all Lorentzian metrics $g$ are time-oriented, so that $M = (M,g)$ is a spacetime, and may refer to such a $g$ as a \emph{spacetime metric}.

\vspace{1pc}
Let $M = (M,g)$ be a spacetime. A piecewise smooth curve $\a : I \to M$ is called \emph{future timelike (resp. null, causal)} if $\a'$, including all one-sided tangents at any breaks or endpoints, is always future-pointing timelike (resp. null, causal). Past curves are defined time dually. The \emph{timelike future} $I^+(S)$ of a subset $S \subset M$ is the set of points $q \in M$ reachable by a future timelike curve from some $p \in S$. The \emph{causal future} $J^+(S)$ is defined similarly using future causal curves, and time dually we have the pasts $I^-(S)$ and $J^-(S)$. For any subset $S \subset M$, the set $I^+(S)$ is open, and $S \subset J^+(S) \subset \overline{I^+(S)}$, and similarly for the pasts. For $p, q \in M$, we write $p \le q$ to mean $q \in J^+(p)$, or equivalently $p \in J^-(q)$. We write $p \ll q$ to mean $ q \in I^+(p)$, or equivalently $p \in I^-(q)$. The timelike and causal relations, $\ll$ and $\le$, form the \emph{causal structure} of the spacetime $M$.

\vspace{1pc}
The Lorentzian arc length of a piecewise smooth curve $\a : [a,b] \to M$, in a spacetime $(M,g)$, is given by $L_g(\a) := \int_a^b \sqrt{|g(\a'(s), \a'(s))|}ds$. On a Riemannian manifold, curves which minimize length are geodesics. But given any two points on a Lorentzian manifold, we can always proceed from one to the other along null subsegments, so minimizing length always gives zero. Fixing $p, q \in M$, it turns out that the (only) interesting thing to optimize is the length of \emph{causal} curves from $p$ to $q$, (if there are any), and we look to \emph{maximize} their length. Indeed, the causal geodesics of a spacetime $M$ are locally length-\emph{maximizing} (among \emph{causal} competitors). For $p, q \in M$, let $\Omega_c(p,q)$ be the set of all piecewise smooth, future-directed causal curves from $p$ to $q$. The \emph{Lorentzian distance function} induced by the spacetime metric tensor $g$ is defined by
$$d_g(p,q) := \sup \, \{ \,L_g(\a) : \a \in \Omega_c(p,q) \, \}$$
where the supremum is taken to be 0 when there are no such curves. Hence, $d_g(p,q) = 0$ whenever $p \not \le q$. Note that, despite its name, Lorentzian `distance' is not a true distance function in the sense of metric spaces. It fails to be definite, symmetric, and only satisfies the following \emph{reverse (causal) triangle inequality}:
$$x \le y \le z \; \; \Longrightarrow \; \; d_g(x,y) + d_g(y,z) \le d_g(x,z)$$

\vspace{1pc}
Indeed, we emphasize the following:

\begin{rmk} [No Canonical Distance Function on Spacetime] In contrast to the Riemannian case, a Lorentzian manifold has no known canonical distance function, in the sense of metric spaces. The question of how to usefully turn a Lorentzian manifold into a metric space is, indeed, central to this work and will be explored below.
\end{rmk}

\vspace{1pc}
\subsection{GRW Spacetimes}

\vspace{1pc}
The classical Friedmann–Lemaître–Robertson–Walker (FLRW) cosmological models are all warped product spacetimes of the following form:

\begin{Def} [GRW Spacetimes] \label{DefGRW} By a \emph{generalized Robertson-Walker (GRW) spacetime} we will mean a warped product spacetime of the form
$$(M^{n+1},g) = (I \times S^n, -dt^2 + f^2(t)h) =: I \times_f S$$
where $I = (a,b) \subset \field{R}$ is an open interval, $-\infty \le a < b \le \infty$, $f : I \to (0, \infty)$ is a smooth positive function, and the base fiber, $S = (S^n,h)$, is an arbitrary connected Riemannian manifold. Note: We emphasize that no further (completeness or homogeneity) assumptions are made about $(S,h)$. For $p \in M = I \times S$, we will write $p = (t,x) = (t_p,p_S)$, and thus $M = \{(t,x) : t \in I, x \in S\} = \{(t_p,p_S) : t_p \in I, p_S \in S\}$.
\end{Def}

\begin{figure}[h]
\begin{center}
\includegraphics[width=6cm]{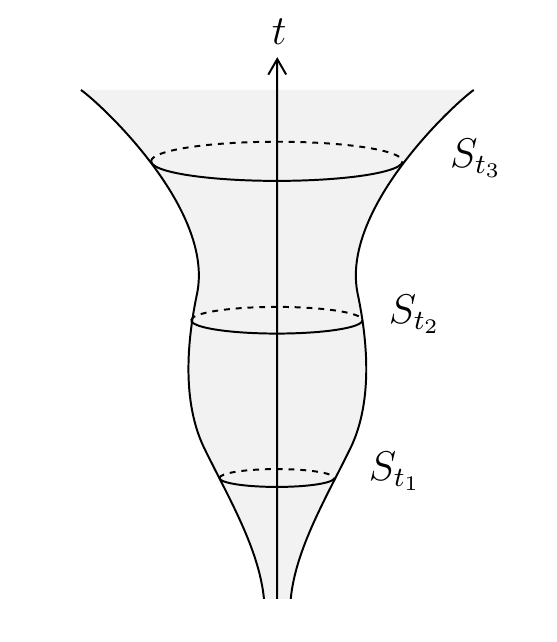}
\vspace{-.5pc}
\caption[]{In a GRW spacetime, the geometry of space is essentially the same at all times, with only the overall scale changing. We also note that, as above, spacetimes are typically visualized with time increasing vertically.} 
\label{GRW_fig}
\end{center}
\end{figure}

\vspace{1pc}
The interval $I$ represents `time', and $S$, roughly, represents `space'. Hence, for $p = (t_p, p_S) \in I \times S = M$, we think of $t_p \in I$ as the location in time, and $p_S \in S$ as the location in space. Fixing any time $t \in I$, the time slice $S_{t} := \{t\} \times S$ represents `space at time $t$'. Each time slice $S_t$ is a smooth spacelike hypersurface in the spacetime $M$, with induced Riemannian metric $f^2(t)h$. Hence, each slice $(S_t, f^2(t)h)$ is homothetic to the base fiber $(S,h)$. In particular, at all times in the universe $M$, space has exactly the same (not only topological, but geometric) shape, but with the overall scale potentially changing over time. If $S$ is compact, we will say $M = I \times_f S$ is \emph{spatially closed}. If $S$ is noncompact, we will say $M = I \times_f S$ is \emph{spatially open}. If $I = (a,b)$ with $- \infty < a$, we will say $M = I \times_f S$ is \emph{past finite}. If $b < \infty$, we will say $M$ is \emph{future finite}.

\vspace{1pc}
Note that the case $f \equiv 1$ means that product spacetimes, $(I \times S, -dt^2 + h)$ are included in the class of GRW spacetimes. The simplest example of the latter is of course Minkowski space, $\field{M}^{n+1} = (\field{R} \times \field{R}^n, -dt^2 + g_{\field{E}^n})$. Another basic example will also be of particular interest below.

\vspace{.5pc}
\begin{exm} [Future Half-Minkowski Space]  Define the \emph{future half-Minkowski space} to be the open subspacetime $\field{M}^{n+1}_+ := \{t > 0\} \subset \field{M}^{n+1}$. More explicitly:
$$\field{M}^{n+1}_+ = (\, (0,\infty) \times \field{R}^n \; , \; -dt^2 + d{\bf x}^2 \, )$$
where, again, $d {\bf x}^2 = d(x^1)^2 + \cdots + d(x^n)^2  = g_{\field{E}^{n}}$ is the standard Euclidean metric on the spatial factor, $\field{R}^n$. Note that this is a product, and hence also a GRW spacetime, (with warping function $f \equiv 1$).
\end{exm}

\vspace{1pc}
Conversely, any GRW spacetime is conformal to a (non-warped) product spacetime: 

\begin{lem} \label{unwarping} Consider any GRW spacetime, $(M,g) = (I \times S, -dt^2 + f^2(t)h)$. Fix any time $c \in I$, and define the function $\psi(t) := \int_{c}^t \, [1/f(w)]\,dw$. Letting $u = \psi(t)$, $\tilde{f}(u) = f(\psi^{-1}(u))$, and $\tilde{g} = -du^2 + h$, then the original GRW metric tensor $g$ is conformal to the (non-warped) product metric tensor $\tilde{g}$, with
$$g \; = \; -dt^2 + f^2(t)h \; = \; \tilde{f}^{\, 2}(u) ( - du^2 + h) \; = \; \tilde{f}^{\, 2}(u) \, \tilde{g}$$
\end{lem}

\vspace{1pc}
\subsection{Time Functions and Causality Conditions}

\vspace{1pc}
Let $M$ be a spacetime. Following \cite{BEE}, we will call a real-valued function $\tau : M \to \field{R}$ a \emph{generalized time function} if $\tau$ is strictly increasing along all (regular) future-directed causal curves. If further $\tau$ is continuous, then $\tau$ is called a \emph{time function}. For example, the standard coordinate function $\tau(t,x^1, ..., x^n) = t$ is a time function on Minkowski space. Indeed, on any GRW spacetime, we have:

\begin{Def} [GRW Standard Time] \label{DefGRW_standard_time} Consider a GRW spacetime $(M,g) = (I \times S, -dt^2 + f^2(t)h)$. Then projection onto the first factor, $\tau : I \times S \to I$, $\tau(t,x) = t$, is a smooth time function, with everywhere past-pointing timelike unit gradient, $\nabla \tau = - \d_t$. We will call this the \emph{standard time function} of $M = I \times_f S$.
\end{Def}

\vspace{1pc}
More generally, we have the following:

\begin{lem} On an arbitrary spacetime $M$, any smooth function $\tau : M \to \field{R}$ with everywhere past-pointing timelike gradient, $\nabla \tau$, is a smooth time function.
\end{lem}

\vspace{1pc}
For $p, q \in M$, we will refer to $I^+(p) \cap I^-(q)$ as a \emph{timelike diamond}, and similarly to $J^+(p) \cap J^-(q)$ as a \emph{causal diamond}. Note that all timelike diamonds are open. A spacetime is \emph{strongly causal} if the set of all timelike diamonds generates the topology of $M$, i.e., if each point admits an arbitrarily small timelike diamond neighborhood. A spacetime $M$ is \emph{globally hyperbolic} if the timelike diamonds generate the topology of $M$, and all causal diamonds are compact.

\vspace{1pc}
A spacetime is \emph{past-distinguishing} if $I^-(p) = I^-(q)$ implies $p = q$. Any strongly causal spacetime is necessarily past-distinguishing. We note the following:

\begin{prop} [\cite{BEE}, \cite{MS}] Any past-distinguishing spacetime admits a generalized time function.
\end{prop}

\vspace{1pc}
Furthermore, we recall roughly that a spacetime $(M,g)$ is \emph{causal} if it contains no closed causal curves, and \emph{stably causal} if this property persists under small perturbations of $g$. Any stably causal spacetime is necessarily strongly causal. We note the following fundamental Lorentzian result:

\begin{thm} [\cite{HE}, \cite{SanchezBernalsplit}, \cite{difftime}] \label{tempiftime} Let $M$ be a spacetime. The following are equivalent:
\ben
\item [(a)] $M$ is stably causal.
\item [(b)] $M$ admits a time function.
\item [(c)] $M$ admits a `temporal function', that is, a smooth time function $\tau : M \to \field{R}$, with everywhere past-pointing timelike gradient $\nabla \tau$.
\een
\end{thm}

\vspace{1pc}
\subsection{Past Causal Boundary}

\vspace{1pc}
Let $(M,g)$ be a spacetime. By a \emph{future set} $F \subset M$, we mean any set of the form $F = I^+(S)$, where $S \subset M$ is any subset. It follows that $F \subset M$ is a future set iff $F = I^+(F)$. Note that future sets are open, and that the union of any collection of future sets is a future set. Time-dual statements hold for \emph{past sets}. By an \emph{indecomposable future set (IF)} we mean a future set $F$ which can not be decomposed as the union of two proper subsets $F_1, F_2 \subset F$, with both $F_1$ and $F_2$ being future sets. 

\vspace{0pc}
\begin{thm} [Indecomposable Future Sets, \cite{GKPidealpoints}] A subset $F \subset M$ is an indecomposable future set iff $F = I^+(\g)$, where $\g$ is a past-directed timelike curve, which either has a past endpoint in $M$, or is past-inextendible in $M$.
\end{thm}

This result means that there are precisely two types of IFs, and the following terminology is used. In the former case above, when $F = I^+(\g)$ and $\g$ has a past endpoint $p \in M$, we have $F = I^+(p)$, and we call $F$ a \emph{proper indecomposable futures (PIF)}. In the latter case, when $F = I^+(\g)$ with $\g$ past-inextendible in $M$, we call $F$ a \emph{terminal indecomposable future set (TIF)}. As sets, the \emph{past causal boundary} $\d^-(M)$ of $M$ is the set of all TIFs in $M$, and the \emph{past causal completion} $M^-$ of $M$ is the union $M^- := M \cup \d^-(M)$. Under appropriate causality conditions, we may identify each point $p \in M$ with its corresponding PIF $F = I^+(p)$. Then $\d^-(M)$ is the set of all TIFs, $M$ is (identified with) the set of all PIFs, and $M^-$ is simply the set of all IFs.

\vspace{1pc}
\pagebreak
\section{Distances on Spacetimes} \label{sec_Distances}

\vspace{1pc}
\subsection{Riemannianizing a Spacetime} 

\vspace{1pc}
Superficially, Euclidean space, $\field{E}^{n+1}$, and Minkowski space, $\field{M}^{n+1}$, differ only by a single minus sign. In total dimension three, for example, we can write each metric tensor on $\field{R}^3 = \{(t,x,y) : t,x,y \in \field{R}\}$, respectively, as
\begin{align*}
g_{\field{E}} \; & = \; \; \; \, \, dt^2 + dx^2 + dy^2 \\[.5pc]
g_{\field{M}} \; & = \; - \, dt^2 + dx^2 + dy^2
\end{align*}
Thus, one space is obtained from the other by simply `flipping the sign' on $dt^2$. Indeed, as is well known, this type of transformation can be done on more general Riemannian and Lorentzian manifolds. See \cite{aazami_ream} for a recent application. We now review this procedure, (in one among other possible forms). Let $M$ be a smooth manifold, and $g$ a metric tensor on $M$ which is either Riemannian or Lorentzian. Then given a vector $V$ on $M$, its \emph{norm} or \emph{length} with respect to $g$ is defined by $\|V\|_g := \sqrt{|g(V,V)|}$. The following generalizes the relationship between Euclidean space, $\field{E}^{n+1}$, and Minkowski space, $\field{M}^{n+1}$, described above.

\vspace{1pc}
\begin{prop} [Flipping a Metric] \label{flipmetric}  Let $M$ be a smooth manifold.
\ben
\item [(1)] Let $g$ be any Riemannian metric on $M$. If $Z$ is a smooth nonvanishing 
vector field on $M$, then the $(0,2)$-tensor $g^L = g^L_Z$ defined by
$$g^L_Z(X,Y) := - \frac{2}{\|Z\|^2_g}g(Z,X)g(Z,Y) + g(X,Y)$$
is a Lorentzian metric on $M$. Moreover, $U := Z/\|Z\|_g$ is also unit with respect to $g^L$ and any $g$-orthonormal frame $\{U, E_1, ..., E_n\}$ is also $g^L$-orthonormal. We may refer to $g^L_Z$ as the `Lorentzification' of $g$ induced by $Z$. In fact, $g^L$ is time-orientable, with the vector field $Z$ being either everywhere future-pointing timelike or everywhere past-pointing timelike, depending on the choice of time orientation. 
\item [(2)] Let $g$ be a Lorentzian metric on $M$. If $T$ is a smooth timelike vector field on $M$, then the $(0,2)$-tensor $g^R = g^R_T$ defined by
$$g^R_T(X,Y) := \frac{2}{\|T\|^2_g}g(T,X)g(T,Y) + g(X,Y)$$
is a Riemannian metric on $M$. Moreover, $U := T/\|T\|_g$ is also unit with respect to $g^R$ and any $g$-orthonormal frame $\{U, E_1, ..., E_n\}$ is also $g^R$-orthonormal. We will refer to $g^R_T$ as the `Riemannianization' of $g$ induced by $T$. Note that since a spacetime always admits such a vector field $T$, every spacetime can be thus Riemannianized.
\een
\end{prop}

\vspace{1pc}
\begin{exm} [Euclidean vs Minkowski Space] Let $\field{R}^{n+1} = \{(t = x^0,x^1, ..., x^n) : x^i \in \field{R}\}$ and consider the vector field $U = \d_t = \d_0$. It is straightforward to verify that the Lorentzification of the standard Euclidean metric induced by $U$ is the Minkowski metric, and conversely that the Riemannianization of the Minkowski metric induced by $U$ is the Euclidean metric. That is, we have $(g_{\field{E}})^L_U = g_{\field{M}}$ and $(g_{\field{M}})^R_U = g_{\field{E}}$. 
\end{exm}

\vspace{1pc}
Our focus here is primarily on the process of `Riemannianizing' spacetimes, that is, turning a spacetime $(M,g)$ into a Riemannian manifold, $(M,g^R)$, which then has an induced Riemannian distance function, $d^R$, and thus produces a metric space structure on the original spacetime, $(M,d^R)$. Note that for $T$ timelike we have $\| T\|^2_g = - g(T,T)$, thus the induced Riemannianized metric $g^R = g^R_T$ can be rewritten in various ways:
\begin{align*}
g^R_T(X,Y) & \; = \; \frac{2}{\|T\|^2_g}g(T,X)g(T,Y) + g(X,Y) \; = \; - 2 \, \frac{g(T,X)g(T,Y)}{g(T,T)} + g(X,Y)
\end{align*}

\vspace{1pc}
We now list a few basic properties. For one, we note first that the Riemannianization formula defined above effectively normalizes the choice of timelike vector field to unit length in the process. 

\begin{lem} [Riemannianization Basic Properties] \label{Riem_basic_props} Let $(M,g)$ be a spacetime. Fix any smooth timelike vector field $T$ on $M$ and consider the corresponding Riemannian metric $g^R_T$. 
\ben
\item [(1)] Letting $U := T / \|T\|_g$, we have: $g^R_T(X,Y) = g^R_{(-T)}(X,Y) = g_U^R(X,Y)$
\item [(2)] Riemannianization respects conformal scaling, that is, for any smooth positive function $f : M \to (0, \infty)$, we have:  $(f g)^R_T = f (g^R_T)$
\item [(3)] For any $p \in M$, and any $X \in T_pM$, $g(T,X) = 0$ iff $g^R_T(T,X) = 0$. In other words, the orthogonal compliment of $T$ in $T_pM$ is the same with respect to either metric, $g$ or $g^R_T$. We may thus unambiguously denote this space by simply $T^\perp$. Furthermore, $g^R_T$ and $g$ agree when restricted to $T^\perp$. That is, for any $X, Y \in T^\perp$, we have $g^R_T(X,Y) = g(X,Y)$.
\item [(4)] Define the $(0,2)$-tensor \; 
$$\omega^R_T(X,Y) := - \frac{g(T,X)g(T,Y)}{g(T,T)} + g(X,Y)$$
For any $p \in M$, and any $X, Y \in T_pM$, we have $\omega^R_T(X,Y) \ge 0$. 
\een
\end{lem}

\vspace{2pc}
Given a spacetime $(M,g)$, we can use any timelike vector field $T$ to `flip' the metric $g$, and produce a corresponding Riemannian metric $g^R_T$ on $M$. Below, we will usually do this using a time function, as in the following:

\begin{cor} [Riemannianizing via Time Function] Consider a spacetime $(M,g)$, and suppose that $\tau : M \to \field{R}$ is a smooth function with $T = \nabla \tau$ everywhere timelike. Then the following defines a smooth Riemannian metric on $M$:
$$g^R_\tau := \dfrac{2}{\; \|\nabla \tau \|_g^2} d\tau^2 + g$$
In particular, if $\nabla \tau$ is timelike unit, then $g^R_\tau = 2d\tau^2 + g$.
(Note: We have just introduced a slight abuse of notation. In fact, the metric defined here is $g^R_{\nabla \tau}$. However, no confusion should arise in practice.) In this setting, we will let $L^R_\tau$ and $d^R_\tau$ denote the Riemannian length and distance with respect to $g^R_\tau$. Note that, as per Proposition \ref{Riem_distance_function}, $d^R_\tau$ induces the manifold topology.
\end{cor}

\vspace{1pc}
We note the following:

\begin{lem} [Riemannianization Under Time Reparameterization] \label{Riem_invar_reparam} Consider a spacetime $(M,g)$. Suppose that for two smooth time functions $\tau_1, \tau_2 : M \to \field{R}$, there is a differentiable function $\phi : I \to J$, where $I$ and $J$ are open intervals in $\field{R}$, such that $\tau_2 = \phi \circ \tau_1$. Then $\nabla \tau_2 = (\phi' \circ \tau_1) \nabla \tau_1$ and $d \tau_2 = (\phi' \circ \tau_1) d\tau_1$. Suppose that $\nabla \tau_1$ is everywhere timelike. Then $\nabla \tau_2$ is everywhere timelike iff $\phi'(u)> 0$ for all $u \in I$, and in this case, the Riemannianized metric tensors agree; 
$$g^R_{\tau_2} \; = \; g^R_{\tau_1}$$
\end{lem}

\vspace{2pc}
The following employs somewhat standard considerations:

\begin{lem} [Riemannianized Distance Lipschitz Continuity] \label{Riem_dist_Lipschitz} Let $(M,g)$ be a spacetime. Suppose that $\tau$ is a smooth time function on $M$, with timelike gradient $\nabla \tau$, and consider the corresponding Riemannian metric $g^R_\tau$. Let $L^R_\tau$ and $d^R_\tau$ denote Riemannian length and distance with respect to $g^R_\tau$. Suppose that the $g$-norm of the gradient $\nabla \tau$ is bounded above by a positive constant, $\|\nabla \tau \|_g \le D$. Then, for any two points $p, q \in M$, we have:
$$\dfrac{1}{D} \cdot |\tau(q) - \tau(p)| \, \le \, d^R_\tau(p,q)$$
Thus, as a map between metric spaces, $\tau : (M, d^R_\tau) \to (\field{R}, |\cdot |)$ is Lipschitz, with Lipschitz constant $\lambda = D$, and in particular, uniformly continuous.
\end{lem}

\begin{proof} Fix $p, q \in M$. Let $\g : [a,b] \to M$ be any piecewise smooth curve from $p = \g(a)$ to $q = \g(b)$. Using the fact that $\omega^R_\tau \ge 0$ as in Lemma \ref{Riem_basic_props}, and then the estimate on the norm of the gradient, we have:
\begin{align*}
L^R_{\tau}(\g) & \; = \; \int_a^b \bigg( \dfrac{2}{\; \|\nabla \tau \|_g^2}\, [d\tau(\g'(u))]^2 + g(\g'(u), \g'(u)) \bigg)^{1/2} \, du \\[1pc]
 & \; \ge \; \int_a^b  \dfrac{1}{\; \|\nabla \tau \|_g}\, |d\tau(\g'(u))|du \\[1pc]
& \; \ge \; \dfrac{1}{D} \cdot \int_a^b |d\tau(\g'(u))|\, du \; \ge \; \dfrac{1}{D} \, |\tau(q) - \tau(p)|
\end{align*}

\vspace{.5pc}
\noindent
Taking an infimum then gives the result.
\end{proof}

\vspace{2pc}
\begin{lem} [Riemannianized Distance on Causal Pairs] \label{Riem_dist_causal_pairs} Let $(M,g)$ be a spacetime. Let $\tau$ be a smooth time function on $M$, with timelike gradient $\nabla \tau$, and consider the Riemannianized metric $g^R_\tau$, and its induced Riemannian distance function $d^R_\tau$. If the norm of $\nabla \tau$ is bounded away from zero by some positive constant, $0 < C \le \|\nabla \tau\|_g$, then we have:
$$p \le q \; \; \implies \; \; d^R_\tau(p,q) \; \le \; \dfrac{\sqrt{2}}{C} \cdot (\tau(q) - \tau(p))$$
\end{lem}

\begin{proof} Fix $p,q \in M$, with $p \le q$. Let $\g = \g(u)$, $a \le u \le b$, be a future causal curve from $p = \g(a)$ to $q = \g(b)$. Since $\g$ is causal, we have $g(\g'(u)), \g'(u)) \le 0$, for all $u \in [a,b]$. Since $\tau$ is a smooth time function, we have $d \tau (\g'(u)) > 0$, for all $u \in [a,b]$. Thus we have:
\begin{align*}
L^R_{\tau}(\g) & \; = \; \int_a^b \bigg( \dfrac{2}{\; \|\nabla \tau \|_g^2}\, [d\tau(\g'(u))]^2 + g(\g'(u), \g'(u)) \bigg)^{1/2} \, du \\[1pc]
 & \; \le \; \int_a^b \bigg( \dfrac{2}{C^2} \, [d\tau(\g'(u))]^2\bigg)^{1/2} \, du \\[1pc]
 & \; = \; \dfrac{\sqrt{2}}{C} \, \int_a^b d\tau(\g'(u)) \, du \; = \; \dfrac{\sqrt{2}}{C} \, (\tau(q) - \tau(p))
\end{align*}
\end{proof}

\vspace{1pc}
\subsection{Null Distance and (Causally) Anti-Lipschitz Functions} \label{nulldistsec}

\begin{figure}[h]
\begin{center}
\includegraphics[width=12cm]{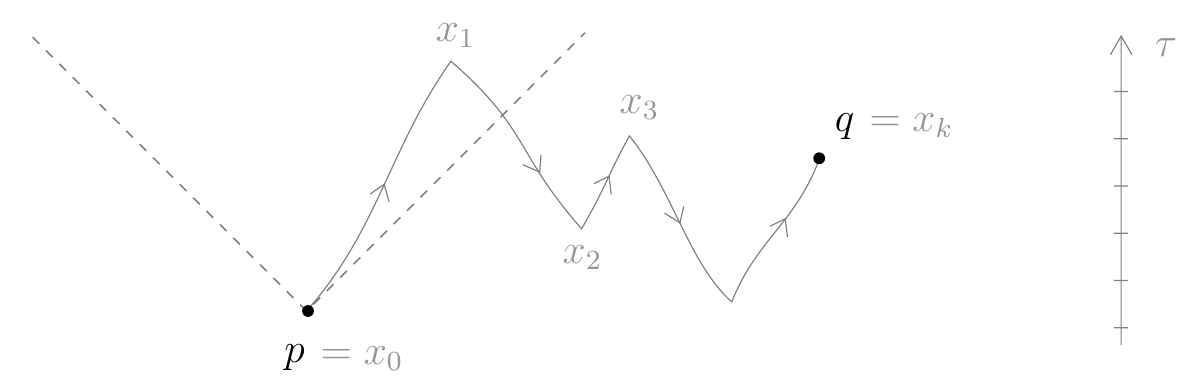}
\caption[]{Measuring the null distance from $p$ to $q$.}
\label{pwcfig}
\end{center}
\end{figure}

\vspace{1pc}
A different way to convert a spacetime into a metric space was introduced in \cite{nulldist}, which we now recall. We start with a particular class of curves:

\begin{Def} [Piecewise Causal Curves, \cite{nulldist}] \label{Def_pwc} Let $(M,g)$ be a spacetime. By a \emph{piecewise causal curve (p.w.c.)} in $M$ we mean a path $\b : [a,b] \to M$, with finitely many breaks $a = s_0 < s_1 < \cdots < s_m = b$, such that each restriction $\b_i := \b|[s_{i-1},s_{i}]$ is a smooth curve which is either future-directed causal, or past-directed causal. Hence, a piecewise causal curve may wiggle backwards and forwards in time, as in Figure \ref{pwcfig}. We will also sometimes write either $\b = \b_1 \cdot \b_2 \cdot \, \cdots \, \cdot \b_m$, or $\b = \b_1 + \b_2 + \cdots + \b_m$. We  count trivial (constant) curves as causal and thus also piecewise causal. 
\end{Def}

\vspace{0pc}
\begin{Def} [Null Length and Distance induced by Time Function, \cite{nulldist}] \label{Def_null_dist} Let $(M,g)$ be a spacetime. Suppose that $\tau : M \to \field{R}$ is a generalized time function on $M$. The corresponding \emph{null length} of a piecewise causal curve $\b : [a,b] \to M$, with break points $x_i = \beta(s_i)$, $0 \le i \le m$, is defined by
$$\hat{L}_\tau(\b) := \sum_{i\, =\, 1}^{m} |\tau(x_{i}) - \tau(x_{i-1})|$$
and the corresponding \emph{null distance function} $\hat{d}_\tau$ on $M$ is defined by
$$\hat{d}_\tau(p,q) := \inf \{ \; \hat{L}_\tau (\b) : \b \textrm{\; is piecewise causal from \,} p \textrm{\, to \,} q \; \}$$
\end{Def}

\vspace{1pc}

\vspace{2pc}
On a given spacetime, $(M,g)$, the null distance function depends highly on the choice of time function $\tau$, as we will see below. When the choice of time function $\tau$ is clear, we may sometimes write simply $\hat{d}$. On the other hand, note that null distance also depends on the spacetime metric tensor, $g$, since it is defined in terms of the collection of piecewise \emph{causal} curves. We may thus write the null distance function in various ways, depending on which quantities we want to indicate explicitly. In addition, we will sometimes write the null distance $\hat{d}$ as $d^N$, as compared to the Riemannianized distance, $d^R$, and similarly $\hat{L} = L^N$, and so on. For example, we may write:
$$\hat{d} = \hat{d}_\tau = \hat{d}(\tau, g) = d^N(\tau, g) = d^N_\tau = d^N$$

\vspace{1pc}
\begin{rmk} [Zig Zag] \label{Lemma_zigzag} A minor point, but note that piecewise causal curves can always be thought of as \emph{alternating} between future-directed and past-directed, as in Figure \ref{pwcfig}, provided we allow the alternating segments to be \emph{piecewise} smooth. 
\end{rmk}

\vspace{1pc}
\begin{exm} [$\tau = t$ on Minkowski Space, \cite{nulldist}] \label{t_on_Mink_example} For the simplest concrete example, consider Minkowski space, $\field{M}^{n+1} = \{(t,{\bf x}) : t \in \field{R}, {\bf x} \in \field{R}^n\}$, together with the standard time function, $\tau(t,{\bf x}) = t$. The induced null distance was computed in \cite{nulldist} and is given by:
$$\hat{d}_t((t_1,{\bf x}_1), (t_2, {\bf x}_2))   \,=  \,\max\{ \, |t_2-t_1|  \, ,  \, \|{\bf x}_2-{\bf x}_1\|  \, \}$$
In particular, the null distance spheres are coordinate cylinders, with axis in the time direction. We revisit this example in Example \ref{exMinkowski} below.
\end{exm}

\vspace{1pc}
Indeed, Example \ref{t_on_Mink_example} was generalized to arbitrary product manifolds in Lemma 4.4 in \cite{nulldistprops}. Using Lemma 3.27 in \cite{nulldist} gives the form presented below.

\begin{lem} [$\tau = t$ on Product Spacetime, \cite{nulldistprops}] \label{t_on_product} Consider a product spacetime, $(M,g) = (I \times S, -dt^2 + h)$, where $I \subset \field{R}$ is an open interval, and $(S,h)$ is a connected Riemannian manifold. Let $d_S$ denote the Riemannian distance function of $(S,h)$. Then using the standard time function $\tau(t, x) = t$ on $M$, the induced null distance between any two points $p = (t_p,p_S)$ and $q = (t_q,q_S)$ is given by:
$$\hat{d}_t(p,q)  \, = \, \max\{\, |t_q - t_p| \, , \, d_S(p_S,q_S) \, \}$$
\end{lem}

\vspace{2pc}
For an arbitrary spacetime $M$ and generalized time function $\tau$, it is easy to see that $\hat{d}_\tau$ is symmetric and satisfies the triangle inequality. What is not so clear is definiteness, and in fact this may fail, depending on the choice of time function $\tau$, (even on Minkowski space), as we now review. What follows is the basic argument given in Proposition 3.4 of \cite{nulldist}. It makes use of what we will here call a `null sawtooth lift', which we will also generalize to GRW spacetimes below. We shall separate the ingredients a bit as follows.

\vspace{1pc}
\begin{exm} [Null Sawtooth Lifts in Minkowski, \cite{nulldist}] \label{example_sawtooth_Mink} Consider the Minkowski plane, $\field{M}^{1+1} = \{(t,x) : t, x \in \field{R} \}$. Fix any time $t = c \in \field{R}$, and any (spatial) length $L > 0$. Consider the `spatial' curve $\s(s) = s$, $0 \le s \le L$, which simply runs from $x = 0$ to $x = L$ on the real line, at unit speed. We can lift this to a `\emph{null sawtooth}' curve $\b$ in the spacetime $\field{M}^{1+1}$, as in Figure \ref{sawtooth0fig}, as follows. For any $k \in \field{N}$, divide $L$ into $2k$ equal pieces of length $\ell = L/(2k)$. Define $\beta_{2j+1}(s) = (c + s, 2j \ell + s)$, $0 \le s \le \ell$, and $\beta_{2j}(s) = (c + \ell - s, (2j-1)\ell + s)$, $0 \le s \le \ell$. Finally, let $\beta$ be the concatenation $\beta = \beta_1 + \b_2 + \cdots + \beta_{2k}$. Note that $\b$ is piecewise null, and zigzags between the two time slices $t = c$ and $t = c + \ell$.

\vspace{2pc}
\begin{figure}[h]
\begin{center}
\includegraphics[width=14cm]{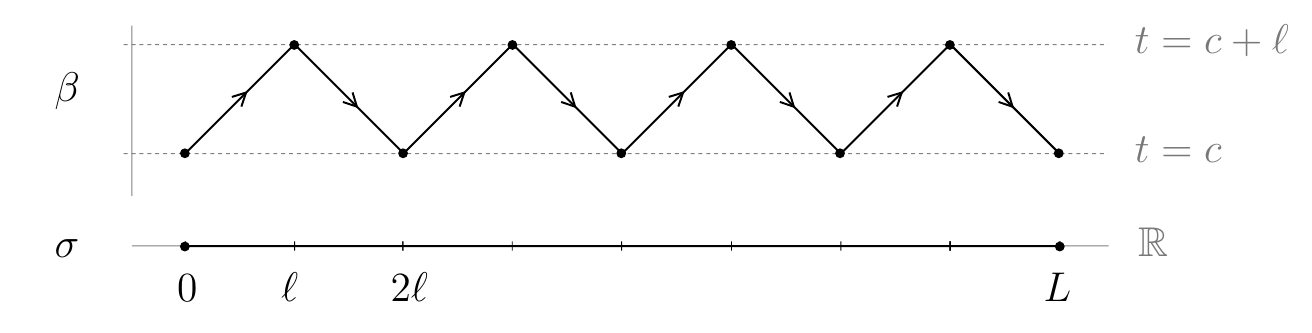}
\vspace{0pc}
\caption[]{A spatial curve $\s$ of length $L$ in $\field{R}$ and its null sawtooth lift $\beta$ to the slab $[c, c + \ell] \times \field{R}$ in the Minkowski plane, as in Example \ref{example_sawtooth_Mink}.}
\label{sawtooth0fig}
\end{center}
\end{figure}
\end{exm}

\vspace{1pc}
The following was established in the proof of Proposition 3.4 in \cite{nulldist}, in a special case. We include the general statement and proof here for completeness and context. This result is generalized to GRW spacetimes in Corollary \ref{nulldistinslice} below.  

\begin{lem} [Null Distance on a Time Slice in Minkowski, \cite{nulldist}] \label{Lemma_Null_on_slice_Mink} Consider any differentiable time function on the Minkowski plane which depends only on the $t$ coordinate, $\tau(t,x) = \phi(t)$. For any two points $p = (c, x_0)$ and $q = (c,x_0 + L)$, in the same time slice $\{t = c\}$, with $L > 0$, we have:
$$\hat{d}_\tau(p,q) \; \le \; \phi'(c) \cdot L$$
\end{lem}

\begin{proof} Let $\beta$ be a null sawtooth lift of $\s(s) = x_0 + s$, $0 \le s \le L$, from $p$ to $q$, as in Example \ref{example_sawtooth_Mink}, and Figure \ref{sawtooth0fig}, which zigzags between the $t = c$ and $t = c + \ell$ slice, where $\ell = L/(2k)$. Then, since $2k = L/\ell$, the null length of $\b$ with respect to $\tau$ is given by
$$\hat{L}_\tau(\beta)  \; = \; ( \, \phi(c + \ell) - \phi(c) \, )\cdot 2k \;  = \; \bigg( \dfrac{ \phi(c + \ell) - \phi(c) }{\ell}\bigg)  \cdot L$$
Taking $k \to \infty$, we have $\ell \to 0$, which gives the estimate
$$\hat{d}_\tau(p,q) \; \le \; \phi'(c) \cdot L$$
\end{proof}

\vspace{1pc}
\begin{exm} [$\tau = t^3$ on Minkowski, \cite{nulldist}] \label{Example_t_cubed} Consider the function $\tau(t,x) = t^3$ on the Minkowski plane. Here $\tau(t,x) = \phi(t) = t^3$ and of course $\phi'(0) = 0$. By Lemma \ref{Lemma_Null_on_slice_Mink}, $\hat{d}_\tau(p,q) = 0$ for all points $p$ and $q$ on the slice $\{t = 0\}$. In particular, definiteness fails.
\end{exm}

\vspace{1pc}
In more general terms, the issue in Example \ref{Example_t_cubed} is that the gradient, $\nabla \tau = -3t^2 \d_t$, does not remain timelike, and indeed vanishes on the entire slice $t = 0$. In light of Lemma \ref{Lemma_Null_on_slice_Mink}, it seems that for smooth time functions $\tau$, we want to keep $\nabla \tau$ timelike to ensure definiteness of $\hat{d}_\tau$. But the null distance construction does not require $\tau$ to be smooth, or even continuous. There is, however, a natural way in fact to quantify the idea of `timelike gradient' for generalized time functions, and it turns out to completely encode the definiteness of null distance. We review this next.

\begin{Def} [(Causally) Anti-Lipschitz Functions, \cite{nulldist}] \label{Def_antiLip} Let $(M,g)$ be a spacetime, and $f : M \to \field{R}$. Given a subset $U \subset M$, we say that $f$ is \emph{(causally) anti-Lipschitz on $U$} if there is some (definite) distance function $d_U$ on $U$ such that, for all $x, y \in U$, we have:
$$x \le y \; \Longrightarrow \; d_U(x,y) \le f(y) - f(x)$$
Roughly, the condition means that, within $U$, the `causal difference quotients' of $f$ are locally bounded away from zero from below. We say $f : M \to \field{R}$ is \emph{locally anti-Lipschitz} if $f$ is anti-Lipschitz on a neighborhood $U$ of each point $p \in M$. This is a `weak version' of having everywhere timelike gradient. We say simply that $f : M \to \field{R}$ is \emph{(globally) anti-Lipschitz} if $f$ is anti-Lipschitz on $U = M$. 
\end{Def}

\vspace{1pc}
\begin{rmk} [\emph{Causally} Anti-Lipschitz] The qualifier `\emph{causally}' was not used in the language in \cite{nulldist}. It is, however, more descriptive and more precise. But it is common to drop such qualifiers in Lorentzian geometry, as in the `reverse triangle inequality', which would, for example, more accurately be described as the `\emph{reverse causal triangle inequality}'. We will continue this practice here, and will mostly omit this qualifier below.
\end{rmk}

\vspace{1pc}
The following is immediate from the definitions:

\begin{lem} Let $(M,g)$ be a spacetime. Then any function $f : M \to \field{R}$ which is locally anti-Lipschitz on $M$ is necessarily a generalized time function on $M$. 
\end{lem}

\vspace{.5pc}
\begin{exm} [$\tau = t$ is Anti-Lipschitz on Minkowski] \label{t_on_Mink_antiLip_example} Consider again Minkowski space, $\field{M}^{n+1} = \{(t,{\bf x}) : t \in \field{R}, {\bf x} \in \field{R}^n\}$, and $\tau(t,{\bf x}) = t$.
There are many ways to see that $\tau = t$ is (globally) anti-Lipschitz on $\field{M}^{n+1}$. Perhaps the most natural way to do this from scratch is as follows. For two points $p = (t_1, {\bf x}_1)$ and $q = (t_2, {\bf x}_2)$,
\begin{align*}
p & \; \le \; q \\[1pc]
\Longrightarrow \hspace{2pc}  t_2 - t_1 & \; \ge \; \| {\bf x}_2 - {\bf x}_1\| \\[1pc]
\Longrightarrow \hspace{2pc}  |t_2 - t_1|^2 & \; \ge \; \| {\bf x}_2 - {\bf x}_1\|^2 \\[1pc]
\Longrightarrow \hspace{2pc} 2 \, |t_2 - t_1|^2 & \; \ge \; |t_2 - t_1|^2 + \| {\bf x}_2 - {\bf x}_1\|^2 \\[1pc]
\Longrightarrow \hspace{2pc} \sqrt{2} \, |t_2 - t_1| & \; \ge \; \sqrt{|t_2 - t_1|^2 + \| {\bf x}_2 - {\bf x}_1\|^2 }\\[1pc]
\Longrightarrow \hspace{2pc} \tau(q) - \tau(p) &\;  \ge \; \frac{1}{\sqrt{2}} \, d_{\field{E}^{n+1}}(p,q)
\end{align*}

\vspace{.5pc}
\noindent
Thus, $\tau(t, {\bf x}) = t$ is (globally) anti-Lipschitz on $U = \field{M}^{n+1}$, using the distance function $d_U := (1/\sqrt{2})d_{\field{E}^{n+1}}$, where $d_{\field{E}^{n+1}}$ is the standard Euclidean distance function of $\field{R}^{n+1} = \{(t,{\bf x}) : t \in \field{R}, {\bf x} \in \field{R}^{n}\}$. 
\end{exm}

\vspace{2pc}
We now collect some general properties established in \cite{nulldist}:

\begin{thm} [Properties of Null Distance, \cite{nulldist}] \label{null_dist_properties_thm} Let $(M,g)$ be a spacetime, and $\tau : M \to \field{R}$ a generalized time function. Let $\hat{d}_\tau$ denote the induced null distance function.

\ben
\item [(1)] $\hat{d}_\tau$ is a pseudometric in general. It is symmetric and satisfies the triangle inequality, but it may be indefinite. 
\item [(2)] Keeping $\tau$ fixed, $\hat{d}_\tau$ is invariant under conformal scaling of the spacetime metric $g$. That is, for any smooth positive function $f: M \to (0, \infty)$,  
$$\hat{d}(\tau, fg) = \hat{d}(\tau, g)$$
\item [(3)] Definiteness can only fail for points in the same $\tau$ level set, a consequence of the fact that for all points $p, q \in M$, we have:
$$\hat{d}_\tau(p,q) \ge |\tau(q) - \tau(p)|$$
\item [(4)] $\hat{d}_\tau$ is definite iff $\tau$ is locally anti-Lipschitz. Moreover, in this case, $\hat{d}_\tau$ induces the manifold topology iff $\tau$ is continuous (wrt the manifold toplogy).
\een
\end{thm}

\vspace{2pc} 
For easier reference, we have also separated off the following:

\begin{prop} [Null Distance on Causal Pairs, \cite{nulldist}] \label{null_dist_causal_pairs} Let $(M,g)$ be a spacetime, and $\tau : M \to \field{R}$ a generalized time function. Let $\hat{d}_\tau$ denote the induced null distance function. Causal curves are minimal, and for all $p, q \in M$,
$$p \le q \; \implies \hat{d}_\tau(p,q) = \tau(q) - \tau(p)$$
\end{prop}

\vspace{2pc}
We note the following consequence of Corollary 4.16 in \cite{nulldist}:

\begin{cor} \label{timelikegrad} If $\tau : M \to \field{R}$ is a smooth time function with everywhere timelike gradient, $\nabla \tau$, then $\tau$ is locally anti-Lipschitz. In particular, $\tau$ induces a definite null distance function, $\hat{d}_\tau$.
\end{cor}

\vspace{2pc}
We consider a very minor modification of Example 3.9 in \cite{nulldistprops}, which nicely illustrates property (4) in Theorem \ref{null_dist_properties_thm}. 

\begin{exm} (Anti-Lipschitz but Discontinuous $\tau$, c.f. \cite{nulldistprops}) \label{antiLip_but_discont} Consider Minkowski space, $\field{M}^{n+1} = \{(t,{\bf x}) : t \in \field{R}, {\bf x} \in \field{R}^n\}$, and the function
\begin{displaymath}
   \tau(t,{\bf x}) := \left\{
     \begin{array}{lr}
       t - 1  & t  < 0\\[.25pc]
       t  & t  = 0\\[.25pc]
       t+0.9  \hspace{1.5pc} &  t > 0
     \end{array}
   \right.
\end{displaymath} 
Since $t$ strictly increases on future causal curves, it follows that so does $\tau$. So $\tau$ is generalized time function. But $\tau$ is not a time function, since it is clearly discontinuous. By considering cases, an argument as in Example \ref{t_on_Mink_antiLip_example} shows that $\tau$ is (globally) anti-Lipschitz, with respect to the same distance function $d = (1/\sqrt{2})d_{\field{R}^{n+1}}$. Thus, $\hat{d}_\tau$ is definite, and $(M, \hat{d}_\tau)$ is a metric space. But the topology induced by $\hat{d}_\tau$ is not the same as the manifold topology. To explore this example in more detail, we now display the null distance function $\hat{d}_\tau$.

\vspace{1pc}
First note that we have $\hat{d}_\tau(p,p) = 0$. But note that this now follows critically from the fact that we count trivial curves as piecewise causal. For $p_0 \in \{t = 0\}$, the infimum of the null lengths of \emph{nontrivial} piecewise causal curves from $p_0$ to $p_0$ does not give zero. Indeed, for $p_0 \in \{t = 0\}$, we have:
$$\inf \, \{ \, \hat{L}_\tau(\b) : \b \textrm{ is a \emph{nontrivial} piecewise causal curve from } p_0 \textrm{ to } p_0 \, \} \; = \; 1.8$$ 
Now for $q \in M$, $q \ne p$, write $p = (t_1, {\bf x}_1) = (t_p,p_S)$ and $q = (t_2, {\bf x}_2) = (t_q, q_S)$. Then for $t_p \le t_q$, we have:
\begin{displaymath}
   \hat{d}_\tau(p,q) = \left\{
     \begin{array}{lr}
       \hat{d}_t(p,q) \hspace{1pc} & t_q < 0\\[.75pc]
       \hat{d}_t(p,q) + 1  & t_p < 0 = t_q\\[.75pc]
       \hat{d}_t(p,q) + 1.8 \hspace{2pc} & t_p = 0 = t_q\\[.75pc]
       \hat{d}_t(p,q) + 0.9 \hspace{1pc} & t_p = 0 < t_q\\[.75pc]
       \hat{d}_t(p,q) \hspace{1pc} & 0 <  t_p
     \end{array}
   \right.
\end{displaymath} 

\vspace{.25pc}
\noindent
In the case $t_p = 0 = t_q$, note that we have $\hat{d}_\tau(p,q) = \hat{d}_t(p,q) + 1.8 = \|{\bf x}_2 - {\bf x}_1\| + 1.8$, for $p \ne q$. In particular, the metric space $(M, \hat{d}_\tau)$ is disconnected. The subset $\{ t < 0 \}$ is connected, and so is $\{ t > 0 \}$. The subset $\{ t = 0\}$ is totally disconnected. 
\end{exm}

\vspace{2pc}
The following is immediate from property (3) in Theorem \ref{null_dist_properties_thm}:

\begin{cor} [Null Distance Lipschitz Continuity] \label{null_dist_Lipschitz} When $\hat{d}_\tau$ is definite, then as a map between metric spaces, $\tau : (M, \hat{d}_\tau) \to (\field{R}, |\cdot |)$ is Lipschitz, with Lipschitz constant 1. In particular, this map is uniformly continuous. (But see Remark \ref{nulldist_topologies} below.)
\end{cor} 

\vspace{0pc}
\begin{rmk} [Null Distance Topologies] \label{nulldist_topologies} We caution the reader to keep in mind that when $\hat{d}_\tau$ is definite, it induces a (Hausdorff) topology on $M$, but this topology coincides with the manifold topology if and only if $\tau$ is continuous (with respect to the manifold topology). See Example \ref{antiLip_but_discont} above.
\end{rmk}

\vspace{2pc}
We now introduce a variation of the anti-Lipschitz condition above. 

\begin{Def} [(Causally) Anti-Lipschitz with Constant] \label{lambda_antiLip} Let $(M,g)$ be a spacetime. Fix any (definite) distance function $d$ on $M$. We will say that a function $f : M \to \field{R}$ is \emph{(causally) anti-Lipschitz (on $M$) with respect to $d$, with anti-Lipschitz constant $\lambda$}, or more simply, \emph{$\lambda$-anti-Lipschitz with respect to $d$}, if there is a positive constant $\lambda > 0$, such that for all $p, q \in M$, we have:
$$p \le q \; \; \Longrightarrow \; \; \lambda \, d(p,q) \, \le \, f(q) - f(p)$$
\end{Def}

\vspace{1pc}
Lemma \ref{Riem_dist_causal_pairs} may thus be rephrased as:

\begin{cor} [Anti-Lipschitz with respect to Riemannianized Distance] \label{Riem_antiLip} Let $(M,g)$ be a spacetime. Let $\tau$ be a smooth time function on $M$, with timelike gradient $\nabla \tau$. If the norm of $\nabla \tau$ is bounded away from zero by some positive constant, $0 < C \le \|\nabla \tau\|_g$, then $\tau$ is anti-Lipschitz on $M$ with respect to the Riemannianized distance function $d^R_\tau$, with anti-Lipschitz constant $\lambda = C/\sqrt{2}$, 
$$p \le q \; \; \implies \; \; \dfrac{C}{\sqrt{2}} \cdot d^R_\tau(p,q) \; \le \; \tau(q) - \tau(p)$$
\end{cor}

\vspace{1pc}
Similarly, Proposition \ref{null_dist_causal_pairs} gives:

\begin{cor} [Anti-Lipschitz with respect to Null Distance] \label{null_antiLip} Let $(M,g)$ be a spacetime. Let $\tau$ be a generalized time function on $M$. If the induced null distance function $\hat{d}_\tau$ is definite, then $\tau$ is anti-Lipschitz on $M$ with respect to $\hat{d}_\tau$, with anti-Lipschitz constant $\lambda = 1$, since we have, in fact, that:
$$p \le q \; \; \implies \; \; \hat{d}_\tau(p,q) \; = \; \tau(q) - \tau(p)$$
\end{cor}

\vspace{1pc}
\begin{rmk} [Ambiguity in Anti-Lipschitz Terminology] We have introduced some ambiguity in our terminology for anti-Lipschitz functions. To be concrete, consider again the standard time function $\tau(t,{\bf x}) = t$ on Minkowski space, $\field{M}^{n+1}$. In Example \ref{t_on_Mink_antiLip_example} we showed that 
$$p \le q \; \; \Longrightarrow \; \; \dfrac{1}{\sqrt{2}} \cdot d_{\field{E}^{n+1}}(p,q) \; \le \; \tau(q) - \tau(p)$$
In the original language of \cite{nulldist}, as in Definition \ref{Def_antiLip} above, we would then simply say that $\tau$ is (globally) anti-Lipschitz, or perhaps that $\tau$ is (globally) anti-Lipschitz with respect to $d = (1/\sqrt{2})d_{\field{E}^{n+1}}$. In terms of the new Definition \ref{lambda_antiLip}, we could say either, that $\tau$ is anti-Lipschitz with respect to $(1/\sqrt{2})d_{\field{E}^{n+1}}$, with anti-Lipschitz constant $\lambda = 1$, or alternatively, that $\tau$ is anti-Lipschitz with respect to $d_{\field{E}^{n+1}}$, with anti-Lipschitz constant $\lambda =1/\sqrt{2}$. Effort will be taken below to avoid any consequential ambiguities. In particular, the more explicit language in Definition \ref{lambda_antiLip} should always be clear.
\end{rmk}

\vspace{2pc}
Lipschitz functions take Cauchy sequences to Cauchy sequences, as noted in Lemma \ref{metric_space_maps}. We now give a version of this fact for (causally) anti-Lipschitz functions. 

\vspace{0pc}
\begin{Def} [Causal Sequences] \label{Def_causal_seq} Let $(M,g)$ be a spacetime. Let $\{p_k\}$ be a sequence in $M$. We will say that $\{p_k\}$ is \emph{future causal} if $p_k \le p_{k+1}$, for all $k$. Time dually, we will say that $\{p_k\}$ is \emph{past causal} if $p_{k+1} \le p_k$, for all $k$. We will say that $\{p_k\}$ is \emph{causal} if it is either future causal, or past causal.
\end{Def}

\vspace{0pc}
\begin{lem} [Anti-Lipschitz Functions and Cauchy Sequences] \label{antiLip_Cauchy} Let $(M,g)$ be a spacetime, and fix a distance function $d$ on $M$. Suppose that $\tau : M \to \field{R}$ is $\l$-anti-Lipschitz with respect to $d$, and thus a generalized time function on $M$. Suppose that $\{p_k\}$ is a causal sequence in $M$, (as in Definition \ref{Def_causal_seq}), such that $\{\tau(p_k)\}$ converges to a finite value $\ell \in \field{R}$. Then $\{p_k\}$ is a Cauchy sequence in $(M,d)$. Moreover, let $(\overline{M}, \overline{d})$ be the metric completion of $(M,d)$. If $\tau : M \to \field{R}$ is continuous, and admits a continuous extension $\overline{\tau} : \overline{M} \to \field{R}$, then we have $[\{p_k\}] \in (\overline{\tau})^{-1}(\ell) \subset \overline{M}$.
\end{lem}

\begin{proof} The $\l$-anti-Lipschitz condition means that $ \lambda \cdot d(p,q) \le \tau(q) - \tau(p)$, for all $p \le q$. Suppose first that $\{p_k\}$ is a past causal sequence in $M$, with $p_j \le p_i$, for all $i \le j$. Fix any $\e > 0$. Since $\{\tau(p_k)\}$ converges in $\field{R}$, it is a Cauchy sequence in $(\field{R}, |\cdot |)$. Thus, there is a $m_0 \in \field{N}$ such that $|\tau(p_i) - \tau(p_j)| < \lambda \e $, for all $i, j \ge m_0$. Then for all $m_0 \le i \le j$, we have $d(p_i,p_j) = d(p_j,p_i) \le (\tau(p_i) - \tau(p_j))/\lambda  < \e$. Hence, $\{p_k\}$ is Cauchy in $(M,d)$. Moreover, if $\tau$ is continuous, and admits a continuous extension $\overline{\tau}$, then as in Proposition \ref{extendfunctionprop}, we have:
$$\overline{\tau}([\{p_k\}]) = \lim_{ k \, \to \, \infty} \tau(p_k) = \ell$$
A similar argument works when $\{p_k\}$ is future causal. 
\end{proof}

\vspace{2pc}
The following is only a slight modification of Lemma 2.1 in \cite{nulldistprops}, and compares null distances under time reparameterization.

\begin{lem} [Null Distance Under Time Reparameterization, c.f. \cite{nulldistprops}] \label{null_reparam} Let $(M,g)$ be a spacetime, and suppose that for two differentiable time functions $\tau_1, \tau_2 : M \to \field{R}$, there is an increasing differentiable function $\phi : I \to J$, where $I$ and $J$ are open intervals in $\field{R}$, such that $\tau_2 = \phi \circ \tau_1$. Define $(\phi')_{\inf} := \inf\{\phi'(t) : t \in I\} \ge 0$ and $(\phi')_{\sup} := \sup\{\phi'(t) : t \in I\} \le \infty$. Then for all $p, q \in M$, we have:
$$(\phi')_{\inf} \cdot \hat{d}_{\tau_1}(p,q) \; \le \; \hat{d}_{\tau_2}(p,q) \; \le \; (\phi')_{\sup} \cdot \hat{d}_{\tau_1}(p,q)$$
\end{lem}

\begin{proof} We simply mimic the proof of Lemma 2.1 in \cite{nulldistprops}. First consider two points $p, q \in M$, with $p \le q$, $p \ne q$, and let $\beta_0$ be a future causal curve from $p$ to $q$. Let $\hat{L}_{\tau_i}$ denote the null length with respect to $\tau_i$. Then by the Mean Value Theorem, there is a number $\tau_1(p) < u_0 < \tau_1(q)$ such that:
\begin{align*}
\hat{L}_{\tau_2}(\beta_0) = \tau_2(q) - \tau_2(p) & = \phi(\tau_1(q)) - \phi(\tau_1(p)) \\[.5pc]
& = \phi'(u_0) (\tau_1(q) - \tau_1(p)) = \phi'(u_0) \hat{L}_{\tau_1}(\beta_0)
\end{align*}
For any piecewise causal curve $\beta$, we then get the following estimate, from which the estimate for the null distances follows:
$$(\phi')_{\inf} \cdot \hat{L}_{\tau_1}(\beta) \; \le \; \hat{L}_{\tau_2}(\beta) \; \le \; (\phi')_{\sup} \cdot \hat{L}_{\tau_1}(\beta)$$
\end{proof}

\vspace{2pc} 
Finally, note that on a fixed spacetime $(M,g)$, a choice of generalized time function $\tau$, fully determines the corresponding null distance function $\hat{d}_\tau$, by definition. The following observation, similar to the considerations in Lemma 3.17 in \cite{nulldist}, is that, conversely, $\tau$ can be determined from $\hat{d}_\tau$, up to a constant:

\begin{lem} [Time Function From Null Distance] \label{time_from_null_dist} Let $(M,g)$ be a spacetime, and $\tau$ any generalized time function on $M$. Then $\tau$ is determined, up to a constant, by the induced null distance function $\hat{d}_\tau$. 
\end{lem} 

\begin{proof} Fix any reference point $p_0 \in M$, and set $\tau_0 := \tau(p_0)$. Let $q \in M$ be any other point. By Lemma 3.5 in \cite{nulldist}, there is a piecewise causal curve $\b = \b_1 + \b_2 + \cdots + \b_m$ from $p_0$ to $q$, with break points $p_0 = x_0$, $x_1$, ..., $x_m = q$, where $\b_i$ is either future causal or past causal, and goes from $x_{i-1}$ to $x_i$. Consider first $\b_1$. If $\b_1$ is future causal, then we have $p_0 = x_0 \le x_1$, and thus $\hat{d}_\tau(p_0,x_1) = \tau(x_1) - \tau_0$. If instead $\b_1$ is past causal, then we have $\hat{d}_\tau(p_0,x_1) = \tau_0 - \tau(x_1)$. That is, we have:
$$\tau(x_1) \; = \; \tau_0 \, \pm \, \hat{d}_\tau(p_0,x_1)$$
where we take the $+$ sign if $\b_1$ is future causal, or the $-$ sign if $\b_1$ is past causal. Similarly, we will get $\tau(x_2) \; = \; \tau(x_1) \, \pm \, \hat{d}_\tau(p_1,x_2)$, where we take the $+$ sign if $\b_2$ is future causal, or the $-$ sign if $\b_2$ is past causal. For $i \in \{1, ..., m\}$, define $c_i = 1$ if $\b_i$ is future causal, and $c_i = -1$ if $\b_i$ past causal. It follows that we have:
$$\tau(q) \; = \; \tau_0 \; + \; \sum_{i \, = \, 1}^m \, c_i \cdot \hat{d}_\tau(x_{i-1},x_i)$$
\end{proof}

\vspace{1pc}

\subsection{Some Comparisons}

\vspace{1pc}
Let $(M,g)$ be a spacetime. If $\tau$ is a suitable time function on $M$, we can consider both, the Riemannianized distance $d^R_\tau$, as well as the null distance $\hat{d}_\tau$. As for how these distances compare, we note the following for now:

\begin{lem} [Riemannianized vs Null Distance: Some Basic Comparisons] \label{null_v_Riem_easy} Let $(M,g)$ be a spacetime, and suppose that $\tau : M \to \field{R}$ is a smooth time function, with $\nabla \tau$ everywhere timelike. Let $L^R_\tau$ and $d^R_\tau$, and $\hat{L}_\tau$ and $\hat{d}_\tau$, denote the corresponding Riemannianized and null lengths and distances, respectively.
\ben
\item [(1)] Let $(\|\nabla \tau\|_g)_{\inf} := \inf \{\|\nabla \tau\|_g(p) : p \in M\}$. Then for all $p, q \in M$, we have
$$(\|\nabla \tau\|_g)_{\inf} \cdot d^R_\tau(p,q) \; \le \; \sqrt{2} \cdot \hat{d}_\tau(p,q)$$
\item [(2)] Suppose that $0 < c \equiv \|\nabla \tau \|_g$ is constant. Then for any piecewise null curve $\beta$, we have the equality:
$$ \hat{L}_\tau(\beta) \; = \; \dfrac{c}{\sqrt{2}} \cdot L^R_\tau(\beta)$$
\een
\end{lem}

\begin{proof} The proof of (1) in the special case $\|\nabla \tau\|_g =1$ was carried out in Lemma 4.1 in \cite{nulldist}. The inequality here holds automatically if $(\|\nabla \tau\|_g)_{\inf} = 0$. Suppose then that $C := (\|\nabla \tau\|_g)_{\inf} > 0$. Fixing any two points $p, q \in M$, let $\b : [a,b] \to M$ be any piecewise causal curve from $p$ to $q$, with break points $x_i = \b(s_i)$, $0 \le i \le m$, where $p = x_0$ and $q = x_m$. For each $1 \le i \le m$, let $\b_i = \b | [s_{i-1},s_i]$ denote the smooth subsegments of $\b$. Then we have:
\begin{align*}
L^R_\tau(\b) \;  = \; \sum_{i \, =\, 1}^m L^R_\tau(\b_i) & \; = \; \sum_{i \, =\, 1}^m \, \int_{s_{i-1}}^{s_i} \, \sqrt{\dfrac{2}{\|\nabla \tau\|^2_g} \, d \tau^2(\b'_i, \b'_i) + g(\b'_i, \b'_i)} \; ds\\[1.25pc]
& \; \le \; \sum_{i \, =\, 1}^m  \, \int_{s_{i-1}}^{s_i} \,\sqrt{\dfrac{2}{\|\nabla \tau\|^2_g} \, d \tau^2(\b'_i, \b'_i)} \; ds\\[1.25pc]
& \; \le \; \dfrac{\sqrt{2}}{C} \, \sum_{i \, =\, 1}^m  \, \int_{s_{i-1}}^{s_i} \,\sqrt{d \tau^2(\b'_i, \b'_i)} \; ds\\[1.25pc]
&\;  = \; \dfrac{\sqrt{2}}{C} \, \sum_{i \, =\, 1}^m  |\tau(x_i) - \tau(x_{i-1})| \; = \; \dfrac{\sqrt{2}}{C} \,\hat{L}_\tau(\b)
\end{align*}

\vspace{.5pc}
\noindent
Since $C \cdot L^R_\tau(\b) \le \sqrt{2} \cdot \hat{L}_\tau(\b)$, this gives $C \cdot d^R_\tau(p,q) \le \sqrt{2} \cdot \hat{L}_\tau(\b)$. Finally, taking the infimum over all piecewise causal curves from $p$ to $q$ gives $C \cdot d^R_\tau(p,q) \le \sqrt{2} \cdot \hat{d}_\tau(p,q)$. The statement in (2) follows similarly.
\end{proof}

\vspace{1pc}

\subsection{Extrinsic vs Intrinsic Measurements} \label{Sec_ext_vs_int}

\begin{figure}[h]
\begin{center}
\includegraphics[width=14cm]{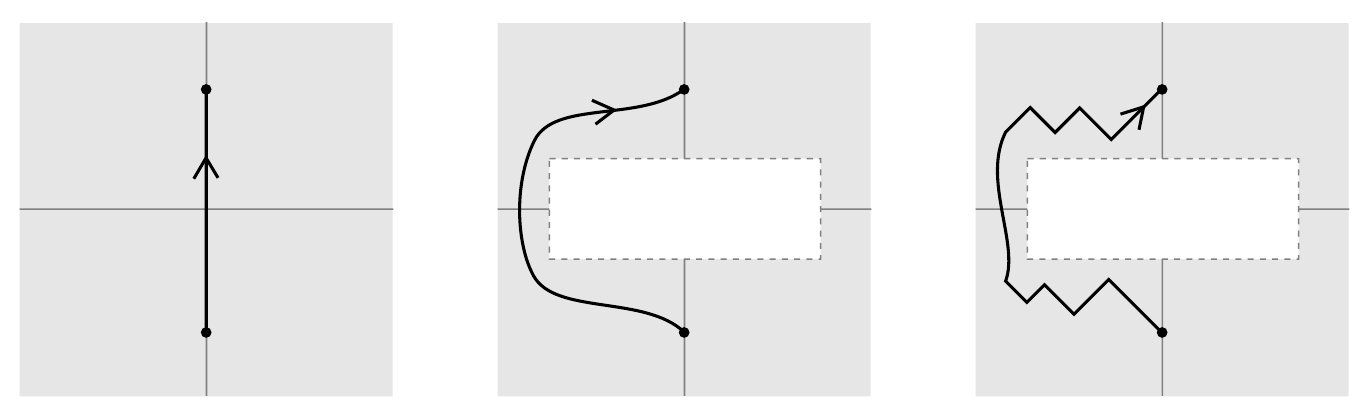}
\caption[]{In general, restricting to a subset reduces the amount of intrinsic curves of any given type, (e.g., piecewise smooth, piecewise causal, etc.).}
\label{rectangle_removed}
\end{center}
\end{figure}

\vspace{.5pc}
We now take a moment to discuss extrinsic vs intrinsic measurements. The basic issue we note here is essentially that `when restricting to a subset, the number of available curves decreases', as in Figure \ref{rectangle_removed}. Note that this is related to Riemannian and Riemannianized distances, null distances, etc.

\vspace{1pc}

\pagebreak
\vspace{1pc}
\begin{exm} [Euclidean Plane with Rectangle Removed] \label{Euclidean_plane_rect_removed} For a simple concrete example, consider the Euclidean plane, $\field{R}^2$. First recall that the standard Euclidean distance function $d_{\field{E}^2}$ on $\field{R}^2$ is given by 
$$d_{\field{E}^2}((x_1,y_1), (x_2,y_2)) = \sqrt{(x_2-x_1)^2 + (y_2-y_1)^2}$$
But recall that $\field{R}^2$ is also a Riemannian manifold, via its standard Riemannian metric tensor $g_{\field{E}^2} = dx^2 + dy^2$, which induces a corresponding Riemannian distance function $d_{g_{\field{E}^2}}$ defined by
$$d_{g_{\field{E}^2}}(p,q) := \inf \{ L_{g_{\field{E}^2}}(\g) : \g \in \Omega_{\field{R}^2}(p,q)\}$$
where $\Omega_{\field{R}^2}(p,q)$ is the set of all piecewise smooth curves in $\field{R}^2$ from $p$ to $q$. As noted in Example \ref{Riem_vs_Euclid_dist} above, these two distance functions coincide:
$$d_{g_{\field{E}^2}} \; = \; d_{\field{E}^2}$$

\vspace{.5pc}
Now consider the closed rectangle $R := [-2,2] \times [-1,1] = \{(x,y) \in \field{R}^2 : -2 \le x \le 2 \, -1 \le y \le 1\}$. We will investigate the compliment, $M := \field{R}^2 - R$, similar to the picture in Figure \ref{rectangle_removed}. Being a subset $M \subset \field{R}^2$, $M$ becomes a metric subspace $(M,d_{\field{E}^2}) = (\field{R}^2,d_{g_{\field{E}^2}})$, via the restriction of the Riemannian distance function $d_{g_{\field{E}^2}}$ of the ambient Riemannian manifold $(\field{R}^2, g_{\field{E}^2})$. That is, the distance between two points $p$ and $q$ in $(M,d_{g_{\field{E}^2}})$ is simply the distance between $p$ and $q$ in $(\field{R}^2, d_{g_{\field{E}^2}})$, which is just the standard Euclidean distance. On the other hand, being an open subset $M \subset \field{R}^2$, $M$ is also a Riemannian manifold itself, via the restriction of the metric tensor $g_{\field{E}^2} = dx^2 + dy^2$ to $M$. Letting $d_M$ denote the corresponding Riemannian distance function of $(M,g_{\field{E}^2})$, note that this is given by
$$d_M(p,q) := \inf \{ L_{g_{\field{E}^2}}(\g) : \g \in \Omega_{M}(p,q)\}$$
where now, critically, $ \Omega_{M}(p,q)$ is the set of all piecewise smooth curves \emph{within} $M$ from $p$ to $q$. Since there are now fewer available connecting curves in the subset $M$, some Riemannian distances will be longer. (Since we must now avoid the rectangular obstacle, it takes longer to walk to certain places.) Thus, we have two distance functions on $M$, the restriction of the Riemannian distance function $d_{g_{\field{E}^2}}$ of the ambient Riemannian manifold $(\field{R}^2, g_{\field{E}^2})$, and the intrinsic Riemannian distance $d_M$ of $(M,g_{\field{E}^2})$ itself, and the key point is that \emph{these two distances functions do not coincide},  
$$d_M \; \ne \; d_{g_{\field{E}^2}} $$
For example, for $p = (0, -2)$ and $q = (0,2)$, we have
$$ d_{g_{\field{E}^2}}(p,q) = d_{\field{E}^2}(p,q) \; = \; 4 \; <  \; 2 + 2\sqrt{5} \; = \; d_M(p,q)$$
\begin{flushright} $\Diamond$\end{flushright}
\end{exm}

\vspace{2pc}
Moving to spacetimes, we first note the following:

\begin{lem} [Time Functions on Subspacetimes] If $(M,g)$ is a spacetime, $\tau : M \to \field{R}$ is a (generalized) time function, and $U \subset M$ is a subspacetime, then the restriction $\tau : U \to \field{R}$ is also a (generalized) time function. 
\end{lem}

\vspace{1pc}
We now consider a spacetime analog of Example \ref{Euclidean_plane_rect_removed}.
\begin{exm} [Minkowski Plane with Rectangle Removed] \label{Mink_plane_rect_removed} Now consider the Minkowski plane, $\field{M}^{1+1} = \{(t,x) : t, x \in \field{R}\}$, with its standard Minkowski metric tensor $g_{\field{M}^{1+1}} = -dt^2 + dx^2$, where as usual, we visualize the $t$-axis running vertically. Consider the closed rectangle $R := [-1,1] \times [-2,2] = \{(t,x) : -1 \le t \le 1, -2 \le x \le 2\}$. We will investigate the compliment $M := \field{M}^{1+1} - R$, which is an open subspacetime of $\field{M}^{1+1}$, similar again to the picture in Figure \ref{rectangle_removed}. Consider the standard time function $\tau(t,x) = t$ on $M$. The Riemannianization of $M$ gives the situation in Example \ref{Euclidean_plane_rect_removed} above. Alternatively, for the same time function $\tau(t,x) = t$, let $\hat{d}_M$ denote the induced null distance function of $M$, and let $\hat{d}_{\field{M}}$ denote the induced null distance function of the full Minkowski plane $\field{M}^{1+1}$. Then, for example, for $p = (0,-2)$ and $q = (0,2)$, it is straightforward to verify that
$$\hat{d}_M(p,q) \; = \; 4 \; < \; 6 \; = \; \hat{d}_{\field{M}}(p,q)$$
\begin{flushright} $\Diamond$\end{flushright}
\end{exm}

\vspace{2pc}
To be able to clearly distinguish between extrinsic and intrinsic measurements, a few preliminaries are in order. As in \cite{burago}, a \emph{length structure} $(M, \mathcal{A}, L)$ is a Hausdorff topological space $M$, a class of `\emph{admissible paths}' $\mathcal{A}$ in $M$, and a `\emph{length (functional)}' $L : \mathcal{A} \to [0, \infty]$. For any subset $W \subset M$, we will set $\mathcal{A}(W)$ to be the subset of paths in $\mathcal{A}$ which lie entirely within $W$, that is,
$$\mathcal{A}(W) := \{\a \in \mathcal{A} : \mathrm{Im}(\a) \subset W\}$$
We have, of course, $\mathcal{A}(M) = \mathcal{A}$. For any subset $W \subset M$, and any two points $p, q \in W$, let $\mathcal{A}_{p,q}(W)$ be the subset of all paths in $\mathcal{A}(W)$ which join $p$ and $q$,
$$\mathcal{A}_{p,q}(W) := \{\a \in \mathcal{A}(W) : \a \textrm{ runs from } p \textrm{ to } q\}$$
The \emph{length metric} $d^L$ associated to $(M, \mathcal{A}, L)$ is then defined for all $p,q \in M$ by
$$d^L(p,q) := \inf \{L(\a) : \a \in \mathcal{A}_{p,q}(M)\}$$
where, if $\mathcal{A}_{p,q}(M)$ is empty, we take $d^L(p,q) := \infty$. 

\vspace{2pc}
First note that this applies to Riemannian/Riemannianized manifolds:

\begin{exm} [Riemannian Manifolds, Riemannianized Spacetimes] Suppose that $(M,g)$ is a Riemannian manifold. Let $\mathcal{A}^{pws}$ be the set of all piecewise smooth curves in $M$, and let $L_g$ be the Riemannian arc length functional of $(M,g)$. Then $(M,\mathcal{A}^{pws}, L_g)$ is a length structure, and its induced length metric is precisely the Riemannian distance function $d_g$. If we begin instead with a spacetime $(M,g)$, then the same applies to any Riemannianization.
\end{exm}

\vspace{1pc}
Theorem 1.1 in \cite{nulldistprops} shows this also applies in the context of null distance:

\begin{thm} [Null Length Structure on a Spacetime, \cite{nulldistprops}] Consider a spacetime $(M,g)$. Let $\hat{\mathcal{A}}$ be the set of all piecewise causal curves in $M$. Fix any generalized time function $\tau : M \to \field{R}$, and let $\hat{L}_\tau$ be the associated null length functional. If $\tau$ is locally anti-Lipschitz, then $(M,\hat{\mathcal{A}}, \hat{L}_\tau)$ is a length structure, and the null distance function $\hat{d}_\tau$ is the induced length metric.
\end{thm}

\vspace{1pc}
Now, if $(M,\mathcal{A}, L)$ is a length structure, and $W \subset M$ is any subset, viewed as a topological subspace, then $(W, \mathcal{A}(W), L)$ is also a length structure, with its own length metric, given for all $p,q \in W$ by
$$d^L_W(p,q) := \inf \{L(\a) : \a \in \mathcal{A}_{p,q}(W)\}$$

\vspace{1pc}
To formally distinguish between extrinsic and intrinsic measurements, we make the following definitions.

\begin{Def} [Extrinsic and Intrinsic Distances Between Points in a Subset] \label{ext_v_int_dist} Let $(M, \mathcal{A}, L)$ be a length structure, with length metric $d^L$. Fix any subset $W \subset M$. Then for any two points $p,q \in W$, there are two ways to measure their distance:

\ben
\item [(1)] We define their \emph{extrinsic distance (in $M$)} to be 
$$d^L_{\mathrm{ext}}(p,q) := \inf \{L(\a) : \a \in \mathcal{A}_{p,q}(M)\} = d^L_M(p,q) = d^L(p,q)$$
\item [(2)] We define their \emph{intrinsic distance (in $W$)} to be  
$$\hspace{-4.5pc} d^L_{\mathrm{int}}(p,q) := \inf \{L(\a) : \a \in \mathcal{A}_{p,q}(W)\} = d^L_W(p,q)$$
\een
\end{Def}

\vspace{1pc}
We now turn attention to measuring the size of subsets. We first recall that, given any metric space $(X,d)$, the \emph{diameter} of a subset $W \subset X$ is defined by
$$\mathrm{diam}(W) = \sup \{d(x,y) : x, y \in W\}$$
But here again, in our context, we will want to distinguish between extrinsic vs intrinsic measurements, and thus we make the following definitions.

\begin{Def} [Extrinsic and Intrinsic Diameters of Subsets] Let $(M, \mathcal{A}, L)$ be a length structure, with length metric $d^L$. Then there are two ways to measure the diameter of a subset $W \subset M$:

\ben
\item [(1)] We define the \emph{extrinsic diameter} of $W$ (in $M$) by
$$\mathrm{diam}^L_{\mathrm{ext}}(W)  :=  \sup \{d^L_M(p,q) : p, q \in W\}  =:  \mathrm{diam}^L_M(W) = \mathrm{diam}^L(W)$$
\item [(2)] We define the \emph{intrinsic diameter} of $W$ by
$$\hspace{-6.5pc} \mathrm{diam}^L_{\mathrm{int}}(W)  := \sup \{d^L_W(p,q) : p, q \in W\} =: \mathrm{diam}^L_W(W)$$
\een
\end{Def}

\vspace{1pc}
Since the number of admissible curves decreases when restricting to a subset, we have the following:

\begin{lem} [Extrinsic vs Intrinsic Measurements] Let $(M, \mathcal{A}, L)$ be a length structure, with length metric $d^L = d^L_M$. Fix any subset $W \subset M$, and let $d^L_W$ denote intrinsic distance in $W$. Fix any two points $p, q \in W$. Then we have:
\ben
\item [(1)] \; $d^L_M(p,q) \, \le \, d^L_W(p,q)$
\item [(2)] \; $\mathrm{diam}^L_{M}(W) \, \le \, \mathrm{diam}^L_{W}(W)$
\een
\end{lem}

\vspace{1pc}

\subsection{Metric Completions}

\vspace{1pc}
We first illustrate the metric completion construction in a basic concrete example.

\begin{exm} [Euclidean Half-Space] \label{Euclidean_halfspace} Consider the Euclidean Half-Space
$$M = \field{R}^{n+1}_+ := (0, \infty) \times \field{R}^n = \{(t,{\bf x}) : t \in (0,\infty) \, , \, {\bf x} \in \field{R}^n\}$$
The representation above is chosen for compatibility with the spacetime context, and we can think of the $t$ factor  as running `vertically'. Let $d_{\field{E}^{n+1}}$ be the standard Euclidean distance function on the full Euclidean space $\field{R}^{n+1}$. Note that the restriction of $d_{\field{E}^{n+1}}$ to the upper half-space $M$ coincides with the intrinsic Riemannian distance function $d^R$ of $(M, g^R)$, where $g^R$ is the restriction of the standard Euclidean metric tensor $g_{\field{E}^{n+1}}$ on $\field{R}^{n+1}$, i.e., we have $d^R = d_{\field{E}^{n+1}}$ on $M$. 

\vspace{1pc}
Let $(\overline{M}, \overline{d})$ be the metric completion of the metric space $(M,d^R)$. Consider an arbitrary element $[\{(t_k, {\bf x}_k)\}] \in \overline{M}$. Recall that, in particular, this means $\{(t_k, {\bf x}_k)\}$ is a Cauchy sequence in $(M, d^R)$. Since $\{(t_k, {\bf x}_k)\}$ is thus also Cauchy in $(\field{R}^{n+1}, d_{\field{E}^{n+1}})$, which is complete, this sequence converges to some point $(t_0, {\bf x}_0) \in [0, \infty) \times \field{R}^n$. Consider any other element $[\{(u_k, {\bf y}_k)\}] \in \overline{M}$, and let $(u_0, {\bf y}_0)$ be the corresponding limit in $[0, \infty) \times \field{R}^n$. Then we have:
\begin{align*}
\overline{d} \, ( \, [\{(u_k, {\bf y}_k)\}] \, , \,  [\{(t_k, {\bf x}_k)\}] \, ) & \; = \; \lim_{k \, \to \, \infty} \, d^R \, (  (u_k, {\bf y}_k)  ,  (t_k, {\bf x}_k) ) \\[.5pc]
& \; = \; \lim_{k \, \to \, \infty} \, d_{\field{E}^{n+1}} \, (  (u_k, {\bf y}_k)  ,   (t_k, {\bf x}_k)  ) \\[.5pc]
& \; = \; d_{\field{E}^{n+1}} \, ( (u_0, {\bf y}_0)  ,  (t_0, {\bf x}_0) )
\end{align*}

\vspace{.5pc}
\noindent
It follows that $[\{(u_k, {\bf y}_k)\}] = [\{(t_k, {\bf x}_k)\}]$ in $\overline{M}$ iff $(u_0, {\bf y}_0) = (t_0, {\bf x}_0)$ in $[0, \infty) \times \field{R}^n$. Moreover, since we identify $M$ with the subset of $\overline{M}$ of equivalence classes of Cauchy sequences in $M$ which converge to a point in $M$ itself, it follows that $[\{(t_k, {\bf x}_k)\}] \in M \subset \overline{M}$ iff $t_0 > 0$, and $[\{(t_k, {\bf x}_k)\}] \in \overline{M} - M$ iff $t_0 = 0$. Also, fixing any two such `boundary elements' $[\{(a_k, {\bf z}_k)\}] , [\{(b_k, {\bf w}_k)\}] \in \overline{M} - M$, with corresponding limits $(0, {\bf z}_0), (0, {\bf w}_0) \in [0, \infty) \times \field{R}^n$, note that
\begin{align*}
\overline{d} \, ( \, [\{(a_k, {\bf z}_k)\}] \, , \,  [\{(b_k, {\bf w}_k)\}] \, ) & \; = \; d_{\field{E}^{n+1}} \, ( \, (0, {\bf z}_0) \, , \,  (0, {\bf w}_0) \, )  \; = \; d_{\field{E}^{n}} \, ({\bf z}_0 ,  {\bf w}_0 )
\end{align*}
A bit more work shows that, using the obvious restrictions of each distance function, we have the expected metric space isometries (and thus also topological homeomorphisms):
\begin{align*} 
( \, \overline{M} \, , \, \overline{d}  \, ) &  \; \approx \; ( \, [0, \infty) \times \field{R}^n \, ,  \, d_{\field{E}^{n+1}} \, )\\[.5pc]
( \, \overline{M} - M \, , \, \overline{d}  \, ) & \; \approx \; ( \, \{0\} \times \field{R}^n \, ,  \, d_{\field{E}^{n+1}} \, ) \; \approx \; (\, \field{R}^n \, , \, d_{\field{E}^n}\, )
\end{align*}
\begin{flushright} $\Diamond$\end{flushright}
\end{exm}

\vspace{2pc}
We now introduce notation for the metric completions of a spacetime with respect to the distance functions discussed above. 

\begin{Def} [Null (Distance) Completion] \label{Defnullcompletion} Let $(M,g)$ be a spacetime. If $\tau$ is a generalized time function on $M$ such that the null distance $\hat{d}_\tau$ is definite, (i.e., if $\tau$ is locally anti-Lipschitz), then we set:
$$(\, \overline{M}_N \, , \, \overline{d}_N\, ) :=  \textrm{the metric completion of the metric space } (M, \hat{d}_\tau)$$
We will refer to this as the \emph{null (distance) completion} of $M$. If there is a need to specify the time function, we can write the completion more explicitly as, say, $(\overline{M}_{N,\tau}, \overline{d}_{N,\tau})$. As usual, we will identify $M$ with its corresponding subset in $\overline{M}_N$.
\end{Def}

\vspace{1pc}
\begin{Def} [Riemannianized (Distance) Completion] \label{DefRiemcompletion} Let $(M,g)$ be a spacetime. Suppose that $\tau : M \to \field{R}$ is smooth time function, with $\nabla \tau$ everywhere timelike. Let $g^R_\tau$ be the corresponding Riemannianized metric tensor, and $d^R_\tau$ the corresponding Riemannian distance function. We set: 
$$(\, \overline{M}_R \, , \, \overline{d}_R\, ) :=  \textrm{the metric completion of the metric space } (M, d^R_\tau)$$
We will refer to this as the \emph{Riemannianized (distance) completion} of $M$. Again, if there is a need to specify the time function, we can write the completion more explicitly as, say, $(\overline{M}_{R,\tau}, \overline{d}_{R,\tau})$, and as usual, we identify $M$ with its corresponding subset in $\overline{M}_R$.
\end{Def}

\vspace{1pc}
\begin{cor} [Extending $\tau$ to $\overline{M}_N$] \label{extending_tau_wrt_nulldist} Suppose that $\tau : M \to \field{R}$ is a locally anti-Lipschitz (continuous) time function on a spacetime $(M,g)$. Let $(\overline{M}_N, \overline{d}_N)$ be the metric completion of $(M,\hat{d}_\tau)$. Then $\tau$ has a unique continuous extension to $\overline{M}_N$, which we will denote by $\overline{\tau}_N : \overline{M}_N \to \field{R}$.
\end{cor}

\begin{proof} By Corollary \ref{null_dist_Lipschitz}, the map $\tau : (M, \hat{d}_\tau) \to (\field{R}, |\cdot|)$ is Lipschitz, with Lipschitz constant $\lambda = 1$. The unique continuous extension to the completion $\overline{M}_N$ then follows as in Proposition \ref{extendfunctionprop}. 
\end{proof}

\vspace{1pc}
\begin{cor} [Extending $\tau$ to $\overline{M}_R$]  \label{extending_tau_wrt_Riem} Suppose that $\tau : M \to \field{R}$ is a smooth time function on a spacetime $(M,g)$, with timelike gradient. Let $(\overline{M}_R, \overline{d}_R)$ be the metric completion of $(M,d^R_\tau)$. If there is a positive constant $D$ such that, $\|\nabla \tau \|_g \le D$, then $\tau$ has a unique continuous extension to $\overline{M}_R$, which we will denote by $\overline{\tau}_R : \overline{M}_R \to \field{R}$. 
\end{cor}

\begin{proof} By Lemma \ref{Riem_dist_Lipschitz}, the map $\tau : (M, d^R_\tau) \to (\field{R}, |\cdot|)$ is Lipschitz, with Lipschitz constant $\lambda = D$. The unique continuous extension to the completion $\overline{M}_R$ then follows again from Proposition \ref{extendfunctionprop}. 
\end{proof}

\pagebreak
\section{GRW Spacetimes} \label{sec_GRW}

\vspace{1pc}
We now turn attention to GRW spacetimes, 
$$(M,g) = (I \times S, -dt^2 + f^2(t)h) = I \times_f S$$
where $I = (a,b) \subset \field{R}$ is an open (time) interval, with $-\infty \le a < b \le \infty$, $(S,h)$ is a connected (spatial) Riemannian manifold, and $f : I \to (0,\infty)$ is a smooth positive function. We emphasize again that, unless stated otherwise, we do not make any further (completeness, homogeneity) assumptions about $(S,h)$. As above, for $p \in M = I \times S$, we will write $p = (t,x) = (t_p,p_S)$, and thus 
$$M = \{(t,x) : t \in I, x \in S\} = \{(t_p,p_S) : t_p \in I, p_S \in S\}$$
Letting $| \cdot |$ denote absolute value in $\field{R}$, we will consider the corresponding metric space $(I, |\cdot|) = (I, d_I)$, where $d_I(t_1,t_2) = |t_2 - t_1|$. Note that this is a metric subspace of $(\field{R}, |\cdot |) = (\field{R}, d_\field{E})$, where again $d_\field{E}(t_1,t_2) = |t_2 - t_1|$. For the base spatial distance, we write:
$$d_S \; = \textrm{ the intrinsic Riemannian distance function of } (S,h)$$
As for the warping function, for any subinterval $J \subset I$, we shall consider:
\begin{align*}
A_J & := \inf \{f(t) : t \in J\} \; \; \; \; , \; \; \; f_{\inf}  := \inf \{f(t) : t \in I\} = A_I \ge 0\\[.5pc]
B_J & := \sup \{f(t) : t \in J\} \; \; \; , \; \; \; f_{\sup} := \sup \{f(t) : t \in I\} = B_I \le \infty
\end{align*}

\vspace{1pc}

\subsection{GRW Riemannianized Distance}

\vspace{1pc}
We begin by considering Riemannianizations, and first note the following:

\begin{cor} [Riemannianizing GRW Spacetimes] \label{corGRWRiemannize} Consider a GRW spacetime $(M,g) = (I \times S, -dt^2 + f^2(t)h)$. 

\ben
\item [(1)] The Riemannianized metric induced by any smooth timelike vector field of the form $T = A(t,x) \d_t$ is given by simply flipping the sign on $dt^2$, as below. 
\begin{align*}
g & \; = \, -dt^2 + f^2(t)h \hspace{5pc} \\[.5pc] 
\longrightarrow \; \; \; \; g^R_T & \; = \; \; \, \, dt^2 + f^2(t)h \; \; = \; g^R_{\d_t}
\end{align*}

\item [(2)] The Riemannianized metric induced by any smooth time function of the form $\tau(t,x) = \phi(t)$, with $\phi'(t) > 0$, is again given by simply flipping the sign on $dt^2$ in $g$, as below. In particular, we have $g^R_\tau = g^R_t$, $L^R_{\tau} = L^R_t$, and $d^R_\tau = d^R_t$.
\begin{align*}
g & \; = \, -dt^2 + f^2(t)h \hspace{5pc}\\[.5pc]
\longrightarrow \; \; \; \; g^R_\tau & \; = \; \; \, \, dt^2 + f^2(t)h \; \; = \; g^R_t
\hspace{3pc}
\end{align*}
\een
\end{cor}

\vspace{1pc}
\begin{Def} [GRW Standard Riemannianization] Consider a GRW spacetime $(M,g) = (I \times S, -dt^2 + f^2(t)h)$. We will refer to the Riemannianization $(M,g^R_t) = (I \times S, dt^2 + f^2(t)h)$ induced by the standard time function $\tau(t,x) = t$ as the \emph{standard Riemannianization}, and to the corresponding Riemannianized distance function $d^R = d^R_t$, as the \emph{standard Riemannianized distance function} of $M$. 
\end{Def}

\vspace{1pc}
\begin{rmk} We note that the standard Riemannianized distance function on a GRW spacetime, $I \times_f S$, (e.g., any Riemannianized distance as in Corollary \ref{corGRWRiemannize}), is simply the Riemannian distance function of the Riemannian warped product manifold $(M,g^R) = (I \times S, dt^2 + f^2(t)h)$, about which much is surely known. Nonetheless, we will establish most of the properties we will need here from scratch. (Conversely, the properties established here also serve as Riemannian results.)
\end{rmk}

\vspace{1pc}
We first note the following basic properties:

\begin{lem} [GRW Standard Riemannianized Distance: Basic Properties] \label{RiemdistGRW_basic_estimates} Consider a GRW spacetime, $(M,g) = ((a,b) \times S, -dt^2 + f^2(t)h)$, with $- \infty \le a < b \le \infty$, and $M = \{(t_p,p_S) : t_p \in I, p_S \in S\}$. Let $d_S$ denote the Riemannian distance function of $(S,h)$. Consider the standard Riemannianized distance function $d^R_t$ on $M$, induced by $\tau(t,x) = t$. For any subinterval $J \subset (a,b)$, set $A_J := \inf \{f(t) : t \in J\}$. Let $f_{\inf} := \inf\{f(t) : a < t < b \} = A_{(a,b)}$.

\ben
\item [(1)] Vertical distances are vertical displacements:
$$p_S = q_S \; \; \Longrightarrow \; \; d^R_t(p,q) = |t_q - t_p|$$
\item [(2)] For any $c \in I$, we have:
$$d^R_{\, t \,}(p,q) \;  \le \; |t_p - c| \; + \; f(c) \cdot  d_S(p_S,q_S) \; + \; |t_q-c|$$
\item [(3)] For $t_p \le t_q$, we have the general upper bound:
$$d^R_t(p,q) \;  \le  \; |t_q -t_p| \; + \; A_{[t_p,t_q]} \cdot  d_S(p_S,q_S)$$
\item [(4)] For all $p,q \in M$, we have:
$$f_{\inf} \cdot  d_S(p_S,q_S)  \; \le \; d^R_{\, t \,}(p,q)$$
\item [(5)] For any subinterval $(t_1, t_2) \subset (a,b)$, and any $p, q \in (t_1,t_2) \times S$, we have:
$$\min \, \{ \, \, E_{(t_1,t_2)}(t_p,t_q)  \, \, , \, \, A_{(t_1,t_2)} \cdot d_S(p_S,q_S)\,  \,\} \; \le \; d^R_t(p,q)$$
where the `\emph{vertical exit cost}' $E_{(t_1,t_2)}(t_p,t_q)$ is defined by
\begin{displaymath}
   E_{(t_1,t_2)}(t_p,t_q) :=  \left\{
     \begin{array}{lr}
       \min \{ \, t_p + t_q - 2t_1 \, , \, 2t_2 - t_p - t_q\, \} & \hspace{1pc} a < t_1 < t_2 < b\\[.5pc]
       t_p + t_q - 2t_1 & a < t_1 < t_2 = b\\[.5pc]
       2t_2 - t_p - t_q & a = t_1 < t_2 < b\\[.5pc]
       \infty & a = t_1 < t_2 = b
     \end{array}
   \right.
\end{displaymath}
\item [(6)] For $t_p \le t_q$, we have the general lower bound:
$$\min \{\, |t_q -t_p| \; , \; A_{[t_p,t_q]} \cdot  d_S(p_S,q_S) \,\} \; \le \; d^R_t(p,q)$$
\een
\end{lem}

\begin{proof} (1) is a standard Riemannian fact. To verify it, note that the purely vertical curve $\a : [0,1]  \to M$, $\a(u) = (t_p + u(t_q-t_p), p_S)$ is smooth and joins $p$ to $q$, and has length $|t_q-t_p|$. Now consider any other piecewise smooth curve, $\g(u) = (\mu(u), \s(u))$, $u_1 \le u \le u_2$, from $\g(u_1) = p$ to $\g(u_2) = q$. Then we have:
\begin{align*}
L^R_t(\g) & \; = \; \int_{u_1}^{u_2} \sqrt{[\mu'(u)]^2 + f^2(\mu(u)) h(\s'(u), \s'(u))} \, du \\[1pc]
 & \; \ge \; \int_{u_1}^{u_2} |\mu'(u)| \, du  \; \ge \; \bigg|\int_{u_1}^{u_2} \mu'(u) \, du  \bigg|  \; = \; |t_q - t_p|
\end{align*}
Since $\g$ was arbitrary, we have $d^R_t(p,q) \ge |t_q-t_p|$, and thus $\a$ is minimal and distance-realizing. (In fact, the converse should also hold in (1), though we will not prove nor need it here.) To prove (2), first fix $\e > 0$. Let $\s_\e : [0,1] \to S$ be a piecewise smooth curve in $S$, from $\s_\e(0) = p_S$ to $\s_\e(1) = q_S$, with $L_S(\s_\e) \le d_S(p_S,q_S) + \e$. For any $c \in I$, consider the lift $\g_{c, \e} : [0,1] \to M$, $\g_{c, \e}(u) := (c,\s_\e(u))$. Also, let $\g_{t_p,c} : [0,1] \to M$ be the curve $\g_{t_p,c}(u) := (t_p + (c-t_p)u, p_S)$, and let $\g_{c,t_q} : [0,1] \to M$ be the curve $\g_{c,t_q}(u) := (c + (t_q-c)u, q_S)$. Then the concatenation $\g_{t_p,c} + \g_{c,\e} + \g_{c,t_q}$ is a piecewise smooth curve in $M$, from $p$ to $q$, and we have:
\begin{align*}
d^R_t(p,q) & \; \le \; L^R_t(\g_{t_p,c} + \g_{c,\e} + \g_{c,t_q}) \\[1pc]
& \; = \; L^R_t(\g_{t_p,c}) \, + \, L^R_t(\g_{c,\e}) \, + \, L^R_t(\g_{c,t_q}) \\[1pc]
& \; \le \; \; |t_p-c| \; + \; f(c) \cdot (\, d_S(p_S, q_S) + \e \,) \; + \; |t_q-c|
\end{align*}

\vspace{.5pc}
\noindent
Since $\e > 0$ was arbitrary, this verifies (2). For (3), note that for $c \in [t_p,t_q]$, we have $|t_p - c| + |t_q - c| = (c-t_p) + (t_q - c) = t_q - t_p$. We may then choose $c \in [t_p,t_q]$ in (2) to be the value at which $f(c) = \min \{f(t) : t_p \le t \le t_q\} = A_{[t_p,t_q]}$. To verify (4), fix any $p,q \in M$. Let $\g(u) = (\mu(u), \s(u))$, $u_1 \le u \le u_2$, be any piecewise smooth curve from $p$ to $q$. Then we have:
\begin{align*}
L^R_t(\g) & \; = \; \int_{u_1}^{u_2} \sqrt{[\mu'(u)]^2 + f^2(\mu(u)) h(\s'(u), \s'(u))} \, du \\[1pc]
 & \; \ge \; f_{\inf} \int_{u_1}^{u_2} \sqrt{h(\s'(u), \s'(u))} \, du \\[1pc]
 & \; = \;  f_{\inf} \cdot L_S(\s) \; \ge \;  f_{\inf} \cdot d_S(p_S,q_S)
\end{align*}

\vspace{.5pc}
\noindent
To verify (5), fix $(t_1, t_2) \subset (a,b)$. Fix two points $p, q \in (t_1, t_2) \times S$. Let $\g(u) = (\mu(u), \s(u))$, $u_1 \le u \le u_2$, be any piecewise smooth curve from $\g(u_1) = p$ to $\g(u_2) = q$. First note that if $\g$ remains within the slab $(t_1, t_2) \times S$, that is, if $t_1 \le \mu(u) \le t_2$ for all $u_1 \le u \le u_2$, then we have:
\begin{align*}
L^R_t(\g) & \; = \; \int_{u_1}^{u_2} \sqrt{[\mu'(u)]^2 + f^2(\mu(u)) h(\s'(u), \s'(u))} \, du \\[1pc]
 & \; \ge \; \int_{u_1}^{u_2} f(\mu(u)) \sqrt{h(\s'(u), \s'(u))} \, du \\[1pc]
 & \; \ge \; \int_{u_1}^{u_2} A_{(t_1,t_2)} \sqrt{h(\s'(u), \s'(u))} \, du\\[1pc]
 & \; = \;  A_{(t_1,t_2)}\cdot L_S(\s) \; \ge \;  A_{(t_1,t_2)}  \cdot d_S(p_S,q_S)
 \end{align*}

\vspace{.5pc}
\noindent
Otherwise, if $\g$ leaves the slab $(t_1, t_2) \times S$ at some time $u = u_0$, we have: 
\begin{align*}
L^R_t(\g) & \; = \; \int_{u_1}^{u_2} \sqrt{[\mu'(u)]^2 + f^2(\mu(u)) h(\s'(u), \s'(u))} \, du \\[1pc]
& \; \ge \; \int_{u_1}^{u_2} |\mu'(u)| du \\[1pc]
&\;  = \;  \int_{u_1}^{u_0} |\mu'(u)| du \; + \; \int_{u_0}^{u_2} |\mu'(u)| du\\[1pc]
&\;  \ge \; |\mu(u_0) - t_p| \; +\;  |\mu(u_0) - t_q|
\end{align*}

\vspace{.5pc}
\noindent
If $\mu(u_0) \le t_1$, then we must have $t_1 > a$, and the above gives $L^R_t(\g) \ge t_p + t_p - 2t_1$. If $\mu(u_0) \ge t_2$, then we must have $t_2 < b$, we have $L^R_t(\g) \ge 2t_2 - t_p - t_q$. Hence, in either case, if $\g$ leaves the slab $[t_p, t_q] \times S$, it follows that $L^R_t(\g) \; \ge \; E_{(t_1,t_2)}(t_p,t_q)$. Putting the above together, we have shown that, whether $\g$ leaves the slab or not,
$$L^R_t(\g) \; \ge \; \min \, \{ \, E_{(t_1,t_2)}(t_p,t_q) \, , \, A_{(t_1,t_2)}  \cdot d_S(p_S,q_S) \}$$
Taking the infimum over all piecewise smooth curves $\g$ from $p$ to $q$ gives (5). Finally, for (6), fix $p, q \in M$, with $t_p \le t_q$. Fix $\delta > 0$ small enough so that $t_1 := t_p - \delta$ and $t_2 := t_q + \delta$ satisfy $(t_1, t_2) \subset (a,b)$. Then we have:
\begin{align*}
E_{(t_1,t_2)}(t_p,t_q) & \; = \; \min \{ \, (t_p + t_q - 2(t_p - \delta)) \, , \, (2(t_q + \delta) - t_p - t_q) \, \}\\[1pc]
& \; = \; \min \{ \, (t_q - t_p +2\delta) \, , \, (t_q -t_p +2\delta) \, \}\\[1pc]
& \; = \; t_q - t_p +2\delta
\end{align*}

\noindent
Thus, by part (5), we have:
$$\min \{ \, (t_q - t_p + 2\delta) \, , \, A_{(\, t_p \, - \, \delta\, , \, t_q \, + \, \delta\, )} \cdot d_S(p_S,q_S) \, \} \; \le \; d^R_t(p,q)$$
The result then follows by taking $\delta \to 0$.
\end{proof}

\vspace{2pc} 
We note the following consequence of Lemma \ref{RiemdistGRW_basic_estimates}:

\begin{lem} [GRW Standard Riemannianized Distance and Spatial Pinching] \label{Riem_pinching} Consider a GRW spacetime $(M,g) = ((a,b) \times S, -dt^2 + f^2(t)h)$, $- \infty \le a < b \le \infty$. Consider the standard Riemannianized distance function $d^R_t$ on $M$. Suppose there is a sequence $a_j \to a^+$ along which $f(a_j) \to 0$, (keeping in mind that $a \ge - \infty$). Then for any two points $p, q \in M$, $p = (t_p,p_S)$, $q = (t_q,q_S)$, we have:
$$d^R_t(p,q) \; \le \;  t_p + t_q - 2 a$$
Consequently, if $a = 0$, then for all $p, q \in M$,
$$d^R_t(p,q) \; \le \; t_p + t_q$$
More generally, if $a > - \infty$ is finite, then any sequence $p_k = (t_k, x_k)$ with $t_k \to a^+$ is a Cauchy sequence in $(M,d^R_t)$, (regardless of the spatial motion of $\{x_k\}$). Now let $(\overline{M}_R, \overline{d}_R)$ be the metric completion of $(M,d^R_t)$, and let $\overline{\tau}_R : \overline{M}_R \to \field{R}$ be the unique continuous extension of $\tau(t,x) = t$, as in Corollary \ref{extending_tau_wrt_Riem}. If $q_k = (u_k,y_k)$ is any other sequence, with $u_k \to a^+$, then $\{p_k\}$ and $\{q_k\}$ are equivalent, giving the same element $[\{p_k\}] = [\{q_k\}] =: \overline{p}_0 \in \overline{M}_R$ in the metric completion, and $(\overline{\tau}_R)^{-1}(a) = \{\overline{p}_0\}$ consists of a single point.
\end{lem}

\begin{proof} Fix any two points $p = (t_p, p_S)$ and $q = (t_q, q_S)$ in $M$. Applying part (2) of Lemma \ref{RiemdistGRW_basic_estimates} with $c = a_j$, then for all sufficiently large $j$, we have:
$$d^R_t(p,q)  \; \le \; t_p - a_j \; + \; f(a_j) \cdot d_S(p_S,q_S)  \; + \;  t_q - a_j$$
The main estimate then then follows by taking $j \to \infty$. Now suppose that $a > - \infty$ is finite. Consider a sequence $p_k = (t_k,x_k)$ with $t_k \to a^+$. Fix $\e > 0$. Let $k_1 \in \field{N}$ such that $0 < t_k - a < \e/2$, for all $k \ge k_1$. Then for any indices $\ell, m \ge k_1$, the main estimate gives $d^R_t(p_\ell,p_m) \le (t_\ell - a) + (t_m - a) < \e$. Thus $\{p_k\}$ is a Cauchy sequence in $(M, d^R_t)$. Moreover, consider another sequence $q_k = (u_k, y_k)$ with $u_k \to a^+$. Let $k_2 \in \field{N}$ be such that $u_k - a < \e/2$ for all $k \ge k_2$. Then, for all $k \ge \max \{k_1, k_2\}$, we have $d^R_t(p_k,q_k) \le (t_k - a) + (u_k - a) < \e$. It follows that:
$$\overline{d}_R([\{p_k\}], [\{q_k\}]) \; = \; \lim_{k \, \to \, \infty} \, d^R_t(p_k,q_k) \; = \; 0$$
Hence, $[\{p_k\}] = [\{q_k\}] =: \overline{p}_0 \in \overline{M}_R$, and $(\overline{\tau}_R)^{-1}(a) = \{\overline{p}_0\}$.
\end{proof}

\vspace{1pc}
\subsection{GRW Elevator Distance}

\vspace{1pc}
On any metric product, one can consider a natural `taxicab distance'. On a GRW spacetime, there is a natural `warped taxicab distance', which we formalize here as follows.

\vspace{1pc}
\begin{Def} [Piecewise Vertical-or-Horizontal Curves in GRW Spacetimes] \label{Def_pvh_curves} Consider a GRW spacetime, $(M,g) = (I \times S, -dt^2 + f^2(t)h)$. Let $\pi_I : I \times S \to I$ and $\pi_S : I \times S \to S$ be the projections onto each factor, $\pi_I(t,x) = t$, and $\pi_S(t,x) = x$. Call a path $\g$ in $M$ \emph{(purely) vertical} if $\pi_S(\g)$ is constant. Call a path $\g$ in $M$ \emph{(purely) horizontal} if $\pi_I(\g)$ is constant. Note that a path is then trivial (constant) iff it is both vertical and horizontal. By a \emph{piecewise vertical-or-horizontal (p.v.h.)} curve we will mean a path $\g : [a,b] \to M$, such that there is a partition $a = s_0 < s_1 < \cdots < s_m = b$, with each subsegment $\g_i := \g|_{[s_{i-1}, s_i]}$ being smooth, and, (if nontrivial), either purely vertical or purely horizontal. In this case, we will also sometimes write either $\g = \g_1 \cdot \g_2 \cdot \, \cdots \, \cdot \g_m$, or $\g = \g_1 + \g_2 + \cdots + \g_m$. Let $\mathcal{A}^{pvh}$ be the set of all p.v.h. curves in $M$.
\end{Def}

\vspace{1pc}
\begin{Def} [Elevator Distance on GRW Spacetimes] \label{Def_elevator} Consider a GRW spacetime, $(M,g) = (I \times S, -dt^2 + f^2(t)h)$. For any two points $p, q \in M$, let $\mathcal{A}^{pvh}(p,q)$ denote the set of all p.v.h curves in $M$ from $p$ to $q$. Suppose that $\tau : M \to \field{R}$ is a smooth time function, with everywhere timelike gradient $\nabla \tau$. Define the induced \emph{elevator length} of a p.v.h. curve $\g \in \mathcal{A}^{pvh}$ to be its corresponding  Riemannianized length, $L^E_\tau(\g) := L^R_\tau(\g)$, and define the induced \emph{elevator distance} between any two points $p,q$ by:
$$d^E_\tau(p,q) := \inf \, \{L^E_\tau(\g) : \g \in \mathcal{A}^{pvh}(p,q)\}$$
\end{Def}

\vspace{2pc}
Note that by construction, every piecewise vertical-or-horizontal curve $\g$ is piecewise smooth, and we have $L^E_\tau(\g) = L^R_\tau(\g)$. Consequently, elevator distance is bounded below by Riemannianized distance, and is thus a definite distance function:

\begin{lem} [GRW Elevator Distance is a Distance] \label{GRW_Riem_vs_elevator_basic} Consider a GRW spacetime $(M,g) = (I \times S, -dt^2 + f^2(t)h)$, and a smooth time function $\tau$ on $M$ with timelike gradient $\nabla \tau$. For any two points $p, q \in M$, we have:
$$d^R_\tau(p,q) \; \le \; d^E_\tau(p,q)$$
In particular, $d^E_\tau$ defines a distance function on $M$, (which is definite, symmetric, and satisfies the triangle inequality), and any lower bounds on the Riemannianized distance are inherited by the elevator distance. Indeed, $(M, \mathcal{A}^{pvh}, L^E_\tau)$ defines a length structure, and $d^E_\tau$ is its length metric.
\end{lem}

\vspace{1pc}
\begin{Def} [GRW Standard Elevator Distance] \label{Def_std_elevator} Consider a GRW spacetime $(M,g) = (I \times S, -dt^2 + f^2(t)h)$. We will refer to the elevator distance function $d^E = d^E_t$, induced by the standard time function $\tau(t,x) = t$, as the \emph{standard elevator distance function} of $M$.
\end{Def}

\vspace{1pc}
By Corollary \ref{corGRWRiemannize}, we have:
\begin{cor}  \label{corGRWelevator} Consider a GRW spacetime $(M,g) = (I \times S, -dt^2 + f^2(t)h)$. Then the elevator distance $d^E_\tau$ induced by any smooth time function of the form $\tau(t,x) = \phi(t)$, with $\phi'(t) > 0$, coincides with the standard elevator distance $d^E_t$, induced by the standard coordinate time function $t$.
\end{cor}

\vspace{1pc}
\begin{rmk} [GRW Elevator Distance Heuristics] \label{elevator_heuristics} Consider the standard elevator distance $d^E_t$ on a GRW spacetime $(M,g) = (I \times S, -dt^2 + f^2(t)h)$. First note that $d^E_t$ is really a Riemannian construction, applying to Riemannian warped products of the form $(M,g^R) = (I \times S, dt^2 + f^2(t)h)$, (and may very well have been studied elsewhere). We have nonetheless chosen to present this tailored to own focus here, which is on spacetimes. Now recall that in a GRW spacetime, space at each time is geometrically the same, except for an overall scale factor. As usual, we can think of time as running vertically, and space being horizontal. Then think of the universe $M$ as a large `building', with each `floor' corresponding to space at a fixed time. All the floors in the building look identical, except they differ in overall scale. Imagine that at every point in the building, you can take an elevator straight up or down to any other floor; these correspond to the vertical curves in $M$. You can also move around within a fixed floor; this corresponds to the horizontal curves. To get from one point $p$ in the building to another point $q$, you are free to embark on any finite sequence of such movements, by moving within a fixed floor, and/or by taking any of the elevators; this corresponds to the p.v.h. curves. Roughly speaking, the shortest such path then corresponds to the elevator distance between $p$ and $q$. Note that, even if two points $p$ and $q$ are not separated by many floors, or even if they are on the same floor, it may, for example, still be more economical to go far up or down to a distant but  `smaller' floor, to traverse the spatial displacement from $p_S$ to $q_S$.
\end{rmk}

\vspace{1pc}
\begin{rmk} [More General Elevator Distances] While the elevator distance construction is defined here in the fairly specialized GRW setting, we note that a similar construction can also be considered on more general products, including, for example, globally hyperbolic splittings as in \cite{SanchezBernalsplit}. 
\end{rmk}

\vspace{2pc}
The following basic properties are more or less immediate:

\begin{lem} [GRW Standard Elevator Length: Basic Properties] \label{elevator_basic_properties} Consider a GRW spacetime $(M,g) = (I \times S, -dt^2 + f^2(t)h)$, and its standard Riemannianization $(M,g^R_t) = (I \times S, dt^2 + f^2(t)h)$. Let $L_I$ and $d_I$ denote the Riemannian arc length and distance of $(I, dt^2)$, and let $L_S$ and $d_S$ denote the Riemannian arc length and distance of $(S,h)$. Consider the standard elevator length functional $L^E = L^E_t$. Then we have the following:
\ben
\item [(1)] Given any two points $p, q \in M$, there is a p.v.h. curve $\g$ starting at $p$ and ending at $q$. 
\item [(2)] Any p.v.h. curve $\g$ in $M = I \times S$ projects to a connected piecewise smooth curve in each factor, that is, the curves $\pi_I(\g)$ and $\pi_S(\g)$, defined in the natural way, give connected piecewise smooth curves in $I$ and $S$, respectively.
\item [(3)] For any p.v.h. curve $\g = \g_1 + \g_2 + \cdots + \g_m$, we have
$$L^E(\g) =  L^E(\g_1) + L^E(\g_2) + \cdots + L^E(\g_m)$$
Moreover, denoting the vertical subsegments by $\g^T_1, ..., \g^T_{\ell_T}$, and the horizontal subsegments by $\g^S_1, ..., \g^S_{\ell_S}$, formally set $\g^T := \g^T_1 +  \cdots +  \g^T_{\ell_T}$ to be the `\emph{vertical component}' of $\g$, and $\g^S := \g^S_1 +  \cdots +  \g^S_{\ell_S}$ to be the `\emph{horizontal component}' of $\g$. Then we have $L^E(\g) = L^E(\g^T) +  L^E(\g^S)$, where
\begin{align*}
L^E(\g^T) & \; := \; L(\g^T_1) \, + \, \cdots \, + \, L^E(\g^T_{\ell_T})\\[.5pc]
L^E(\g^S) & \; := \; L(\g^S_1) \, + \, \cdots \, + \, L^E(\g^S_{\ell_S})
\end{align*}
\item [(4)] If $\g^T$ is (purely) vertical, from $p = (t_p, x_0)$ to $q = (t_q,x_0)$, then
$$L^E(\g^T) \; = \; L_I(\pi_I(\g^T)) \; \ge \; d_I(t_p,t_q) \; = \; |t_q-t_p|$$
\item [(5)] For any p.v.h. curve $\g$ from $p$ to $q$, we have:
$$L^E(\g) \;  \ge \; L^E(\g^T) \; \ge \; |t_q - t_p|$$
\item [(6)] If $\g^S$ is (purely) horizontal, from $p = (t_0, p_S)$ to $q = (t_0,q_S)$, then
$$L^E(\g^S) \; = \; f(t_0) \cdot L_S(\pi_S(\g^S)) \; \ge \; f(t_0) \cdot d_S(p_S,q_S)$$
\een
\end{lem}

\vspace{2pc}
Lemma \ref{RiemdistGRW_basic_estimates} has an exact analog for elevator distance:

\begin{lem} [GRW Standard Elevator Distance: Basic Estimates] \label{elevator_basic_estimates} Consider a GRW spacetime, $(M,g) = ((a,b) \times S, -dt^2 + f^2(t)h)$, $- \infty \le a < b \le \infty$, with $M = \{(t_p,p_S) : t_p \in I, p_S \in S\}$. Let $d_S$ denote the Riemannian distance function of $(S,h)$. Consider the standard elevator distance function $d^E = d^E_t$ on $M$, induced by $\tau(t,x) = t$. For any subinterval $J \subset (a,b)$, set $A_J := \inf \{f(t) : t \in J\}$. Let $f_{\inf} := \inf\{f(t) : a < t < b \} = A_{(a,b)}$.

\ben
\item [(1)] Vertical distances are vertical displacements:
$$p_S = q_S \; \; \Longrightarrow \; \; d^E(p,q) = |t_q - t_p|$$
\item [(2)] For any $c \in I$, we have:
$$d^E(p,q) \;  \le  \; |t_p - c| \;  + \; f(c) \cdot  d_S(p_S,q_S) \;+ \; |t_q - c|$$
\item [(3)] For $t_p \le t_q$, we have the general upper bound:
$$d^E(p,q) \;  \le  \; |t_q -t_p| \; + \; A_{[t_p,t_q]} \cdot  d_S(p_S,q_S)$$
\item [(4)] For all $p,q \in M$, we have:
$$f_{\inf} \cdot  d_S(p_S,q_S)  \; \le \; d^E(p,q)$$
\item [(5):] For any subinterval $(t_1, t_2) \subset (a,b)$, and any $p, q \in (t_1,t_2) \times S$, we have:
$$\min \, \{ \, \, E_{(t_1,t_2)}(t_p,t_q)  \, \, , \, \, A_{(t_1,t_2)} \cdot d_S(p_S,q_S)\,  \,\} \; \le \; d^E(p,q)$$
where the `\emph{vertical exit cost}' $E_{(t_1,t_2)}(t_p,t_q)$ defined exactly as in Lemma \ref{RiemdistGRW_basic_estimates}. 
\item [(6)] We have the general lower bound:
$$\min \{\, |t_q -t_p| \; , \; A_{[t_p,t_q]} \cdot  d_S(p_S,q_S) \,\} \; \le \; d^E(p,q)$$
\een
\end{lem}

\begin{proof} The proof of (1) here is analogous to that in Lemma \ref{RiemdistGRW_basic_estimates}, for example using Lemma \ref{elevator_basic_properties}. The proof of (2) in Lemma \ref{RiemdistGRW_basic_estimates} uses a p.v.h. curve, so it applies here as well. Part (3) follows from (2) exactly as in Lemma \ref{RiemdistGRW_basic_estimates}. Finally, the lower bounds in (4), (5), (6) are all inherited from the Riemannianized distance, by Lemma \ref{GRW_Riem_vs_elevator_basic}. 
\end{proof}

\vspace{2pc}
First note that Lemma \ref{elevator_basic_estimates} gives the following:

\begin{cor} [GRW Standard Elevator Distance Induces Manifold Topology] \label{elevator_topology} Consider a GRW spacetime $M = I \times_f S$, and the standard elevator distance $d^E = d^E_t$.
\ben
\item [(1)] Fix any point $p = (t_p,p_S) \in M$. For any $\delta > 0$ with $[t_p - \delta, t_p + \delta] \subset I$, set  $B_\delta := \max \{f(t) : |t - t_p| \le \delta \}$, and note that $0 < B_\delta < \infty$. Then for all points $z = (t_z, z_S) \in M$ with $|t_z - t_p| \le \delta$, we have:
$$d^E(p,z) \;  \le  \; \delta \; + \; B_\delta \cdot  d_S(p_S,z_S)$$
\item [(2)] Consequently, the topology induced by the standard elevator distance $d^E$ coincides with the manifold topology. 
\een
\end{cor}

\begin{proof} (1) Fix $z = (t_z,z_S)$, with $|t_z - t_p| \le\delta$. Choosing $c = t_p$ in (2) from Lemma \ref{elevator_basic_estimates}, we get the upper bound:
$$d^E(p,z) \;  \le  \; |t_z -t_p| \; + \; f(t_p) \cdot  d_S(p_S,z_S) \; \le \; \delta \; + \; B_\delta \cdot d_S(p_S,z_S)$$
To verify (2), i.e., that $d^E$ induces the manifold topology, let $\mathscr{T}^M$ denote the collection of all sets which are open in the manifold topology, and let $\mathscr{T}^E$ denote the collection of all sets which are open in the topology induced by the elevator distance, $d^E$. We want to show that these two collections of open sets coincide, i.e., $\mathscr{T}^E = \mathscr{T}^M$. Fix any point $p = (t_p,p_S) \in M$. Fix any $r > 0$, and consider the elevator ball $B^E_r(p) = \{(t_z,z_S) \in M : d^E(z,p) < r\}$. Set $\delta := r/2$, and $\rho := \min \{r/2, r/(2B_\delta)\}$. We will show that the ball product 
$$B_{\rho}^I(t_p) \times B_{\rho}^S(p_S) = \{ \,(t_z,z_s) \in M : |t_z-t_p| < \rho \, , \, d_S(z_S,p_S) < \rho \,\}$$
is contained in $B^E_r(p)$. Fix any $z \in B_{\rho}^I(t_p) \times B_{\rho}^S(p_S)$. Then $|t_z - t_p| < \rho \le r/2 = \delta$. Thus, by the estimate in (1),
\begin{align*}
d^E(z,p) & \; \le  \; \delta \; + \; B_\delta \cdot  d_S(z_S,p_S)\\[1pc]
 & \; = \; \tfrac{r}{2} \; + \; B_{r/2} \cdot  d_S(z_S,p_S) \\[1pc]
 & \; < \; \tfrac{r}{2} \; + \; B_{r/2} \cdot  \rho \; \le \; \tfrac{r}{2} \; + \; B_{r/2} \cdot  (\, \tfrac{r}{2B_{r/2}} \, ) \; = \; r
\end{align*}

\vspace{.5pc}
\noindent
Since such ball products are open in the manifold topology, it follows that $\mathscr{T}^E \subset \mathscr{T}^M$. To show the reverse inclusion, recall that the standard Riemannianized distance function $d^R = d^R_t$ also generates the manifold topology. Let $B^R_r(p) = \{z \in M : d^R(z,p) < r\}$ denote the Riemannianized ball centered at $p$ of radius $r$. The basic inequality $d^R \le d^E$ means that, if $d^E(z,p) < r$, we have also $d^R(z,p) < r$, that is, if $z \in B^E_r(p)$, then also $z \in B^R_r(p)$. Thus, we have $B^E_r(p) \subset B^R_r(p)$, which implies that every set which is open in the manifold topology is also open in the elevator topology, i.e., $\mathscr{T}^M \subset \mathscr{T}^E$.
\end{proof}

\vspace{2pc}
The following analog of Lemmas \ref{Riem_dist_Lipschitz} and \ref{null_dist_Lipschitz}, and Corollaries \ref{extending_tau_wrt_nulldist} and \ref{extending_tau_wrt_Riem}, is immediate from Lemmas \ref{Riem_dist_Lipschitz} and \ref{GRW_Riem_vs_elevator_basic}, and Proposition \ref{extendfunctionprop}:

\begin{cor} [GRW Elevator Lipschitz Continuity and Completion] \label{elevator_Lipschitz} Consider a GRW spacetime $M = I \times_f S$. Fix any smooth time function $\tau$ on $M$, with timelike gradient, and consider the induced elevator distance $d^E_\tau$. If there is a positive constant $D$ such that $\|\nabla \tau\|_g \le D$, then for all $p, q \in M$, we have:
$$\dfrac{1}{D} \cdot |\tau(q)- \tau(p)| \; \le \; d^E_\tau(p,q)$$
In other words, the map $\tau : (M, d^E_\tau) \to (\field{R}, | \cdot |)$ is Lipschitz, with Lipschitz constant $\lambda = D$. In particular, it is uniformly continuous. Consequently, letting $(\overline{M}_E, \overline{d}_E)$ be the metric completion of $(M,d^E_\tau)$, then $\tau$ admits a unique continuous extension $\overline{\tau}_E : \overline{M}_E \to \field{R}$. 
\end{cor}

\vspace{2pc}
We also have the following analog of Lemma \ref{Riem_dist_causal_pairs} and Proposition \ref{null_dist_causal_pairs}, and Corollaries \ref{Riem_antiLip} and \ref{null_antiLip}:

\begin{lem} [GRW Standard Elevator Distance on Causal Pairs] \label{elevator_dist_causal_pairs} Consider a GRW spacetime, $M = I \times_f S$. The standard GRW elevator distance $d^E_t$ satisfies:
$$p \le q \; \; \Longrightarrow \; \; d^E_t(p,q) \; \le \; 2 \cdot (t_q - t_p)$$
In other words, the standard time function $\tau(t,x) = t$ is anti-Lipschitz with respect to $d^E_t$, with anti-Lipschitz constant $\lambda = 1/2$.
\end{lem} 

\begin{proof} Suppose that $p \le q$. Let $A_{[t_p,t_q]} = \min \{f(t) : t_p \le t \le t_q\} > 0$. Then, using Lemma 3.27 in \cite{nulldist}, we have: 
$$d_S(p_S, q_S) \; \le \; \int_{t_p}^{t_q} \dfrac{1}{f(w)} \, dw \; \le \; \int_{t_p}^{t_q} \dfrac{1}{A_{[t_p,t_q]}} \, dw \; = \; \dfrac{t_q - t_p}{A_{[t_p,t_q]}}$$
Thus, we have $A_{[t_p,t_q]} \cdot d_S(p_S, q_S) \; \le \; t_q - t_p$. The result then follows from part (3) of Lemma \ref{elevator_basic_estimates}.
\end{proof}

\vspace{2pc}
Finally, we note:

\begin{lem} [Standard Elevator Distance on Product] \label{elevator_on_product} Consider any product spacetime of the form, $(M,g) = (I \times S, -dt^2 + h)$, that is, a GRW spacetime with warping function $f \equiv 1$. Then we have the following:
\ben
\item [(1)] The standard elevator distance $d^E = d^E_t$ is given by:
$$d^E(p,q) \; = \; |t_q - t_p| \, + \, d_S(p_S,q_S)$$
\item [(2)] For arbitrary (connected) spatial fiber $(S,h)$ we have:
$$q \in \overline{I^+(p)} \; \Longleftrightarrow \; d^E(p,q) \, \le \, 2\, (t_q - t_p)$$
\item [(3)] If $(S,h)$ is complete, then we have:
$$p \le q \; \Longleftrightarrow \; d^E(p,q) \, \le \, 2 \, (t_q - t_p)$$
\een
\end{lem}

\begin{proof} To verify (1), fix any two points $p = (t_p,p_S)$ and $q = (t_q,q_S)$ in $M$. Let $\g^T : [0,1] \to M$ be the vertical curve $\g^T(u) := (t_p + u(t_q - t_p), p_S)$. Fix any $\e > 0$. Let $\s_\e : [1,2] \to S$ be a piecewise smooth spatial curve in $S$, from $\s_\e(1) = p_S$ to $\s_\e(2) = q_S$, with $L_S(\s_\e) \le d_S(p_S,q_S) + \e$. Consider the horizontal lift $\g^S_\e : [1,2] \to M$, $\g^S_\e(u) := (t_q, \s_\e(u))$. Then $\g_\e := \g^T + \g^S_\e$ is a p.v.h. curve in $M$ from $p$ to $q$, and we have:
$$d^E(p,q) \; \le \; L^E(\g_\e)  \; = \; |t_q - t_p| + L_S(\s_\e) \; \le \; |t_q - t_p| + d_S(p_S,q_S) + \e$$
Taking $\e \to 0$ gives $d^E(p,q) \le |t_q - t_p| + d_S(p_S,q_S)$. On the other hand, letting $\g$ now be any p.v.h. curve from $p$ to $q$, basic lower bounds give:
$$L^E(\g) = L^E(\g^T) + L^E(\g^S) \ge |t_q - t_p| + d_S(p_S,q_S)$$
This in turn gives $d^E(p,q) \ge |t_q - t_p| + d_S(p_S,q_S)$.
Parts (2) and (3) then follow from Lemmas 3.23 and 3.27 in \cite{nulldist}.
\end{proof}

\vspace{1pc}
\subsection{Minkowski and Future Half-Minkowski}

\vspace{1pc}
We take a moment now to explore and illustrate the constructions discussed so far in two model cases, Minkowski space, and future half-Minkowski space, both through the lens of the standard time function $\tau = t$.

\vspace{1pc}
\begin{exm} [$\tau = t$ on Minkowski Space] \label{exMinkowski} Consider Minkowski space, 
$$M \; = \; \field{M}^{n+1} \; = \; (\, \field{R} \times \field{R}^n \,,\, -dt^2 + g_{\field{E}^n} \,)$$
where we will write $\field{R} \times \field{R}^n = \{(t_p,p_S) : t_p \in \field{R}, p_S \in \field{R}^n\}$, where $t_p$ is the location in time, and $p_S$ is the location in space, and where $g_{\field{E}^{n}}$ is the standard Euclidean metric tensor of $\field{R}^n$. Let $\| \cdot \| = \| \cdot \|_{\field{E}^n}$ denote the standard Euclidean norm of $\field{R}^n$.

\vspace{1pc}
Using the standard time function $\tau(t,{\bf x}) = t$, it follows from Corollary \ref{corGRWRiemannize}, Lemma \ref{elevator_on_product}, and Example \ref{t_on_Mink_example} above, (the latter coming from Proposition 3.3 in \cite{nulldist}), that the corresponding Riemannianized, null, and elevator distance functions, $d^R = d^R_t$, $d^N = \hat{d}_t$, $d^E = d^E_t$, respectively, are given as below. See Figure \ref{Mink_REN_spheres} for pictures of the corresponding spheres.
\begin{align*}
d^R(p,q) & \; = \; \sqrt{|t_q-t_p|^2 + \| q_S - p_S\|^2} \; = \; d_{\field{E}^{n+1}}(p,q)\\[1pc]
d^N(p,q) & \; = \; \max \, \{ \, |t_q-t_p| \, , \| q_S - p_S\| \, \} \\[1pc]
d^E(p,q) & \; = \; |t_q-t_p| + \| q_S - p_S\|
\end{align*}

\vspace{1pc}
\begin{figure}[h]
\begin{center}
\includegraphics[width=12cm]{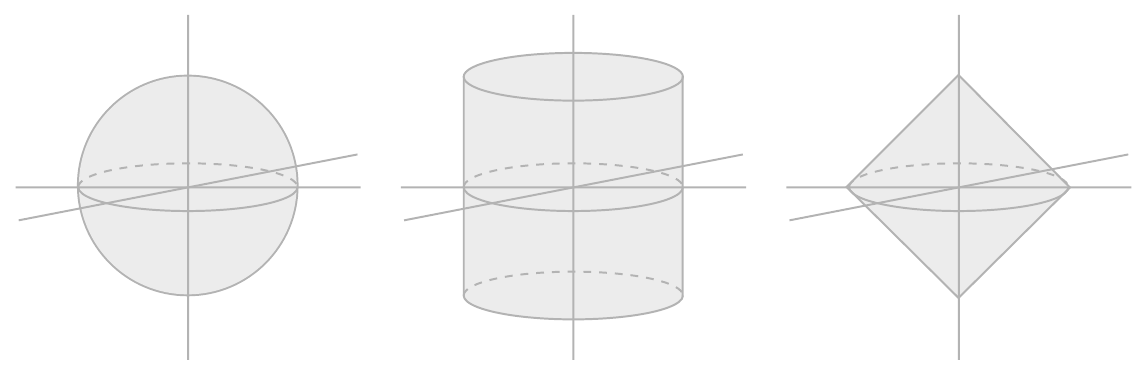}
\end{center}
\vspace{-.5pc}
\caption[]{Spheres of the same radius are shown in Minkowski space, for the standard Riemannianized distance on the left, null distance in the middle, and elevator distance on the right.}
\label{Mink_REN_spheres}
\end{figure}

\vspace{1pc}
Note then, for example, that we have:
\begin{align*}
d^N(p,q) & \; = \; \max \, \{ \, |t_q-t_p| \, , \| q_S - p_S\| \, \}\\[1pc]
& \; \le \; \sqrt{|t_q-t_p|^2 + \| q_S - p_S\|^2}\\[1pc]
& \; \le \; \sqrt{\, 2 \, (\max\{|t_q-t_p|, \|q_S-p_S\| \})^2\, } \; = \; \sqrt{2} \, \, d^N(p,q)
\end{align*}

\vspace{.5pc}
\noindent
Hence, $d^R$ and $d^N$ are Lipschitz equivalent, with $d^N \le d^R \le \sqrt{2} \, d^N$. Indeed, similar considerations show that all three distances are pairwise Lipschitz equivalent, with:
\begin{align*}
d^N & \; \le \; d^R \; \le \; \sqrt{2} \, d^N \\[.5pc]
d^R & \; \le \; d^E \; \le \;  2 \, d^R \\[.5pc]
d^N & \; \le \; d^E \; \le \;  2 \, d^N
\end{align*}

\vspace{2pc}
Now for $M = \field{M}^{n+1}$, consider the metric completions:
\begin{align*}
(\overline{M}_R, \overline{d}_R) & \; = \; \textrm{the metric completion of } (M, d^R)\\[.5pc]
(\overline{M}_N, \overline{d}_N) &  \; = \; \textrm{the metric completion of } (M, d^N)\\[.5pc]
(\overline{M}_E, \overline{d}_E) & \; = \; \textrm{the metric completion of } (M, d^E)
\end{align*}

\vspace{.5pc}
\noindent
Since $d^R$ is the standard Euclidean metric, it is complete. The Lipschitz equivalences then show that both $d^N$ and $d^E$ are also complete. Thus, as in Propositions \ref{Lipequivcompletionprop} and \ref{completion_prop}, we have set identifications, $\overline{M}_R \, = \, \overline{M}_N \, = \, \overline{M}_E \, \approx \, M$, and the metrics $\overline{d}_R$, $\overline{d}_N$, $\overline{d}_E$ are also Lipschitz equivalent, (with the same constants). In particular, they induce the same topology. That is, the identifications above are also homeomorphisms of topological spaces,
$$\overline{M}_R \; \approx \; \overline{M}_N \; \approx \; \overline{M}_E \; \approx \; M$$
While not exactly an appropriate comparison in this case, we note that Minkowski space $M = \field{M}^{n+1}$ has nonempty past causal boundary, $\d^-(M) \ne \emptyset$. In particular, the past causal completion $M^-$, differs from the metric completions above. For example, as sets, we have $M^- \; \not \approx  \; M \; \approx \;  \overline{M}_R \; \approx \; \overline{M}_N \; \approx \; \overline{M}_E$.

\vspace{2pc}
As for causality, let $p = (t_p,p_S)$ and $q = (t_q,q_S)$. First note that we have:
$$p \le q  \; \; \;  \Longleftrightarrow \; \; \;\|q_S-p_S\| \, \le \, t_q-t_p$$
It follows then that the following conditions are all equivalent:
\begin{align*} 
p \le q & \; \Longleftrightarrow \; d^N(p,q) \, = \, t_q - t_p\\[.5pc]
& \; \Longleftrightarrow \; d^R(p,q) \, \le \, \sqrt{2} \, (t_q - t_p)\\[.5pc]
& \; \Longleftrightarrow \; d^E(p,q) \, \le \, 2 \, (t_q - t_p)
\end{align*}

\vspace{.5pc}
\noindent
The \emph{equality} for $d^N$ stands out compared to the \emph{inequalities} for $d^R$ and $d^E$. It turns out that this is a consequential distinction. This is discussed further in Example \ref{Riem_does_not_encode}.
\begin{flushright} $\Diamond$\end{flushright}
\end{exm}

\vspace{2pc}
\begin{exm} [$\tau = t$ on Future Half-Minkowski] \label{exfuturehalfMinkowski} Building on Example \ref{exMinkowski} above, we now consider the future half-Minkowski space,
$$M = \field{M}^{n+1}_+ = (\, (0,\infty) \times \field{R}^n \; , \; -dt^2 + g_{\field{E}^n} \, )$$
Consider again the standard time function $\tau(t, {\bf x}) = t$. While not quite automatic, it is true that the Riemannianized distance function induced by $\tau = t$ on $M$ is precisely the restriction of that for the full Minkowski space, and similarly for the null and elevator distances. We shall thus continue to denote these by $d^R$, $d^N$, $d^E$. Since these distance functions are Lipschitz equivalent on all of $\field{M}^{n+1}$, they are also Lipschitz equivalent on the subset $M = \field{M}^{n+1}_+$. 

\vspace{1pc}
For $M = \field{M}^{n+1}_+$, consider the metric completions:
\begin{align*}
(\overline{M}_R, \overline{d}_R) & \; = \; \textrm{the metric completion of } (M, d^R)\\[.5pc]
(\overline{M}_N, \overline{d}_N) &  \; = \; \textrm{the metric completion of } (M, d^N)\\[.5pc]
(\overline{M}_E, \overline{d}_E) & \; = \; \textrm{the metric completion of } (M, d^E)
\end{align*}

\noindent
We have in fact already studied $(\overline{M}_R, \overline{d}_R)$, in Example \ref{Euclidean_halfspace} above. Using the restriction of the Euclidean distance $d_{\field{E}^{n+1}}$, we noted the metric space isometries:
$$( \, \overline{M}_R \, , \, \overline{d}_R  \, )  \; \approx \; ( \, [0, \infty) \times \field{R}^n \, ,  \, d_{\field{E}^{n+1}} \, )\; \approx \; ( \, [0, \infty) \times \field{R}^n \, ,  \, d^R \, )$$
Now consider $(\overline{M}_N, \overline{d}_N)$. Fix any Cauchy sequence $\{(u_k, {\bf x}_k)\}$ in $(\field{M}^{n+1}_+, d^N)$. It follows that $\{(u_k, {\bf x}_k)\}$ is also Cauchy in the ambient metric space $(\field{M}^{n+1}, d^N)$. But in Example \ref{exMinkowski} we saw that $(\field{M}^{n+1}, d^N)$ is complete, and that the metric space topology is the standard Euclidean topology. It follows then that $\{(u_k, {\bf x}_k)\}$ has a limit $(u_0, {\bf x}_0) \in [0, \infty) \times \field{R}^n$. Now fix any two Cauchy sequences $\{(u_k, {\bf x}_k)\}$ and $\{(v_k, {\bf y}_k)\}$ in $(M, d^N)$, with corresponding limits $(u_0, {\bf x}_0)$ and $(v_0, {\bf y}_0)$ in $[0, \infty) \times \field{R}^n$, and consider the corresponding elements $[\{(u_k, {\bf x}_k)\}]_N$ and $[\{(v_k, {\bf y}_k)\}]_N$ in $\overline{M}_N$. Then we have:
\begin{align*}
\overline{d}_N \, (\, [\{(u_k, {\bf x}_k)\}]_N \, , \, [\{(v_k, {\bf y}_k)\}]_N \, ) & \; = \; \lim_{k \, \to \, \infty} \, d^N ( (u_k, {\bf x}_k) , (v_k, {\bf y}_k)) \\[1pc]
& \; = \; d^N ((u_0, {\bf x}_0) , (v_0, {\bf y}_0)) \\[1pc]
&\;  = \; \max\{|u_0-v_0|, \|{\bf x}_0-{\bf y}_0\| \}
\end{align*}

\noindent
Thus, with the usual identifications, $\overline{d}_N$ is simply the restriction of the standard null distance function on the full Minkowski space, $d^N$, and we have the following metric space isometry:
$$( \, \overline{M}_N \, , \, \overline{d}_N  \, )  \; \approx \; ( \, [0, \infty) \times \field{R}^n \, ,  \, d^N \, )$$

\vspace{.5pc}
\noindent
The analysis of $(\overline{M}_E, \overline{d}_E)$ is now essentially identical. Fix any Cauchy sequence $\{(u_k, {\bf x}_k)\}$ in $(\field{M}^{n+1}_+, d^E)$. It follows that $\{(u_k, {\bf x}_k)\}$ is also Cauchy in the ambient metric space $(\field{M}^{n+1}, d^E)$. But in Example \ref{exMinkowski} we saw that $(\field{M}^{n+1}, d^E)$ is complete, and that the metric space topology is the standard Euclidean topology, thus $\{(u_k, {\bf x}_k)\}$ has a limit $(u_0, {\bf x}_0) \in [0, \infty) \times \field{R}^n$. Now fix any two Cauchy sequences $\{(u_k, {\bf x}_k)\}$ and $\{(v_k, {\bf y}_k)\}$ in $(M, d^E)$, with corresponding limits $(u_0, {\bf x}_0)$ and $(v_0, {\bf y}_0)$ in $[0, \infty) \times \field{R}^n$, and consider the corresponding elements $[\{(u_k, {\bf x}_k)\}]_E$ and $[\{(v_k, {\bf y}_k)\}]_E$ in $\overline{M}_E$. Then we have:
\begin{align*}
\overline{d}_E \, (\, [\{(u_k, {\bf x}_k)\}]_E \, , \, [\{(v_k, {\bf y}_k)\}]_E \, ) & \; = \; \lim_{k \, \to \, \infty} \, d^E ( (u_k, {\bf x}_k) , (v_k, {\bf y}_k)) \\[1pc]
& \; = \; d^E ((u_0, {\bf x}_0) , (v_0, {\bf y}_0)) \\[1pc]
&\;  = \; |u_0-v_0| \; + \;  \|{\bf x}_0-{\bf y}_0\| 
\end{align*}

\noindent
Thus, with the usual identifications, $\overline{d}_E$ is again simply the restriction of the standard elevator distance function on the full Minkowski space, $d^E$, and we have the following metric space isometry:
$$( \, \overline{M}_E \, , \, \overline{d}_E  \, )  \; \approx \; ( \, [0, \infty) \times \field{R}^n \, ,  \, d^E \, )$$
Viewed as distance functions on $[0, \infty) \times \field{R}^{n+1}$, $\overline{d}_R$, $\overline{d}_N$, $\overline{d}_E$ are thus Lipschitz equivalent, and in particular, we have homeomorphisms:
$$[0, \infty) \times \field{R}^n \; \approx \; \overline{M}_R \; \approx \; \overline{M}_N \approx \; \overline{M}_E$$
As for the past causal completion $M^-$ of $M = \field{M}^{n+1}_+$, it is straightforward to verify that in this case we do indeed have set identifications:
$$M^- \; \approx \; [0,\infty) \times \field{R}^n \; \approx \; \overline{M}_R \; \approx \; \overline{M}_N  \; \approx \; \overline{M}_E$$
Moreover, note that the unique continuous extensions $\overline{\tau}_N : \overline{M}_N \to \field{R}$,  $\overline{\tau}_R : \overline{M}_R \to \field{R}$,  $\overline{\tau}_E : \overline{M}_E \to \field{R}$ of $\tau : M \to \field{R}$ are again simply $\overline{\tau}_E = \overline{\tau}_N = \overline{\tau}_R = \tau = t$. This then gives set identifications:
\begin{align*} 
\field{R}^n \; \approx \; \{0\} \times \field{R}^n & \; \approx \; (\overline{\tau}_R)^{-1}(0) \; \approx \; \overline{M}_R - M\\[1pc]
& \; \approx \; (\overline{\tau}_N)^{-1}(0) \; \approx \; \overline{M}_N - M \\[1pc] 
& \; \approx \; (\overline{\tau}_E)^{-1}(0) \; \approx \; \overline{M}_E - M \\[1pc] 
& \; \approx \; \; \;  \d^-(M) \; \; \; \approx \; M^- - M
\end{align*}
\begin{flushright} $\Diamond$\end{flushright}
\end{exm}

\vspace{1pc}
\subsection{GRW Null Distance}

\vspace{1pc}
We now turn to null distance on GRW spacetimes. We first devote some attention to lifting spatial curves to piecewise null curves in GRW spacetimes. deSitter space, $\field{R} \times_f \field{S}^n$, $f(t) = \cosh t$, gives a simple example of a GRW spacetime in which not every `spatial curve' can be (fully) lifted to a causal curve. However, the following observation, which is straightforward to verify, is that sufficiently short spatial curve segments can always be lifted.

\begin{lem} [Null Lifts in GRW spacetimes]  \label{GRW_null_lifts} Consider a GRW spacetime $(M,g) = ((a,b) \times S, -dt^2 + f^2(t)h)$, with $- \infty \le a < b \le \infty$. Let $\s = \s(s)$, $0 \le s \le L$, be a piecewise smooth curve segment in $(S,h)$, with unit $h$-speed, $\|\s'(s)\|_h = 1$. 

\ben
\item [(1)] (Future Null Lifts) For any $c \in (a,b)$, define $\psi_c^+ : (a-c, b-c) \to \field{R}$ by
\begin{align}
\psi_c^+(u) := \int_{c}^{c \, +\, u} \dfrac{1}{f(w)} \, dw \label{future_null_reparam}
\end{align}
Note that $\psi_c^+$ is an increasing function, with supremum given by
$$\ell^+_c  \; := \; \int_c^{b} \dfrac{1}{f(w)}dw \; \in \; (0, \infty]$$
For any $\ell \le L$, with $\ell \; <  \; \ell^+_c$, there is a unique $\delta > 0$ such that $\psi_c^+(\delta) = \ell$. Then for $0 \le u \le \delta$, the reparameterization $\tilde{\s}_c(u) := \s(\psi_c^+(u))$ has (new) $h$-speed $\| \tilde{\s}_c'(u)\|_h  = 1/f(c + u)$, and the curve $\beta(u) = (c + u,\s(\psi_c^+(u)))$ is a piecewise smooth future-directed null curve in $M$, which lifts the initial segment $\s(s)$, $0 \le s \le \ell$. 
\item [(2)] (Past Null Lifts) For any $c \in (a,b)$, define $\psi_c^- : (c-b, c-a) \to \field{R}$ by
\begin{align}
\psi_c^-(u) := \int_{c \, -\, u}^c \dfrac{1}{f(w)} \, dw \label{past_null_reparam}
\end{align}
Note that $\psi_c^-$ is an increasing function, with supremum given by
$$\ell^-_c  \; := \; \int_a^c \dfrac{1}{f(w)}dw \; \in \; (0, \infty]$$
For any $\ell \le L$, with $\ell \; <  \; \ell^-_c$, there is a unique $\delta > 0$ such that $\psi_c^-(\delta) = \ell$. Then for $0 \le u \le \delta$, the reparameterization $\tilde{\s}_c(u) := \s(\psi_c^-(u))$ has (new) $h$-speed $\| \tilde{\s}_c'(u)\|_h  = 1/f(c - u)$, and the curve $\beta(u) = (c-u,\s(\psi_c^-(u)))$ is a piecewise smooth past-directed null curve in $M$, which lifts the initial segment $\s(s)$, $0 \le s \le \ell$. 
\een
\end{lem}

\vspace{2pc}
\begin{figure}[h]
\begin{center}
\includegraphics[width=14cm]{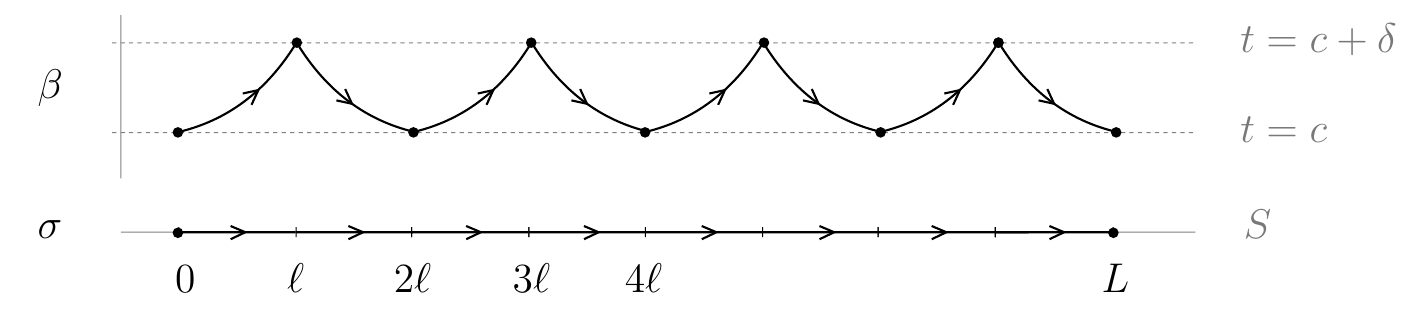}
\caption[]{A spatial curve $\s$ of length $L$ in $S$ and its null sawtooth lift $\beta$ to the slab $[c, c + \delta] \times S$ in the GRW spacetime $I \times_f S$}
\label{sawtoothfigure}
\end{center}
\end{figure}

\vspace{1pc}
Now, given any spatial curve $\s$ in $(S,h)$, we want to construct a `null sawtooth' lift of $\s$ to the spacetime $I \times_f S$, as depicted schematically in Figure \ref{sawtoothfigure}. In other words, we want to generalize Example \ref{example_sawtooth_Mink} and Figure \ref{sawtooth0fig} to arbitrary GRW spacetimes. This is still straightforward though a bit more tedious in the general GRW setting.

\begin{lem} [Null Sawtooth Lifts in GRW spacetimes] \label{sawtoothlemma} Consider a GRW spacetime $(M,g) = ((a,b) \times S, -dt^2 + f^2(t)h)$, where $- \infty \le a < b \le \infty$. Let $\s = \s(s)$, $0 \le s \le L$, be any piecewise smooth curve segment in $(S,h)$, with unit $h$-speed, $\|\s'(s)\|_h = 1$. Fix any $c \in (a,b)$, and as in Lemma \ref{GRW_null_lifts}, let
$$\ell^+_c  \; = \; \int_c^b \dfrac{1}{f(w)}dw$$
Let $k_c \in \field{N}$ be the smallest positive integer for which $L/(2k_c) < \ell_c^+$. Then, fixing any $k \ge k_c$, let $\ell = L/(2k)$. Since $\ell < \ell_c^+$, there exists a unique $\delta > 0$ such that
$$\int_c^{c \,+ \,\delta} \dfrac{1}{f(w)} dw \; = \; \ell$$
The `spatial curve' $\s$ then lifts to a piecewise causal curve $\beta = \beta_1 + \beta_2 + \cdots + \beta_{2k}$ within the slab $[c,c+\delta] \times S$, with $\beta_{2i-1}$ future-directed null, rising from the $t = c$ slice up to the $t = c + \delta$ slice, $\beta_{2i}$ past-directed null, descending from the $t = c + \delta$ slice down to the $t = c$ slice, as in Figure \ref{sawtoothfigure}. Moreover, the null length of $\beta$, with respect to any generalized time function of the form $\tau(t,x) = \phi(t)$ is given by 
$$\hat{L}_\tau(\beta) = [\phi(c+\delta) - \phi(c)] \cdot (2k)  = [\phi(c+\delta) - \phi(c)] \cdot (L/\ell)$$
\end{lem}

\begin{proof} It follows immediately from Lemma \ref{GRW_null_lifts} that $\b_1$ can be constructed as desired, as a future null lift of the initial portion $\s(s)$, $0 \le s \le \ell$. Indeed, by the same argument (and simple reparameterizing), $\b_{2i+1}$ can similarly be constructed as a future null lift of the portion $\s(s)$, $2i \ell \le s \le (2i+1)\ell$. To get $\b_2$, we need to lift down from the $t = c + \delta$ slice. By Lemma \ref{GRW_null_lifts}, we need only verify that $\ell < \ell^{\,-}_{c \, + \delta}$. And, indeed, note that we have:
$$\ell \; = \; \int_c^{c\, +\, \delta} \dfrac{1}{f(w)}dw \; < \; \int_a^{c \, + \, \delta} \dfrac{1}{f(w)}dw \; = \; \ell^{\,-}_{c \,+ \,\delta} $$
$\b_{2i}$ can similarly be constructed as a lift of the subsegment $\s(s)$, $(2i-1)\ell \le s \le 2i \ell$. Finally, we let $\beta := \beta_1 + \beta_2 + \cdots + \beta_{2k}$. The null length of $\beta$ with respect to $\tau(t,x) = \phi(t)$ is clearly given by the formula above.
\end{proof}

\vspace{2pc}
The following generalizes Lemma \ref{Lemma_Null_on_slice_Mink} to GRW spacetimes:

\begin{lem} [Null Distance Estimates on a GRW Spacetime] \label{nulldistinslice} Consider a GRW spacetime $(M,g) = (I \times S, -dt^2 + f^2(t)h)$, and let $d_S$ denote the Riemannian distance function of $(S,h)$. Consider a differentiable time function of the form $\tau(t,x) = \phi(t)$, and let $\hat{d}_\tau$ denote its induced null distance function on $M$.
\ben
\item [(1)] For any two points $p, q \in M$ in the same time slice, $p = (c,p_S)$, $q = (c,q_S)$, 
$$\hat{d}_\tau(p,q) \;  \le \; \phi'(c) \cdot f(c) \cdot d_S(p_S,q_S)$$
\item [(2)] For any two arbitrary points $p, q \in M$, $p = (t_p,p_S)$ and $q = (t_q,q_S)$, and any $c \in I$, we have:
$$\hat{d}_\tau(p,q) \; \le \; |\phi(t_p) - \phi(c)| \; + \; \phi'(c) \cdot f(c) \cdot d_S(p_S,q_S)  \; + \;  |\phi(t_q) - \phi(c)|$$
\een
\end{lem}

\begin{proof} To prove (1), it suffices to assume $p_S \ne q_S$. Let $\s$ be any piecewise smooth curve in $S$ from $p_S$ to $q_S$. We may take $\s$ to be parameterized with respect to $h$-arc length, with $\s = \s(s)$, $0 \le s \le L$. As in Lemma \ref{sawtoothlemma}, let $k_c \in \field{N}$ be the smallest positive integer so that $L/(2k_c) < \ell^+_c$. Then, for any $k \ge k_c$, let $\ell = L/(2k)$, and $\delta > 0$ such that $\psi_c^+(\delta) = \ell$, as above, and let $\b$ be the corresponding null sawtooth lift of $\s$ as in Figure \ref{sawtoothfigure} and Lemma \ref{sawtoothlemma}, which bounces between the slices $t = c$ and $t = c + \delta$, and connects $p = (c,p_S)$ to $q = (c,q_S)$. Then, by Lemma \ref{sawtoothlemma}, 
\begin{align*}
\hat{d}_\tau(p,q) \; \le \; \hat{L}(\b) & \; = \; [\phi(c+\delta) - \phi(c)] \cdot (L/\ell)\\[1pc]
& \; = \; \bigg(\dfrac{\phi(c+\delta) - \phi(c)}{\delta} \bigg) \cdot \bigg(\dfrac{\delta}{\ell}\bigg) \cdot  L\\[1pc]
& \; = \; \bigg(\dfrac{\phi(c+\delta) - \phi(c)}{\delta} \bigg) \cdot \bigg(\; \dfrac{\delta}{\int_c^{c\, +\, \delta} \tfrac{1}{f(w)}dw}\; \bigg) \cdot L
\end{align*}

\vspace{1pc}
\noindent
Taking $k \to \infty$ means $\delta \to 0$ and the above estimate gives
$$\hat{d}(p, q) \; \le \; \phi'(c) \cdot f(c) \cdot L$$
Finally, taking the infimum over all piecewise smooth curves $\s$ in $S$ from $p_S$ to $q_S$ gives (1). For (2), fix any two points $p = (t_p, p_S)$ and $q = (t_q, q_S)$ in $M$. Fix any $c \in I$, and consider the points $p_c = (c, p_S)$ and $q_c = (c, q_S)$. Then by the triangle inequality, Proposition \ref{null_dist_causal_pairs}, and part (1), we have:
\begin{align*}
\hat{d}_\tau(p,q) & \; \le \; \hat{d}_\tau(p,p_c) \; + \; \hat{d}_\tau(p_c,q_c) \;  + \; \hat{d}_\tau(q_c,q)\\[1pc]
 & \; = \; |\phi(t_p) - \phi(c)|  \; + \; \hat{d}_\tau(p_c,q_c) \; + \;  |\phi(t_q) - \phi(c)|\\[1pc]
 & \; \le \; |\phi(t_p) - \phi(c)| \; + \; \phi'(c) \cdot f(c) \cdot d_S(p_S,q_S)  \; + \;  |\phi(t_q) - \phi(c)|
\end{align*}
\end{proof}

\vspace{1pc} 
We note the following consequence of Lemma \ref{nulldistinslice}:

\begin{lem} [GRW Null Distance: Pinching in Space or Time] \label{null_pinching} Consider a GRW spacetime $(M,g) = ((a,b) \times S, -dt^2 + f^2(t)h)$, where $- \infty \le a < b \le \infty$. Consider a differentiable time function of the form $\tau(t,x) = \phi(t)$, and let $\hat{d}_\tau$ denote its induced null distance function on $M$. 
\ben
\item [(1)] If $c \in (a,b)$, and $\phi'(c) = 0$, then for any two points $p, q \in \{t = c\}$,
$$\hat{d}_\tau(p,q) = 0$$
\item [(2)] Suppose that there is a sequence $a_j \to a^+$ along which $[\phi'(a_j) \cdot f(a_j)] \to 0$. Set $\phi_{\inf}:= \inf \{\phi(t) : a < t < b\} = \lim_{t \to a^+} \phi(t) \ge - \infty$. Then for any two points $p, q \in M$, $p = (t_p,p_S)$, $q = (t_q,q_S)$, we have:
$$\hat{d}_\tau(p,q) \; \le \; \phi(t_p) + \phi(t_q) - 2 \phi_{\inf}$$
In particular, if $\phi_{\inf} = 0$, then for all $p, q \in M$,
$$\hat{d}_\tau(p,q) \; \le \; \phi(t_p) + \phi(t_q)$$
More generally, if $\phi_{\inf} > - \infty$ is finite, then any sequence $p_k = (t_k, x_k)$ with $t_k \to a^+$ is a Cauchy sequence in $(M,\hat{d}_\tau)$, (regardless of the spatial motion of $\{x_k\}$). Now let $(\overline{M}_N, \overline{d}_N)$ be the metric completion of $(M,\hat{d}_\tau)$, and $\overline{\tau}_N : \overline{M}_N \to \field{R}$ the unique continuous extension of $\tau$, as in Definition \ref{Defnullcompletion} and Corollary \ref{extending_tau_wrt_nulldist}. If $q_k = (u_k,y_k)$ is any other sequence with $u_k \to a^+$, then $\{p_k\}$ and $\{q_k\}$ are equivalent, giving the same element $[\{p_k\}] = [\{q_k\}] =: \overline{p}_0 \in \overline{M}_N$ in the metric completion, and $(\overline{\tau}_N)^{-1}(\phi_{\inf}) = \{\overline{p}_0\}$.
\een
\end{lem}

\begin{proof} (1) follows immediately from Lemma \ref{nulldistinslice}. For (2), fix any two points $p = (t_p, p_S)$ and $q = (t_q, q_S)$ in $M$. Consider first the sequences $p_j:= (a_j, p_S)$ and $q_j := (a_j, q_S)$. Then by the triangle inequality, Proposition \ref{null_dist_causal_pairs}, and Lemma \ref{nulldistinslice}, we have, for all sufficiently large $j$, 
\begin{align*}
\hat{d}_\tau(p,q) & \; \le \; \hat{d}_\tau(p,p_j) + \hat{d}_\tau(p_j,q_j) + \hat{d}_\tau(q_j,q)\\[1pc]
 & \; \le \; \phi(t_p) - \phi(a_j) \; + \; \phi'(a_j) \cdot f(a_j) \cdot d_S(p_S,q_S)  \; + \;  \phi(t_q) - \phi(a_j)
 \end{align*}

\vspace{.5pc}
\noindent
The main estimate then follows by taking $j \to \infty$. The rest of (2) follows from considerations like those in the analogous Lemma \ref{Riem_pinching}.
\end{proof}

\vspace{2pc}
In the context of null distance on GRW spacetimes, there is evidently some interplay between time functions and the warping function. To help clarify this relationship, we introduce the following language:

\begin{Def} [TSW Spacetimes] Let $I \subset \field{R}$ be an open interval, and $(S,h)$ a connected Riemannian manifold. Let $f_T : I \to (0, \infty)$ and $f_S : I \to (0, \infty)$ be any smooth, positive functions. We will define the corresponding \emph{time and space warped (TSW) spacetime} to be 
$$(M, g) \; := \; \bigg(\, I \times S \, , \; - \, f_T^2(t) \, dt^2 + f_S^2(t) \, h \, \bigg) \; =: \; I \times_{(\, f_T \, ,\, f_S\, )} S$$
We will refer to $f_T(t)$ as the (TSW) \emph{time/temporal warping function}, and to $f_S(t)$ as the (TSW) \emph{space/spatial warping function}. 
\end{Def}

\vspace{2pc}
Superficially, TSW spacetimes generalize GRW spacetimes. However, every TSW is also GRW spacetimes, as in the following:

\begin{lem} [TSW vs GRW] \label{TSW_v_GRW} Let $I \subset \field{R}$ be an open interval, and $(S,h)$ a connected Riemannian manifold. 
\ben
\item [(1)] Any GRW spacetime $M = I \times_f S$ is a TSW spacetime, with time warping function $f_T \equiv 1$ and space warping function $f_S = f$. Indeed, we will often refer to $f$ as the `\emph{spatial warping function}' of $I \times_f S$; $$I \times_f S \; = \; I \times_{(\,1\,,\,f\,)} S$$ 
\item [(2)] Now consider a TSW spacetime, $M = I \times_{(\,f_T\,,\,f_S\,)} S$. Fix any time $c \in I$, and let $\varphi(t) := \int_{c}^t f_T(w)dw$. Letting $J = \mathrm{Im}(\varphi)$, and $\tau = \varphi(t)$, we have:  
$$M \; = \; \bigg( \, I \times S \, , \, -f_T^2(t)dt^2 + f_S^2(t)h \, \bigg) \; = \bigg( \, J \times S \, , \, -d\tau^2 + f_S^2(\varphi^{-1}(\tau)) h \, \bigg)$$
In other words, the TWS spacetime is also a GRW spacetime,
$$I \times_{(\, f_T \, ,\, f_S\, )} S \; = \; J \times_{(f_S \, \circ \, \varphi^{-1})} S$$
\item [(3)] Note that any TSW metric is also clearly conformal to a GRW metric via: 
$$-f_T^2(t)dt^2 + f_S^2(t)h \; = \; f_T^2(t) \, \bigg( \, - dt^2 + \bigg(\dfrac{f_S(t)}{f_T(t)}\bigg)^2 h \, \bigg)$$ 
\een
\end{lem}

\vspace{2pc}
While Lemma \ref{TSW_v_GRW} shows that TSW spacetimes are nothing new, they are somewhat conceptually useful here. For example,  the following interprets $\phi'(t)$ in terms of an effective TSW time warping function, i.e., a lapse function. 

\begin{lem} [Effective TSW Warping Functions] \label{effective_time_warp} Consider a GRW spacetime, $M = I \times_f S$. Consider a smooth function $\phi : I \to \field{R}$, with $\phi'(t) > 0$ for all $t \in I$, and set $J = \mathrm{Im}(\phi)$. Define $\tau : I \times S \to J$ by $\tau(t,x) := \phi(t)$. Then $\tau$ is a smooth time function on $M$, and the original GRW spacetime can be expressed alternatively as a TSW spacetime
$$M \, = \, \bigg( \, I \times S \, , \, -dt^2 + f^2(t)h \, \bigg) \, = \, \bigg( \, J \times S \, , \, - \dfrac{1}{[\phi'(\phi^{-1}(\tau))]^2}d\tau^2 + f^2(\phi^{-1}(\tau))h \, \bigg)$$
In other words, we have $M \; = \; I \times_f S \; = \; J \times_{(\, 1/(\phi' \, \circ \, \phi^{-1}) \, , \, f \, \circ \, \phi^{-1} \, )} S$. In this context, we may refer to $f_T = 1/\phi'$ as the `\emph{effective (TSW) time warping function}', and to $f_S = f$ as the `\emph{(effective) (TSW) spatial warping function}', where more explicitly we emphasize that $f_T(\tau) = 1/[\phi'(\phi^{-1}(\tau))]$ and $f_S(\tau) = f(\phi^{-1}(\tau))$.
\end{lem}

\vspace{2pc}
Let $M$ be a manifold, with spacetime metric tensor $g$, and generalized time function $\tau$. Recall that we can use the notation $\hat{d}(\tau, g)$ to denote the null distance function on $M$ induced by $\tau$ and $g$. The following addresses the relationship between time functions and the (spatial) warping function of a GRW spacetime: 

\begin{prop} [GRW Null Distance: Standard Time vs Standard Space] \label{null_time_vs_space} Consider a GRW spacetime, $(M,g) = (I \times S, -dt^2 + f^2(t)h)$.  
\ben
\item [(1)] \emph{(Standard Space)} Fix any generalized time function $\tau = \tau(t,x)$ on $M$. For any time $c \in I$, define the function $\psi(t) = \int_{c}^t [1/f(w)]\,dw$, and let $u = \psi(t)$. Then the null distance function induced by $\tau = \tau(t,x)$ and the GRW metric $g = -dt^2 + f^2(t)h$ is the same as that induced by $\tilde{\tau} = \tilde{\tau}(u,x) = \tau(\psi^{-1}(u), x)$ and the (non-warped) product metric $\tilde{g} = -du^2 + h$, that is:
$$\hat{d} \, \bigg(\; \tau(t,x) \; , \; -dt^2 + f^2(t)h \; \bigg) \; = \; \hat{d} \, \bigg(\; \tau(\psi^{-1}(u),x) \; , \; -du^2 + h \; \bigg) $$

\item [(2)] \emph{(Standard Time)} Now consider any smooth time function on $M$ of the form $\tau = \tau(t,x) = \phi(t)$, with $\phi' > 0$. Then the null distance function induced by the original GRW metric $g = -dt^2 + f^2(t)h$ and $\tau = \tau(t,x)$ is the same as that induced by the new GRW metric $\tilde{g} = -d\tau^2 + [\phi'(\phi^{-1}(\tau))]^2f^2(\phi^{-1}(\tau))h$ and the resulting standard time function $\tau = \tau$, that is:
$$\hat{d} \, \bigg(\; \phi(t) \; , \; -dt^2 + f^2(t)h \; \bigg) \; = \; \hat{d} \, \bigg(\; \tau \; , \; -d\tau^2 + [\phi'(\phi^{-1}(\tau))]^2f^2(\phi^{-1}(\tau))h \; \bigg)$$
Thus, the time function $\tau = \phi(t)$ can be interpreted alternatively as a new effective spatial warping function $\tilde{f}(\tau)$;
$$\tau = \phi(t) \; \; \longrightarrow \; \; \tilde{f}(\tau) = \phi'(\phi^{-1}(\tau)) f(\phi^{-1}(\tau))$$
In particular, when the original spacetime $M$ is a (non-warped) product, with $f \equiv 1$, then the choice of a non-standard time function can be reinterpreted instead as introducing a spatial warping function via:
$$\tau = \phi(t) \; \; \longrightarrow \; \; \tilde{f}(\tau) = \phi'(\phi^{-1}(\tau))$$
In this context, for any (original) spatial warping function $f$, we may also refer to $\tilde{f} = \phi' \cdot f$ as the new `\emph{effective (GRW) spatial warping function}'.
\een
\end{prop}

\begin{proof} Both statements rely on the fact that null distance is invariant under conformal changes of the original spacetime metric. The statement in (1) uses Lemma \ref{unwarping}. The underlying statement in (2) was established in the proof of Theorem 3.25 in \cite{nulldist}.
\end{proof}

\vspace{2pc}
Hence, note that in the GRW setting, Lemma \ref{effective_time_warp} interprets a choice of time function $\tau(t,x) = \phi(t)$ on $M = I \times_f S$ in terms of an effective TSW time warping (lapse) function $f_T = 1/\phi'$. Alternatively, part (2) of Proposition \ref{null_time_vs_space} interprets $\tau(t,x) = \phi(t)$ as giving a new effective GRW spatial warping function $\tilde{f} = \phi' \cdot f$. For example, we may now essentially recast the first part of Lemma \ref{nulldistinslice} as follows:

\begin{cor} [GRW Null Distance and Effective Warping on a Time Slice] \label{nulldistinsliceTSW} Consider a GRW spacetime $(M,g) = (I \times S, -dt^2 + f^2(t)h)$, and let $d_S$ denote the Riemannian distance function of $(S,h)$. Consider a smooth time function $\tau(t,x) = \phi(t)$ with $\phi'(t) > 0$, and let $\hat{d}_\tau$ denote its induced null distance function on $M$. Let $f_S = f$ denote the (original) space warping function. Let $f_T = 1/\phi'$ be the effective TSW time warping function, as in Lemma \ref{effective_time_warp}. Alternatively, as in Proposition \ref{null_time_vs_space}, let $\widetilde{f} = \phi' \cdot f$ be the effective GRW spatial warping function. Then for any $c \in I$, and any two points $p = (c,p_S)$, $q = (c,q_S)$, we have:
$$\hat{d}_\tau(p,q) \;  \le \; \dfrac{f_S(c)}{f_T(c)} \cdot d_S(p_S,q_S) = \widetilde{f}(c) \cdot d_S(p_S,q_S)$$
\end{cor}

\vspace{2pc}
Combining Theorems 3.25 and 3.28 in \cite{nulldist}, we have:

\begin{thm} [Null Distance on GRW Spacetimes, \cite{nulldist}] \label{nulldistGRW} Consider a GRW spacetime, $(M,g) = (I \times S, -dt^2 + f^2(t)h)$. Consider a smooth time function of the form $\tau(t,x) = \phi(t)$, with $\phi'(t) > 0$ for all $t \in I$. Then we have the following:
\ben
\item [(1)] The null distance function $\hat{d}_\tau$ is definite.
\item [(2)] The causality of $M$ is encoded in $\hat{d}_\tau$ in the sense that:
$$q \in \overline{I^+(p)} \; \Longleftrightarrow \; \hat{d}_\tau(p,q) =  \tau(q) - \tau(p)$$
\item [(3)] If $(S,h)$ is a complete Riemannian manifold, we have:
$$q \in J^+(p) \; \Longleftrightarrow \; \hat{d}_\tau(p,q) =  \tau(q) - \tau(p)$$
\een
\end{thm}

\vspace{2pc}
\begin{Def} [GRW Standard Null Distance] Consider a GRW spacetime $(M,g) = (I \times S, -dt^2 + f^2(t)h)$. We will refer to the null distance function $\hat{d} _t$, induced by the standard time function $\tau(t,x) = t$, as the \emph{standard null distance function} of $M$.
\end{Def}

\vspace{1pc}
The following is essentially Lemma 2.1 in \cite{nulldistprops}. (Recall that the latter was also recast as the somewhat more general Lemma \ref{null_reparam} above.)

\begin{lem} [GRW Null Distance: Standard vs Nonstandard Time, \cite{nulldistprops}] \label{GRW_null_reparam} Consider a GRW spacetime, $M = I \times_f S$. Suppose that $\tau(t,x) = \phi(t)$ is a differentiable time function on $M$. Let $(\phi')_{\inf} := \inf \{\phi'(t) : t \in I\}$ and $(\phi')_{\sup} := \sup\{\phi'(t) : t \in I\}$. Letting $\hat{d}_\tau$ be the null distance induced by $\tau$, and $\hat{d}_t$ the standard null distance induced by the standard coordinate time function $t$, then for all $p,q \in M$, we have:
$$(\phi')_{\inf} \cdot \hat{d}_{\, t \,}(p,q) \; \le \; \hat{d}_{\, \tau}(p,q) \; \le \; (\phi')_{\sup} \cdot \hat{d}_{\, t \,}(p,q)$$
\end{lem}

\vspace{2pc}
Recall that by Proposition \ref{null_dist_causal_pairs}, the null distance is always known on causal pairs, given by $\hat{d}_\tau(p,q) = \tau(q) - \tau(p)$, if $p \le q$. The following, which addresses non-causal pairs, is again only a slight restatement of Lemma 4.9 in \cite{nulldistprops}. In \cite{nulldistprops}, the assumptions are effectively that $f_{\inf} > 0$ and $f_{\sup} < \infty$. Below, we simply note that the respective inequalities also hold automatically if either, $f_{\inf} = 0$, or $f_{\sup} = \infty$. 

\begin{lem} [GRW Standard Null vs Spatial Distance, \cite{nulldistprops}] \label{GRW_inf_sup_spatial_warp} Consider a GRW spacetime, $M = I \times_f S$, and let $d_S$ denote the Riemannian distance function of $(S,h)$. Let $f_{\inf} := \inf \{f(t) : t \in I\}$ and $f_{\sup} := \sup \{f(t) : t \in I\}$. Then for any two points $p = (t_p, p_S) \in M$, $q = (t_q,q_S) \in M$, which are not causally related, i.e., $q \not \in J^{\pm}(p)$, the standard null distance function $\hat{d}_t$ satisfies:
$$f_{\inf} \cdot d_S(p_S, q_S) \, \le \, \hat{d}_{\, t \,} (p,q) \, \le \, f_{\sup} \cdot d_S(p_S, q_S)$$
\end{lem}

\vspace{2pc}
Combining Lemma \ref{GRW_inf_sup_spatial_warp} and Proposition \ref{null_time_vs_space}, we have the following:

\begin{cor} [GRW Null vs Spatial Distance] \label{GRW_null_bounded_warping} Consider a GRW spacetime, $M = I \times_f S$,  and let $d_S$ denote the Riemannian distance function of $(S,h)$. Consider a smooth time function $\tau(t,x) = \phi(t)$, with $\phi' > 0$, and let $\hat{d}_\tau$ be the induced null distance function on $M$. Fix any two points $p = (t_p, p_S) \in M$, $q = (t_q,q_S) \in M$, with $q \not \in J^{\pm}(p)$. We the obvious definitions below, we have:
\begin{align*}
(\phi')_{\inf} \cdot f_{\inf} \cdot d_S(p_S, q_S) & \, \le \, (\phi' \cdot f)_{\inf} \cdot d_S(p_S, q_S)\\[.5pc]
& \, \le \, \hat{d}_{\,\tau}(p,q) \\[.5pc]
&\, \le \, (\phi' \cdot f)_{\sup} \cdot d_S(p_S, q_S) \, \le \, (\phi')_{\sup} \cdot f_{\sup} \cdot d_S(p_S, q_S)
\end{align*}
\end{cor}

\begin{proof} By Proposition \ref{null_time_vs_space}, we may consider $\tau$ as the standard time function on the metric $\widetilde{g} = -d\tau^2 + (\,\widetilde{f}\, )^2(\tau)h$, where the new effective spatial warping function is $\widetilde{f}(\tau) = \phi'(\phi^{-1}(\tau)) \cdot f(\phi^{-1}(\tau))$, in other words, $\widetilde{f} = \phi' \cdot f$. We may then apply Lemma \ref{GRW_inf_sup_spatial_warp} to get:
$$(\phi' \cdot f)_{\inf} \cdot d_S(p_S, q_S)  \, \le \, \hat{d}_{\,\tau}(p,q) \, \le \, (\phi' \cdot f)_{\sup} \cdot d_S(p_S, q_S)$$
where $(\phi' \cdot f)_{\inf} = \inf \{\, [\phi'(t) \cdot f(t)] : t \in I\}$ and $(\phi' \cdot f)_{\sup} = \sup \{\, [\phi'(t) \cdot f(t)] : t \in I\}$. The outer estimates follow from the fact that $(\phi')_{\inf} \cdot f_{\inf} \le (\phi' \cdot f)_{\inf}$, and similarly for the suprema.
\end{proof}

\vspace{1pc}

\subsection{Some GRW Comparisons}

\vspace{1pc}
Another consequence of Lemma 4.9 from \cite{nulldistprops} is the following:

\begin{cor} [GRW Standard Riemannianized vs Null Distances] \label{GRW_Riem_v_null_std_time} Consider a GRW spacetime, $(M,g) = (I \times S, -dt^2 + f^2(t)h)$. Let $d^R_t$ and $\hat{d}_t$ denote the standard Riemannianized and null distance functions, induced by $\tau(t,x) = t$. Let $f_{\inf} := \inf \{f(t) : t \in I\}$ and $f_{\sup} := \sup \{f(t) : t \in I\}$. Then, interpreting $1/ \infty = 0$ if necessary, we have: 
$$\bigg(\dfrac{f_{\inf}}{f_{\sup}}\bigg) \cdot \hat{d}_t \; \le \; d^R_t \; \le \; \sqrt{2} \cdot \hat{d}_t$$
In particular, if $0 < f_{\inf}$ and $f_{\sup} < \infty$, then $d^R_t$ and $\hat{d}_t$ are Lipschitz equivalent distance functions on $M$.
\end{cor}

\begin{proof} The second inequality holds, by Lemma \ref{null_v_Riem_easy}, since $\nabla \tau = -\d_t$ is everywhere unit timelike. Note that the first inequality holds trivially if either $f_{\inf} = 0$, or $f_{\sup} = \infty$, (where we interpret $1/\infty = 0$). Hence, we may assume that $0 < f_{\inf}$ and $f_{\sup} < \infty$. To prove the first inequality in this case, fix $p, q \in M$, write $p = (t_p, p_S)$, $q = (t_q,q_S)$. Let $\g(s) = (\mu(s) , \s(s))$, $u_1 \le s \le u_2$ be any piecewise smooth curve in $M$ from $p$ to $q$. Let $L_S$ and $d_S$ denote the Riemannian arc length functional and distance function, respectively, of $(S,h)$. Let $L^R_t$ denote the Riemannian arc length functional of $(M,g^R_t)$. We then have:
\begin{align*}
L^R_t(\g) & \; = \; \int_{u_1}^{u_2} \sqrt{\, |\mu'(s)|^2 + f^2(\mu(s)) \, h(\s'(s), \s'(s)) \,} \, ds \\[.75pc]
& \; \ge \; \max \bigg \{ \; \int_{u_1}^{u_2} |\mu'(s)|ds \; , \; \int_{u_1}^{u_2} f(\mu(s)) \sqrt{h(\s'(s), \s'(s)) } \, ds  \; \bigg\}\\[1pc]
& \; \ge \; \max \, \{\,  |t_q - t_p| \, , \, f_{\inf} \cdot \, L_S(\s) \, \}
\end{align*}

\vspace{1pc}
If $p \le q$, then $\g$ can be chosen causal, which gives $L^R_t(\g) \ge |t_q - t_p| = \hat{d}_t(p,q)$. Taking an infimum then gives
$$d^R_t(p,q) \; \ge \; \hat{d}_t(p,q) \; \ge \; \bigg(\dfrac{f_{\inf}}{f_{\sup}}\bigg) \cdot \hat{d}_t(p,q)$$
It remains then to consider the case $q \not \in J^{\pm}(p)$. Since $L^R_t(\g) \ge f_{\inf}  \cdot  L_S(\s)$, we get $L^R_t(\g) \ge f_{\inf}  \cdot  d_S(p_S,q_S)$, and this in turn gives $d^R_t(p,q) \ge f_{\inf} \cdot d_S(p_S, q_S)$. Using this estimate, together with Lemma \ref{GRW_inf_sup_spatial_warp}, then gives
$$f_{\sup}  \cdot  d^R_t (p,q) \; \ge \; f_{\sup}  \cdot  f_{\inf}  \cdot  d_S(p_S,q_S)  \; \ge \; f_{\inf}  \cdot  \hat{d}_t(p,q)$$
\end{proof}

\vspace{1pc}
More generally, we have:

\begin{cor} [GRW Riemannianized vs Null Distances] \label{GRW_Riem_v_null_nonstd_time} Consider a GRW spacetime, $(M,g) = (I \times S, -dt^2 + f^2(t)h)$. Consider a smooth time function of the form $\tau(t,x) = \phi(t)$, with $\phi' > 0$. Let $d^R_{\, \tau} = d^R_t$ and $\hat{d}_{\, \tau}$ be the induced Riemannianized and null distance functions on $M$. Then, interpreting $1/\infty = 0$ and $1/(0^+) = \infty$, as necessary, we have:
$$\bigg(\dfrac{f_{\inf}}{(\phi')_{\sup} \cdot f_{\sup}}\bigg) \cdot \hat{d}_{\, \tau} \; \le \; d^R_{\, t} \; \le \; \bigg(\dfrac{\sqrt{2}}{\; (\phi')_{\inf}} \bigg) \cdot \hat{d}_{\, \tau}$$
\end{cor}

\begin{proof} First recall that by Corollary \ref{corGRWRiemannize}, we indeed have $d^R_{\, \tau} = d^R_{\,t}$. For the standard Riemannianized and null distance functions, Corollary \ref{GRW_Riem_v_null_std_time} gives $(f_{\inf}/f_{\sup}) \, \hat{d}_{\, t\,}  \le d^R_{\, t\,}$. By applying Lemma \ref{null_v_Riem_easy}, we then get: 
$$\bigg( \dfrac{f_{\inf}}{f_{\sup}} \bigg) \, \hat{d}_{\, t\,} \, \le \, d^R_{\, t} \, \le \, \bigg( \dfrac{\sqrt{2}}{(\phi')_{\inf}}\bigg) \, \hat{d}_{\, \tau}$$
Finally, by Lemma \ref{GRW_null_reparam}, we have $[1/(\phi')_{\sup}] \cdot \hat{d}_{\, \tau} \le \hat{d}_{\, t \,}$, and using this on the left gives the result. \end{proof}

\vspace{1pc}
\subsection{Causality}

\vspace{1pc} 
We now turn attention to causality. Recall first that, for any spacetime $(M,g)$, and any generalized time function $\tau$, as in Proposition \ref{null_dist_causal_pairs}, we have:
$$p \le q \; \implies \hat{d}_\tau(p,q) = \tau(q) - \tau(p)$$
Furthermore, recall that by Theorem \ref{nulldistGRW}, on any GRW spacetime, $(M,g) = (I \times S, -dt^2 + f^2(t)h)$, if $(S,h)$ is complete, the null distance function induced by any smooth time function of the form $\tau(t,x) = \phi(t)$, with $\phi'(t) > 0$, satisfies: 
$$p \le q \; \; \Longleftrightarrow \; \; \hat{d}_\tau(p,q) \, = \, \tau(q) - \tau(p)$$
Moreover, for general fiber $(S,h)$, we have:
$$q \in \overline{I^+(p)} \; \; \Longleftrightarrow \; \; \hat{d}_\tau(p,q) \, = \, \tau(q) - \tau(p)$$
In light of Lemma \ref{time_from_null_dist}, the equivalences above say that, at least in the model GRW setting, the causal relation of the original spacetime $(M,g)$ is essentially `\emph{encoded}' in the resulting null distance metric space $(M,\hat{d}_\tau)$. This is clearly a very nice property to have when converting a spacetime into a metric space. And, indeed, the latter relationship above should generalize much more broadly.

\vspace{1pc} 
We now observe, however, that no such `causal encoding' property holds, in general, for the Riemannianized or elevator distances. First recall, from Example \ref{exMinkowski}, that in the nicest possible setting, the standard Riemannianized and elevator distances on Minkowski space satisfy the following causality properties:
\begin{align*}
p \le q  & \; \; \Longleftrightarrow \; \; d^R(p,q) \, \le \, \sqrt{2} \, (t_q - t_p)\\[.5pc]
p \le q &  \;  \; \Longleftrightarrow \; \; d^E(p,q) \, \le \, 2 \, (t_q - t_p)
\end{align*}
Hence, in this simplest model situation, both these distance functions do indeed encode the causal relation of the spacetime. However, a key difference stands out, as compared to the null distance, this being the \emph{inequalities} on the right hand side. As we see in the following example, this difference is highly consequential.

\vspace{1pc} 
\begin{exm} [`Causality Leaking' in Riemannianized, Elevator Distances] \label{Riem_does_not_encode} Fix a small $0 < \delta << 1$. First let $\rho_\delta : \field{R} \to [0,1-\delta]$ be a smooth bump function such that $0 \le \rho_\delta(t) \le 1 - \delta$ for all $t \in \field{R}$, $\rho_\delta(t) = 0$ for $t \le 1$ or $t \ge 1+2\delta$, and $\rho_\delta(1+\delta) = 1 - \delta$. Then define $f_\delta : \field{R} \to \field{R}$, by $f_\delta(t) := 1 - \rho_\delta(t)$. Thus, $\delta \le f_\delta(t) \le 1$ for all $t \in \field{R}$, $f_\delta(t) = 1$ for all $t \le 1$ or $t \ge 1 + 2\delta$, and $f_\delta(1+\delta) = \delta$, as in Figure \ref{encoding_fail2}. 

\vspace{1pc}
\begin{figure}[h]
\begin{center}
\includegraphics[width=12cm]{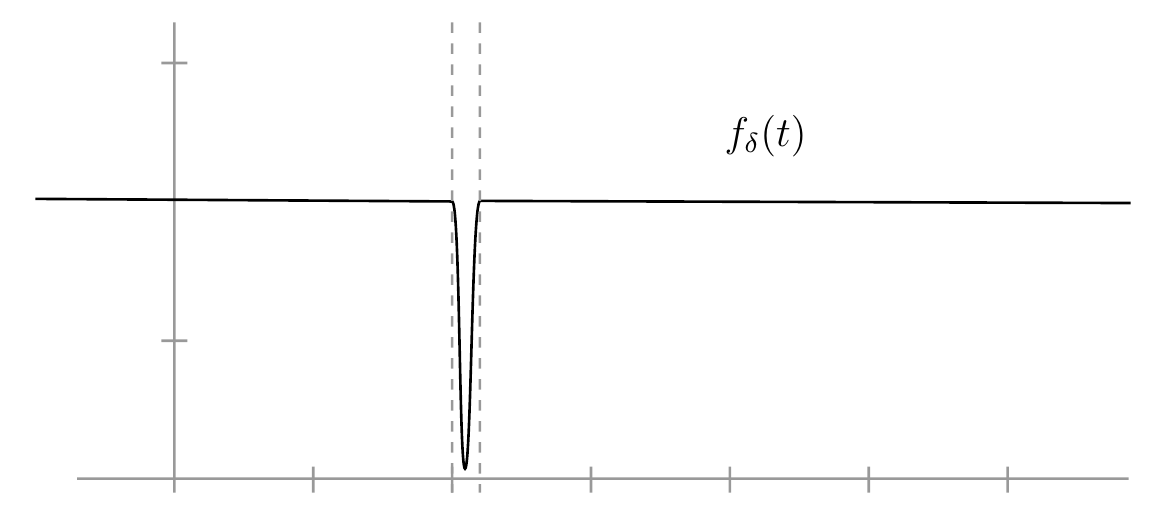}
\end{center}
\vspace{0pc}
\caption[] {The spatial warping function $f_\delta$ in Example \ref{Riem_does_not_encode}.}
\label{encoding_fail2}
\end{figure}

\vspace{1pc}
Now consider the 2-dimensional GRW spacetime $(M,g) = (\field{R} \times \field{R} , -dt^2 + f_\delta^2(t)dx^2)$. Note that outside the strip $1 < t < 1 + 2\delta$, the metric tensor is precisely the Minkowski metric tensor. Consider the point $p = (0,0)$. Fix any $L > 0$, any $0 < a < 1$, and consider point $q = (1 -a, L)$. Consider the standard Riemannianized and elevator distance functions, $d^R = d^R_t$ and $d^E = d^E_t$, on $M$. Let $d^Z \in \{d^R, d^E\}$ denote any one of these. Applying part (2) of either Lemma \ref{RiemdistGRW_basic_estimates} or \ref{elevator_basic_estimates} with $c = 1 + \delta$, we have:
\begin{align*}
d^Z(p,q) & \; \le \; |(0) - (1+\delta)| \; + \; f_\delta(1+\delta) \cdot L \; + \; | (1 - a) - (1 + \delta)|\\[1pc]
& \; = \; 1\; + \; \delta \cdot (L+2)\; + \; a
\end{align*}

\vspace{1pc}
\begin{figure}[h]
\begin{center}
\includegraphics[width=12cm]{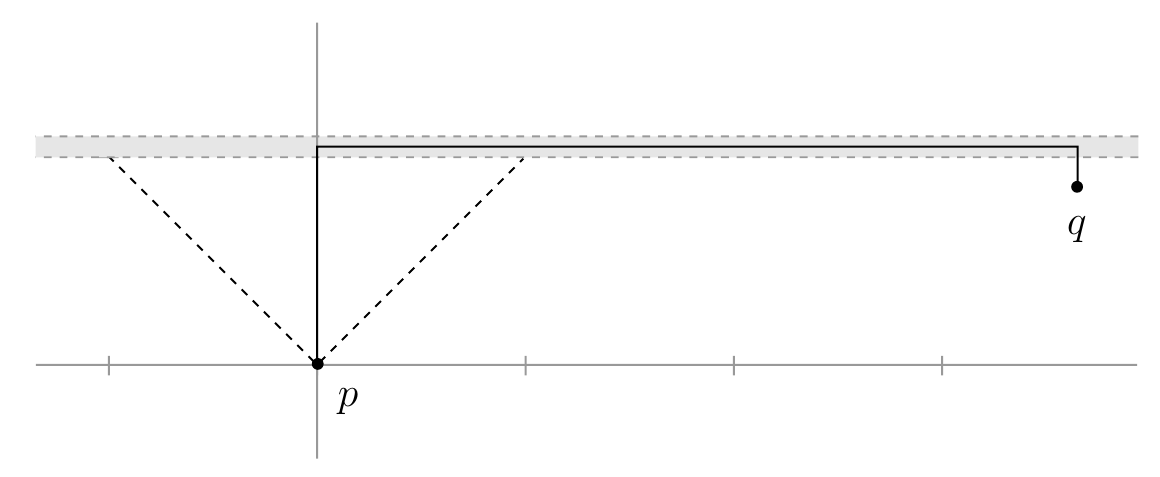}
\end{center}
\vspace{-1pc}
\caption[] {The points $p$ and $q$ in Example \ref{Riem_does_not_encode}.}
\label{encoding_fail2}
\end{figure}

\vspace{2pc}
\noindent
Now take $Z = R$, that is, consider the standard Riemannianized distance, $d^R$. Think of $L$ as fixed. Note that taking both $\delta$ and $a$ sufficiently small makes the right hand side above very close to 1, and thus also:
$$d^R(p,q) \; \le \; \sqrt{2} \cdot (1-a) \; = \; \sqrt{2} \cdot (t_q - t_p)$$
But note that the metric tensor up to time $t = 1$ is precisely the Minkowski metric. This means, for example, that $J^+(p) \cap \{t < 1\} = \{(t,x) : |x| \le t < 1\}$. Since $q = (1-a,L) \in \{t < 1\}$, then $q = (1-a, L) \in J^+(p)$ iff $L \le 1 - a$. Hence, for any choice of $L > 1$, we have $p \not \le q$. That is, with $p = (0,0)$, there are many points $q \in M$, for which 
$$d^R(p,q) \; \le \; \sqrt{2} \cdot (t_p - t_q)\hspace{2pc} \textrm{but} \hspace{2pc} p \; \not \le \; q$$
It follows that the relation $d^R(p,q) \; \le \; \sqrt{2} \cdot (t_p - t_q)$ does not encode the causality relation. And no (simple) modification of the inequality will repair the issue, since, for example, for all $b \in [0,1]$, the point $z := (b,b)$ \emph{does} satisfy both $p \le z$ and $d^R(p,z) = \sqrt{2}\cdot b = \sqrt{2} \cdot (t_z - t_p)$. A similar argument and conclusion applies to the standard elevator distance, $d^E$. 
\end{exm}

\pagebreak
\section{GRW Faithfulness and Singularities} \label{sec_GRW_faith_sing}

\vspace{1pc}
\subsection{GRW Faithfulness}

\vspace{1pc}
On GRW spacetimes, the standard Riemannianized, null, and elevator distance functions share some basic common properties, which we now abstract. 

\vspace{1pc}
\begin{Def} [GRW Faithful Distance Functions] \label{Def_GRWfaithful} Consider a GRW spacetime $(M,g) = (I \times S, -dt^2 + f^2(t)h)$, $I =(a,b)$, with $- \infty \le a < b \le \infty$. For any subinterval $J \subset I$, set $A_J := \inf\{f(t) : t \in J\}$, and $f_{\inf} := \inf \{f(t) : t \in I\} = A_I$. We define the following \emph{GRW faithfulness properties}, (F1)-(F6), for an (abstract) distance function $d$ on $M$.
\ben
\item [(F1):] There is a positive constant $C_1 > 0$, such that:
$$\; p_S = q_S \; \; \Longrightarrow \;\;  C_1 \cdot |t_q - t_p| \; \le \; d(p,q)$$
\item [(F2):] There is a positive constant $C_2 > 0$, such that:
$$\; p_S = q_S \; \; \Longrightarrow \;\;  d(p,q) \; \le \; C_2 \cdot |t_q - t_p|$$
\item [(F3):] There is a positive constant $C_3 > 0$, such that:
$$t_p = t_q \; \; \Longrightarrow \; \; d(p,q) \; \le \; C_3 \cdot  f(t_p) \cdot d_S(p_S,q_S) $$
\item [(F4):] There is a positive constant $C_4 > 0$, such that, for all points $p,q \in M$, and all $c \in I$, we have:
$$d(p,q) \; \le \; C_4 \cdot \bigg(\, |t_p-c| \, + \, f(c) \cdot d_S(p_S,q_S) \, + \, |t_q-c| \, \bigg)$$
\item [(F5):] There is a positive constant $C_5 > 0$, such that, for all points $p,q \in M$,
$$C_5 \cdot f_{\inf} \cdot d_S(p_S,q_S) \; \le \; d(p,q)$$
\item [(F6):] There is a positive constant $C_6 > 0$, such that, for any subinterval $(t_1, t_2) \subset (a,b)$, and all points $p,q \in (t_1,t_2) \times S$, we have:
$$\; C_6 \cdot \min \, \{ \, \, E_{(t_1,t_2)}(t_p,t_q)  \, \, , \, \, A_{(t_1,t_2)} \cdot d_S(p_S,q_S)\,  \,\} \; \le \; d(p,q)$$
where the `\emph{vertical exit cost}' $E_{(t_1,t_2)}(t_p,t_q)$ is defined by
\begin{displaymath}
   E_{(t_1,t_2)}(t_p,t_q) :=  \left\{
     \begin{array}{lr}
       \min \{ \, t_p + t_q - 2t_1 \, , \, 2t_2 - t_p - t_q\, \} & \hspace{1pc} a < t_1 < t_2 < b\\[.5pc]
       t_p + t_q - 2t_1 & a < t_1 < t_2 = b\\[.5pc]
       2t_2 - t_p - t_q & a = t_1 < t_2 < b\\[.5pc]
       \infty & a = t_1 < t_2 = b
     \end{array}
   \right.
\end{displaymath}

\een
\end{Def}

\vspace{2pc}
First note that the following is immediate:

\begin{lem} \label{faithful_Lip_equiv} Consider a GRW spacetime $M = I \times_f S$. Let $d_1$, $d_2$ be two distance functions on $M$, which are Lipschitz equivalent. Then for any $N \in \{1, 2, ..., 6\}$, $d_2$ satisfies property (FN) iff $d_1$ does.
\end{lem}

\vspace{1pc}
To begin investigating these conditions more closely, first note that (F1) and (F2) are weaker versions of the Lipschitz and anti-Lipschitz conditions, respectively:

\begin{lem} [(F1), (F2), and Lipschitz Properties] \label{faithful_v_Lip} Consider a GRW spacetime $M = I \times_f S$, and let $d$ be a distance function on $M$. Consider the standard time function $\tau(t,x) = t$, and the metric space map $\tau : (M,d) \to (\field{R}, |\cdot |)$.
\ben
\item [(1)] If $\tau$ is $\l_1$-Lipschitz, then $d$ satisfies (F1) with $C_1 = 1/\l_1$.
\item [(2)] If $\tau$ is $\l_2$-anti-Lipschitz, then $d$ satisfies (F2) with $C_2 = 1/\l_2$.
\een
\end{lem}

\vspace{2pc}
Similarly, we get the following version of Lemma \ref{antiLip_Cauchy}: 

\begin{lem} [(F2) and Vertical Cauchy Sequences] \label{F2_vertical_Cauchy} Suppose that $d$ is a distance function on a GRW spacetime $M = I \times_f S$, which satisfies (F2). Suppose that $\{t_k\}$ is a Cauchy sequence in $(I,|\cdot|)$, and thus converges to a finite value $c \in \field{R}$. Then fixing any spatial location $p_S \in S$, the sequence $p_k := (t_k,p_S)$ is Cauchy in $(M,d)$. Moreover, if the standard time function $\tau(t,x) = t$ admits a continuous extension $\overline{\tau} : \overline{M} \to \field{R}$, then $[\{p_k\}] \in (\overline{\tau})^{-1}(c) \subset \overline{M}$.
\end{lem}

\begin{proof} Let $C_2$ as in (F2). Fix any $\e > 0$. Since $\{t_k\}$ is a Cauchy sequence in $(I, |\cdot |)$, there is an $N \in \field{N}$ such that, for all $i, j \ge N$, we have $|t_i - t_j| < \e/C_2 $, and thus, by property (F2), $d(p_i,p_j)  \le C_2 \cdot |t_i - t_j|  < \e$. Hence, $\{p_k\}$ is Cauchy in $(M,d)$. Moreover, if a such an extension $\overline{\tau}$ exists, then as in Proposition \ref{extendfunctionprop}, we have $\overline{\tau}([\{p_k\}]) = \lim_{ k \, \to \, \infty} \tau(p_k) = \lim_{ k \, \to \, \infty} t_k = c$. 
\end{proof}

\vspace{2pc}
There are, in fact, several redundancies in the list (F1)-(F6), as we shall now also begin to see. Some of the following arguments mimic those in Lemma \ref{RiemdistGRW_basic_estimates} and similar.

\begin{lem} [Properties (F2), (F3), (F4)] \label{lem_F2_F3_F4} Let $d$ be a distance function on a GRW spacetime. Then we have the following:

\ben
\item [(a)] If $d$ satisfies the both (F2) and (F3), with constants $C_2, C_3$, then it satisfies (F4), with $C_4 = \max \{C_2,C_3\}$. Conversely, if $d$ satisfies (F4) with constant $C_4$, it satisfies both (F2) and (F3), with constants $C_2 = C_3 = C_4$.
\item [(b)] Suppose that $d$ satisfies (F4), with constant $C_4$. For any two points $p, q \in M$, with $t_p \le t_q$, and $A_{[t_p,t_q]} = \min \{f(t) : t_p \le t \le t_q\}$, we have:
\begin{align*}
d(p,q) \; \le \; C_4 \cdot \bigg(\,  |t_q - t_p| \, + \, A_{[t_p,t_q]} \cdot d_S(p_S,q_S)\,  \bigg)
\end{align*}
\een
\end{lem}

\begin{proof} (a): Suppose first that $d$ satisfies (F2) and (F3), with constants $C_2$, $C_3$. Fix any two points $p = (t_p,p_S)$ and $q = (t_q,q_S)$ in $M$. Fix any $c \in I$, and set $p_c := (c,p_S)$ and $q_c:= (c,q_S)$. By the triangle inequality, and properties (F2) and (F3), we have:
\begin{align*}
d(p,q) & \; \le \; d(p, p_c) \; + \; d(p_c, q_c) \; + \; d(q_c,q)\\[1pc] 
& \; \le \; C_2 \cdot |t_p - c| \; + \; C_3 \cdot f(c) \cdot d(p_c, q_c) \; + \; C_2 \cdot |t_q - c| \\[.5pc] 
& \; \le \; \max \{C_2, C_3\} \cdot \bigg( \, |t_p - c| \; + \; f(c) \cdot d(p_c, q_c) \; + \;  |t_q - c| \, \bigg) 
\end{align*}
Thus, $d$ satisfies (F4), with constant $C_4 = \max \{C_2,C_3\}$. Conversely, suppose that $d$ satisfies (F4) with constant $C_4$. Fix any $p, q \in M$. Choosing $c = t_p$ in property (F4) then gives the following:
$$d(p,q) \; \le \; C_4 \cdot \bigg( \, f(t_p) \cdot d_S(p_S,q_S) \; + \; |t_q - t_p| \, \bigg)$$ 
It follows that $d$ satisfies (F2), with constant $C_2 = C_4$, and also (F3), with constant $C_3 = C_4$. (b): Suppose that $d$ satisfies (F4) with constant $C_4$. Fix $p, q \in M$, with $t_p \le t_q$. Note that for any $c \in [t_p,t_q]$, we have: 
$$|t_q - c| + |t_p - c| = (t_q - c) + (c-t_p) = t_q - t_p$$
Pick $c \in [t_p, t_q]$ so that $f(c) = A_{[t_p,t_q]}$. 
\end{proof}

\vspace{2pc}
\begin{lem} [Properties (F1), (F5), (F6)] \label{lem_F1_F5_F6} Let $d$ be a distance function on a GRW spacetime. Then we have the following:
\ben
\item [(a)] If $d$ satisfies property (F6), with constant $C_6 > 0$, then if $t_p \le t_q$, we have:
$$C_6 \cdot \min \{ \, |t_q - t_p| \, , \, A_{[t_p,t_q]} \cdot d_S(p_S,q_S)\,\} \; \le \; d(p,q)$$
\item [(b)] If $d$ satisfies (F6), with constant $C_6$, then it satisfies (F1) and (F5), with constants $C_1 = C_5 = C_6$.
\een
\end{lem}

\begin{proof} The argument for (a) is identical to that for part (6) in Lemma \ref{RiemdistGRW_basic_estimates}. For (b), note that taking $p_S = q_S$ in part (a) gives (F1), with $C_1 = C_6$.  Moreover, taking $(t_1,t_2) = (a,b)$ in (F6), that is, $t_1 = a$ and $t_2 = b$, means $E_{(t_1, t_2)}(t_p,t_q) = E_{(a,b)}(t_p,t_q) = \infty$, and $A_{(t_1,t_2)} = A_{(a,b)} = f_{\inf}$. It follows that (F5) holds with $C_5 = C_6$.
\end{proof}

\vspace{1pc}
The preceding Lemmas \ref{lem_F2_F3_F4} and \ref{lem_F1_F5_F6} show that the full list of GRW faithfulness properties can be reduced to just the pair (F4) and (F6):

\begin{cor} [Properties (F4) and (F6)] \label{F4_F6} Let $d$ be a distance function on a GRW spacetime. If $d$ satisfies properties (F4) and (F6), then it also satisfies (F1), (F2), (F3), (F5). 
\end{cor}

\vspace{2pc}
Further properties of GRW faithful distances will be established below. For now, we note the following:

\begin{thm} [Faithfulness of Riemmianized, Elevator, and Null Distances] \label{RNE_GRW_faithful} Consider a GRW spacetime $(M,g) = (I \times S, -dt^2 + f^2(t)h)$, where $I = (a,b)$, with $- \infty \le a < b \le \infty$, and $M = \{ (t_p,p_S) : t_p \in I, p_S \in S\}$. Consider a smooth time function of the form $\tau(t,x) = \phi(t)$, with $\phi'(t) > 0$, for all $t \in I$. Then, with respect to the GRW faithfulness properties in Definition \ref{Def_GRWfaithful}, we have the following:
\ben
\item [(a)] The Riemannianized distance $d^R_\tau = d^R_t$ satisfies all (F1)-(F6), with constants $C_i = 1$, for all $1 \le i \le 6$.
\item [(b)] The elevator distance $d^E_\tau = d^E_t$ satisfies all (F1)-(F6), with constants $C_i = 1$, for all $1 \le i \le 6$.
\item [(c)] Consider now the null distance $\hat{d}_\tau$.
\ben
\item [(i)] If there are positive constants $C, D$ such that, $C \le \phi'(t) \le D$, for all $t \in I$, then $\hat{d}_\tau$ satisfies all (F1)-(F6), with constants $C_1 = C_5 = C_6 = C$,  $C_2 = C_3 = C_4 = D$. 
\item [(ii)] Conversely, if $\hat{d}_\tau$ satisfies (F1) and (F2), with constants $C_1$, $C_2$, then we have $C_1 \le \phi'(t) \le C_2$, for all $t \in I$.
\een
In particular, $\hat{d}_\tau$ satisfies all (F1)-(F6) iff it satisfies the first two (F1), (F2), and the standard null distance $\hat{d}_t$ satisfies (F1)-(F6), with constants $C_i = 1$, for all $1 \le i \le 6$.
\een
\end{thm}

\begin{proof} For the Riemannianized distance, $d^R_\tau = d^R_t$, this follows immediately from Lemma \ref{RiemdistGRW_basic_estimates}. For the elevator distance, $d^E_\tau = d^E_t$, this follows from the analogous Lemma \ref{elevator_basic_estimates}. Consider now the null distance $\hat{d}_\tau$. Suppose first that there are positive constants $C, D$ such that $C \le \phi'(t) \le D$, for all $t \in I$. To verify (F1) and (F2), recall, from Proposition \ref{null_dist_causal_pairs}, that $\hat{d}_\tau(p,q) = \tau(q) - \tau(p)$, whenever $p \le q$. Suppose that $p, q \in M$ have $p_S = q_S$. We may further suppose that $t_p < t_q$. Then by the Mean Value Theorem, there is a $t^* \in (t_p, t_q)$, such that: 
$$\hat{d}_\tau(p,q) = \tau(q) - \tau(p) = \phi(t_q) - \phi(t_p) = \phi'(t^*) \cdot (t_q - t_p)$$
But then the bounds on $\phi'(t)$ give:
$$C \cdot (t_q - t_p) \; \le \; \hat{d}_\tau(p,q) \; \le \; D \cdot (t_q - t_p)$$
Thus (F1) and (F2) are satisfied, with constants $C_1 = C$ and $C_2 = D$. Note that by (1) of Lemma \ref{nulldistinslice}, we also have:
$$t_p = t_q \; \; \Longrightarrow \; \; \hat{d}_\tau(p,q) \; \le \; D \cdot f(t_p) \cdot d_S(p_S,q_S)$$
which gives property (F3), with $C_3 = D$. Now recall that by Lemma \ref{lem_F2_F3_F4}, (F2) and (F3) imply (F4), with $C_4 = \max \{C_2, C_3\} = D$. By Corollary \ref{GRW_null_bounded_warping}, we have (F5), with $C_5 = C$. To verify (F6), the argument is similar to that in Lemma \ref{RiemdistGRW_basic_estimates}. Fix any subinterval $(t_1, t_2) \subset (a,b)$. Fix any two points $p,q \in (t_1, t_2) \times S$. Consider any piecewise causal curve $\b = \b_1 + \b_2 + \cdots + \b_m$ from $p$ to $q$. Suppose first that $\b$ never leaves the slab $(t_1, t_2) \times S$. Consider an arbitrary subsegment $\b_i$, running from $p_{i-1} = (t_{i-1}, x_{i-1})$ to $p_i = (t_i,x_i)$. Suppose first that $\b_i$ is future-directed. We may parameterize it as $\b_i(u) = (u, \s_i(u))$, with $t_{i-1} \le u \le t_i$. Since $\b_i$ is causal, we have $g(\b_i'(u), \b_i'(u)) \le 0$, and thus $f(u)[h(\s'_i(u), \s'_i(u))]^{1/2} \le 1$. Integrating this from $t_{i-1}$ to $t_i$, we get:
\begin{align*}
t_i - t_{i-1} & \; \ge \; \int_{t_{i-1}}^{t_i} \, f(u) \sqrt{h(\s'_i(u), \s'_i(u))} \, dw \\[1pc]
& \; \ge \; \int_{t_{i-1}}^{t_i} \, A_{(t_1, t_2)}  \sqrt{h(\s'_i(u), \s'_i(u))} \, dw\\[1pc]
& \; = \;  A_{(t_1, t_2)}  \cdot L_S(\s_i) \; \ge \; A_{(t_1, t_2)}  \cdot d_S(x_{i-1}, x_i)
\end{align*}

\vspace{.5pc}
\noindent
If instead $\b_i$ is past-directed, applying the above to its reverse, gives, in any case:
$$|t_i - t_{i-1}| \; \ge \; A_{(t_1, t_2)} \cdot d_S(x_{i-1}, x_i)$$
Then using the Mean Value Theorem again, and the bounds on $\phi'(t)$, we get:
$$\hat{L}_\tau(\b_i) \; = \; |\phi(t_i) - \phi(t_{i-1})| \; \ge \; C \cdot |t_i - t_{i-1}| \; \ge \; C \cdot  A_{(t_1, t_2)}  \cdot d_S(x_{i-1}, x_i)$$
Adding, and using the triangle inequality in $(S, d_S)$, we have:
$$\hat{L}(\b)  \; = \; \sum_{i\, =\, 1}^m \hat{L}_\tau(\b_i) \; \ge \; C \cdot A_{(t_1, t_2)}  \cdot \sum_{i \, =\, 1}^m d_S(x_{i-1}, x_i) \ \; \ge \; C \cdot A_{(t_1, t_2)}  \cdot d_S(p_S, q_S) $$
The estimate above holds if $\b$ does not leave the slab $(t_1, t_2) \times S$. Now suppose otherwise that $\b$ does leave the slab $(t_1, t_2) \times S$, at some point $z_0 = (t_0, x_0) \in \b$, $z_0 \not \in (t_1, t_2) \times S$, i.e., $t_0 \not \in (t_1, t_2)$. Let $\b_0^-$ denote the portion of $\b$ up to $z_0$, and $\b_0^+$ the portion after. Then $\b_0^-$ and $\b_0^+$ are each piecewise causal curves themselves, and it follows, (for example by Lemma 3.6 in \cite{nulldist}), that we have:
$$\hat{L}_\tau(\b) \; = \; \hat{L}_\tau(\b_0^-) \, + \, \hat{L}_\tau(\b_0^+) \; \ge \; |\tau(z_0) - \tau(p)| \, + \, |\tau(z_0) - \tau(q)|$$
By considering cases, as in Lemma \ref{RiemdistGRW_basic_estimates}, and using the Mean Value Theorem and $C \le \phi'(t)$, it follows that we have:
$$\hat{L}_\tau(\b) \; \ge \; C \cdot E_{(t_1,t_2)}(t_p,t_q)$$
Putting the above together, whether $\b$ leaves the slab or not, we see that property (F6) holds, with $C_6 = C$:
\begin{align*}
\hat{L}_\tau(\b) & \; \ge \;  \min \, \{\,  C \cdot E_{(t_1,t_2)}(t_p,t_q) \, , \, C \cdot A_{(t_1,t_2)} \cdot d_S(p_S,q_S) \, \}\\[1pc]
 & \; \ge \;  C \cdot \min \, \{ \, E_{(t_1,t_2)}(t_p,t_q) \, , \, A_{(t_1,t_2)} \cdot d_S(p_S,q_S) \, \}
\end{align*}

\vspace{1pc}
Finally, suppose now, conversely, that $\hat{d}_\tau$ satisfies (F1) and (F2), with constants $C_1$ and $C_2$. Fix any time $t \in I$, any spatial location $x \in S$, and set $p := (t,x)$. For all sufficiently small $h > 0$, such that $t + h \in I$, let $p_h := (t+h, x)$. Since $p \le p_h$, we have $\hat{d}_\tau(p,p_h) = \phi(t+h) - \phi(t)$, and applying (F1) and (F2) to $p$ and $p_h$ gives:
$$C_1 \;  \le \; \dfrac{\phi(t + h) - \phi(t)}{h} \; \le \; C_2$$
Taking the limit as $h \to 0$, we get $C_1 \le \phi'(t) \le C_2$. 
\end{proof}

\vspace{2pc} 
We now generalize Lemmas \ref{Riem_pinching} and \ref{null_pinching} and show that GRW `spatial pinching' is faithfully reflected by any distance function satisfying (F4).

\begin{lem} [(F4) and Spatial Pinching] \label{GRW_faithful_pinching} Consider a GRW spacetime $(M,g) = ((a,b) \times S, -dt^2 + f^2(t)h)$, where $- \infty \le a < b \le \infty$. Suppose that there is some sequence $a_j \to a^+$ along which $f(a_j) \to 0$. Let $d$ be any distance function on $M$, which satisfies GRW faithfulness property (F4) with constant $C_4$. Then for any two points $p, q \in M$, $p = (t_p,p_S)$, $q = (t_q,q_S)$, 
$$d(p,q) \; \le \; C_4 \cdot (\, t_p + t_q - 2 a \,)$$
Moreover, if $a > - \infty$ is finite, then any sequence $p_k = (t_k, x_k)$ with $t_k \to a^+$ is a Cauchy sequence in $(M,d)$, (regardless of the spatial motion of $\{x_k\}$). Morover, letting $(\overline{M}, \overline{d})$ be the metric completion of $(M,d)$, any two sequences $p_k = (t_k, x_k)$ and $q_k = (u_k,y_k)$, with $t_k \to a^+$ and $u_k \to a^+$, are equivalent, giving the same element in the completion, $[\{p_k\}] = [\{q_k\}] =: \overline{p}_0 \in \overline{M}$. In particular, if $\tau(t,x) = t$ admits a continuous extension $\overline{\tau} : \overline{M} \to \field{R}$, then $(\overline{\tau})^{-1}(a) = \{\overline{p}_0\}$ is a singleton.
\end{lem}

\begin{proof} The proof is formally identical to those for Lemmas \ref{Riem_pinching} and \ref{null_pinching}. Fix any two points $p = (t_p, p_S)$ and $q = (t_q, q_S)$ in $M$. By property (F4), we have, for all sufficiently large $j$:
\begin{align*}
d(p,q) 
 & \; \le \; C_4 \cdot [ \, ( t_p - a_j ) \; + \; f(a_j) \cdot d_S(p_S,q_S)  \; + \;  (t_q -a_j) \, ]
 \end{align*}
 
\vspace{.5pc}
\noindent
The estimate then follows by taking $j \to \infty$. Now suppose that $a > - \infty$ is finite. Consider a sequence $p_k = (t_k,x_k)$ with $t_k \to a^+$. Fix $\e > 0$. Let $k_1 \in \field{N}$ such that $t_k - a < \e/(2C_4)$, for all $k \ge k_1$. Now for any $\ell, m \ge k_1$, the main estimate above gives $d(p_\ell,p_m) \le C_4 \cdot (t_\ell - a)  + C_4 \cdot (t_m - a) < \e$. Hence, $\{p_k\}$ is a Cauchy sequence in $(M, d)$. Moreover, let $q_k = (u_k, y_k)$ be another sequence with $u_k \to a^+$. Let $k_2 \in \field{N}$ such that $u_k - a < \e/(2C_4)$ for all $k \ge k_2$. Then, for all $k \ge \max \{k_1, k_2\}$, we have $d(p_k,q_k) \le C_4 \cdot (t_k - a) + C_4 \cdot ( u_k - a)< \e$. Since $\e > 0$ was arbitrary, it follows that we have:
$$\overline{d}([\{p_k\}], [\{q_k\}]) \; = \; \lim_{k \, \to \, \infty} \, d(p_k,q_k) \; = \; 0$$
Hence, $[\{p_k\}] = [\{q_k\}] =: \overline{p}_0 \in \overline{M}$. Finally, if a continuous extension $\overline{\tau} : \overline{M} \to \field{R}$ exists, it follows that we must have $(\overline{\tau})^{-1}(a) = \{\overline{p}_0\}$.
\end{proof}

\vspace{2pc}
\begin{lem} [(F4), (F5), and Minimal Slices: No Pinching] \label{faith_minimal_slice} Consider a GRW spacetime, $M = (I \times S , -dt^2 + f^2(t)h)$, where $I = (a,b)$, $- \infty \le a < b \le \infty$. Suppose that $d$ is a distance function on $M$ which satisfies (F4), (F5), with constants $C_4$, $C_5$. If $f$ achieves a positive minimum value at $t_0 \in I$, i.e., $0 < f(t_0) \le f(t)$, for all $t \in I$, then for any two points in the slice $\{t_0\} \times S$, $p = (t_0,p_S)$, $q = (t_0,q_S)$, we have:
$$C_5 \cdot f(t_0) \cdot d_S(p_S,q_S) \; \le \; d(p,q)  \; \le \; C_4 \cdot f(t_0) \cdot d_S (p_S,q_S)$$ 
In particular, in the case of the standard Riemannianized, elevator, and null distance functions, $d^R = d^R_t$, $d^E = d^E_t$, $d^N = d^N_t$, we have $C_4 = C_5 = 1$, and thus, for any choice $d^Z \in \{d^R,d^E,d^N\}$, 
$$d^Z(p,q)  \; = \; f(t_0) \cdot d_S (p_S,q_S)$$ 
\end{lem}

\begin{proof} Fix any two points $p = (t_0, p_S)$ and $q = (t_0,q_S)$ in the slice $\{t_0\} \times S$. By assumption we have $f(t_0) = f_{\inf} = \inf \{f(t) : t \in I\} > 0$. Then, applying (F5) with $f_{\inf} = f(t_0)$, and (F4) with $c = t_0$, we have:
\begin{align*}
C_5 \cdot f(t_0) \cdot d_S(p_S,q_S) & \; \le \; d(p,q)  \; \le \; C_4 \cdot f(t_0) \cdot d_S(p_S,q_S) 
\end{align*}
\end{proof}

\vspace{1pc}
We now generalize Lemma \ref{faith_minimal_slice}:

\begin{lem} [(F4), (F5), and Completed Slices: Bounded Warping] \label{faith_completed_slices} Consider a GRW spacetime, $M = (I \times S , -dt^2 + f^2(t)h)$, $I = (a,b)$, $- \infty \le a < b \le \infty$. Suppose that $d$ is a distance function on $M$ which satisfies the GRW faithfulness conditions (F4) and (F5), with constants $C_4$ and $C_5$. Let $(\overline{M}, \overline{d})$ be the metric completion of $(M, d)$. Let $(\overline{S}, \overline{d}_S)$ be the metric completion of $(S,d_S)$. Let $(\overline{I}, |\cdot|)$ be the metric completion of $(I, |\cdot|)$. We will also identify $\overline{I}$ with the closure of $I$ in $\field{R}$. Since we are dealing with a few different completions, we can also write $[\{w_k\}]_X$ for elements in the completion $(\overline{X}, \overline{d}_X)$. Suppose that the standard time function $\tau(t,x) = t$ admits a continuous extension $\overline{\tau} : \overline{M} \to \overline{I}$. Let $f_{\inf} = \inf \{f(t) : t \in I \}$ and $f_{\sup} = \sup \{f(t) : t \in I\}$, and suppose that $0 < f_{\inf}$ and $f_{\sup} < \infty$. Fix a time $t_0 \in \overline{I} \cap \field{R}$.

\ben
\item [(1)] Then we have: 
$$(\overline{\tau})^{-1}(t_0) \; = \; \{ \, [\{(t_k, x_k)\}]_M : \, [\{t_k\}]_I = t_0 \in \overline{I} \, , \, [\{x_k\}]_S \in \overline{S} \, \} \; \approx \; \{t_0\} \times \overline{S}$$
\item [(2)] If the $(\overline{\tau})^{-1}(t_0)$ slice is minimal in the sense that $f_0 := \lim_{t \, \to \, t_0} f(t)$ exists, (where the limit is one-sided if $t_0 \in \{a, b\} \cap \field{R}$), and $f_0 = f_{\inf}$, then for any two elements $[\{p_k\}], [\{q_k\}] \in (\overline{\tau})^{-1}(t_0)$, with representative sequences $p_k = (t_k, x_k)$ and $q_k = (u_k, y_k)$, we have:
$$C_5 \cdot f_0 \cdot \overline{d}_S([\{x_k\}], [\{y_k\}]) \; \le \; \overline{d} ([\{p_k\}], [\{q_k\}]) \; \le \; C_4 \cdot f_0 \cdot \overline{d}_S([\{x_k\}], [\{y_k\}])$$ 
\item [(3)] Note that all of the above applies to the standard Riemannianized, elevator, and null distance functions, $d^Z \in \{d^R, d^E, d^N\}$, with $C_4 = C_5 = 1$. Thus, in this setting, we have $(\overline{\tau}_Z)^{-1}(t_0) \approx \{t_0\} \times \overline{S}$, and if the minimality condition in (2) holds,
  we have:
 $$(\, (\overline{\tau}_Z)^{-1}(t_0) \, , \, \overline{d} \, \, ) \; \approx \; (\, \{t_0\} \times \overline{S} \, , \, f_0 \cdot \overline{d}_S \, )$$ 
\een
\end{lem}

\begin{proof} Fix any element $[\{x_k\}]_S \in \overline{S}$. Hence, $\{x_k\}$ is a Cauchy sequence in $(S,d_S)$. Fix any representative Cauchy sequence $\{t_k\}$ in $I$, with $[\{t_k\}]_I = t_0 \in \overline{I}$, that is, with $t_k \to t_0$. Consider the corresponding spacetime sequence $p_k = (t_k, x_k)$. Choosing $c = t_k$ in (F4), then we have:
$$d(p_k,p_\ell)  \; \le \; C_4 \cdot \bigg( |t_k - t_\ell | + f(t_k) \cdot d_S(x_k,x_\ell) \bigg)$$
Since $\{t_k\}$ is Cauchy in $(I, |\cdot |)$, $\{x_k\}$ is Cauchy in $(S,d_S)$, and $\{f(t_k)\}$ is bounded above by $f_{\sup} < \infty$, this shows that $\{p_k\}$ is a Cauchy sequence in $(M,d)$. Since $t_k \to t_0$, we have $[\{p_k\}]_M  \in (\overline{\tau})^{-1}(t_0)$, as in Proposition \ref{extendfunctionprop}. Now let $[\{(u_k, y_k)\}]_M$ be any other such sequence, with $[\{u_k\}]_I = t_0$, and $[\{y_k\}]_S \in \overline{S}$. Setting $q_k := (u_k,y_k)$, then by (F5), we have:
\begin{align*}
\overline{d}_M([\{p_k\}]_M, [\{q_k\}]_M) & \; = \; \lim_{k \, \to \, \infty} \,  d(p_k, q_k)\\[1pc]
& \; \ge \; \lim_{k \, \to \, \infty} \, [ \, C_5 \cdot f_{\inf}  \cdot d_S(x_k,y_k) \, ]\\[1pc]
& \; = \; C_5 \cdot f_{\inf}  \cdot \overline{d}_S([\{x_k\}]_S,[\{y_k\}]_S)
\end{align*}

\noindent
Since $0 < f_{\inf}$, we have $[\{p_k\}]_M \ne [\{q_k\}]_M$ whenever $[\{x_k\}]_S  \ne [\{y_k\}]_S$. Conversely, take an arbitrary element $[\{p_k\}]_M \in (\overline{\tau})^{-1}(t_0)$. Then $p_k = (t_k,x_k)$ is a Cauchy sequence in $(M,d)$, with $t_k \to t_0$. In particular, $[\{t_k\}]_I = t_0 \in \overline{I}$. Now fix any $\e > 0$. Since $\{p_k\}$ is Cauchy in $(M,d)$, and using (F5) again, there is an $N \in \field{N}$ such that for all $k, \ell \ge N$, we have:
$$C_5 \cdot f_{\inf} \cdot d_S(x_k,x_\ell) \; \le \; d(p_k,p_\ell) \; < \; \e$$
Since $0 < f_{\inf}$, this implies that $\{x_k\}$ is Cauchy sequence in $(S,d_S)$, and thus corresponds to an element $[\{x_k\}]_S \in \overline{S}$. We have shown then that: 
$$(\overline{\tau})^{-1}(t_0) \; = \; \{ \, [\{(t_k, x_k)\}]_M : \, [\{t_k\}]_I = t_0 \in \overline{I} \, , \, [\{x_k\}]_S \in \overline{S} \, \} \; \approx \; \{t_0\} \times \overline{S}$$

\vspace{1pc} 
(2) Fix any two elements $[\{p_k\}], [\{q_k\}] \in (\overline{\tau})^{-1}(t_0)$, with representative sequences $p_k = (t_k,x_k)$, $q_k = (u_k,y_k)$. Since $f_0 = f_{\inf}$, and using (F5), and (F4) with $c = t_k$, we have:
\begin{align*}
C_5 \cdot f_0 \cdot \overline{d}_S([\{x_k\}], [\{y_k\}]) & \; = \; \lim_{k \, \to \, \infty} \, [ \, C_5 \cdot f_{\inf} \cdot d_S(x_k,y_k) \,]\\[1pc]
& \; \le \; \lim_{k \, \to \, \infty} \,  d(p_k, q_k)   \; = \; \overline{d}([\{p_k\}], [\{q_k\}])\\[1pc]
&  \; \le \; \lim_{k \, \to \, \infty} \, \bigg[\, C_4 \cdot \bigg(\, |t_k - u_k| \, + \,  f(t_k) \cdot d_S(x_k,y_k) \,\bigg) \, \bigg]\\[1pc]
&  \; = \; C_4 \cdot f_0 \cdot \overline{d}_S([\{x_k\}],[\{y_k\}])
\end{align*}

\end{proof}

\vspace{1pc}
\subsection{Choice of Time Function} 

\vspace{1pc} 
Fix a spacetime $(M,g)$. Given an appropriate time function $\tau$ on $M$, we can Riemannianize to get the metric space $(M,d^R_\tau)$, or we can use the null distance function to get the metric space $(M,\hat{d}_\tau)$. We now explore a few basic concrete examples illustrating how the choice of time function $\tau$ affects the resulting metric spaces.

\vspace{1pc} 
In fact, we will restrict attention here to Minkowski and future half-Minkowski. These have already been analyzed using the standard time function $\tau = t$, in Examples \ref{exMinkowski} and \ref{exfuturehalfMinkowski} above. Here, we will consider nonstandard time functions of the form $\tau(t, {\bf x}) = \phi(t)$. As indicated in Proposition \ref{null_time_vs_space}, this can have a dramatic effect on the resulting null distance functions. However, by Corollary \ref{corGRWRiemannize}, as in Corollary \ref{Riem_of_Mink} below, \emph{this has no effect on Riemannianizations.} Consequently, the focus in the following examples shall be on comparison with the resulting null distance structures. 

\vspace{1pc} 
\begin{cor} \label{Riem_of_Mink} Fix any smooth time function of the form $\tau(t, {\bf x}) = \phi(t)$, with $\phi'(t) > 0$, on Minkowski space. Then the Riemannianized metric tensor $g^R_\tau = dt^2 + d{\bf x}^2 = g_{\field{E}^{n+1}}$ is precisely the standard Euclidean metric tensor, and thus:
\ben
\item [(1)] The induced Riemannianization of Minkowski space, $(\field{M}^{n+1},d^R_\tau)$, is precisely the standard Euclidean metric space $(\field{R}^{n+1}, d_{\field{E}^{n+1}})$. 
\item [(2)] The induced Riemannianization of future half-Minkowski space, $(\field{M}^{n+1}_+, d^R_\tau)$, is precisely the standard Euclidean half-space $(\field{R}^{n+1}_+, d_{\field{E}^{n+1}})$.
\een
\end{cor}

\vspace{2pc} 
When considering the examples below, the following (very) rough null distance analog to Corollary \ref{Riem_of_Mink} can also be kept in mind:

\begin{rmk} Recall that Corollary \ref{null_antiLip} says that, whenever a generalized time function $\tau$ has definite null distance function $\hat{d}_\tau$, then $\tau$ is (globally) anti-Lipschitz with respect to $\hat{d}_\tau$, with anti-Lipschitz constant $\lambda = 1$, \emph{regardless of any other geometric context.} 
\end{rmk}

\vspace{1pc}
Finally, we also note the following:

\begin{lem} \label{Mink_MVT_lemma} Consider a differentiable time function of the form $\tau(t, {\bf x}) = \phi(t)$ on Minkowski space. Fix any two points $p = (t_p,p_S)$ and $q = (t_q,q_S)$, with $t_p < t_q$. Then by the Mean Value Theorem, there is a $t^* \in (t_p, t_q)$ such that $\phi(t_q) - \phi(t_p) = \phi'(t^*)(t_q - t_p)$. Thus, in the case $p_S = q_S$, we have $\hat{d}_\tau(p,q) \; = \; \phi'(t^*) \cdot (t_q - t_p)$.
\end{lem}

\begin{figure}[h]
\begin{center}
\includegraphics[width=8cm]{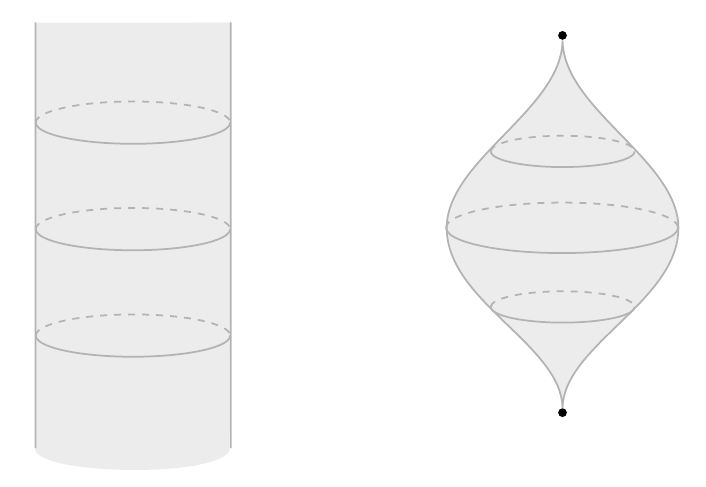}
\end{center}
\vspace{-.5pc}
\caption[]{If we take $\tau = \arctan(t) + \pi/2$ on Minkowski space, $M = \field{M}^{n+1}$, as in Example \ref{ex_arctan_Mink}, the Riemannianization is just Euclidean space, $\field{R}^{n+1}$, which is complete, and is shown on the left. The null distance completion, $\overline{M}_N$, looks much different, and is shown on the right, with its `faux bang point' $\overline{p}_0$ and `faux crunch point' $\overline{q}_0$. (Note: The pictures are visualized schematically as being spatially compactified.)}
\label{arctan_fauxbang_fauxcrunch}
\end{figure}

\vspace{2pc} 
\begin{exm} [Arctangent on Minkowski Space] \label{ex_arctan_Mink} Consider Minkowski space, $M = \field{M}^{n+1} = \{(t,{\bf x}) : t \in \field{R}, {\bf x} \in \field{R}^n\}$, and the time function $\tau : M \to (0, \pi)$, 
$$\tau(t, {\bf x}) = \arctan(t) + \pi/2 = \phi(t)$$

Note first that $\inf \tau = 0$, (even though $M$ has `infinite past', being for example past timelike geodesically complete). Note also that $\tau$ is smooth, with everywhere timelike gradient $\nabla \tau = - [1/(t^2+1)] \d_t$. Hence, by Corollary \ref{timelikegrad} and Theorem \ref{null_dist_properties_thm}, the induced null distance function $\hat{d}_\tau$ is definite, and induces the manifold topology. Let $(\overline{M}_N, \overline{d}_N)$ be the metric completion of $(M,\hat{d}_\tau)$, and as in Corollary \ref{extending_tau_wrt_nulldist}, let $\overline{\tau}_N : \overline{M}_N \to [0,\pi]$ be the unique continuous extension of $\tau : M \to (0, \pi)$. 

\vspace{1pc} 
Since $\tau(t,x) = \phi(t) = \arctan(t) + \pi/2$, we have both $\phi(t) \to 0$ and $\phi'(t) \to 0$ as $t \to - \infty$. Then by Lemma \ref{null_pinching}, we have $\hat{d}_\tau(p,q) \le \tau(p) + \tau(q)$, for all points $p, q \in M$, and as a consequence, any two sequences which approach `past infinity', i.e., any sequences $p_k = (t_k, {\bf x}_k)$ and $q_k = (u_k, {\bf y}_k)$, with $t_k \to -\infty$ and $u_k \to - \infty$, (regardless of their spatial motions), are Cauchy sequences in $(M,\hat{d}_\tau)$ and give the same element $[\{p_k\}] = [\{q_k\}] =: \overline{p}_0$ in $\overline{M}_N$ in the completion. That is, $(\overline{\tau}_N)^{-1}(0) = \{\overline{p}_0\}$ is a singleton. Similar reasoning gives that $(\overline{\tau}_N)^{-1}(\pi) = \{\overline{q}_0\}$ is also a singleton, where we can take, for example, $\overline{q}_0 = [\{(k, {\bf 0})\}]$.

\vspace{1pc}
Indeed, note that $\tau = \arctan(t) + \pi/2 = \phi(t)$ gives $\phi^{-1}(\tau) = \tan(\tau - \pi/2)$, and 
$$\phi'(\phi^{-1}(\tau)) = \dfrac{1}{\tan^2(\tau - \tfrac{\pi}{2}) + 1} = \cos^2(\tau - \tfrac{\pi}{2}) = \sin^2(\tau)$$
Thus, as in Proposition \ref{null_time_vs_space}, with the notation $\hat{d}(\tau, g)$ for the null distance induced on $M$ by the metric tensor $g$ and the (generalized) time function $\tau$, we have:
$$\hat{d} \, \bigg(\; \arctan(t) + \tfrac{\pi}{2} \; , \; -dt^2 + d{\bf x}^2 \; \bigg) \; = \; \hat{d} \, \bigg(\; \tau \; , \; -d\tau^2 + \sin^4(\tau) \, d{\bf x}^2 \; \bigg)$$
where $- \infty < t < \infty$ on the left, and $0 < \tau < \pi$ on the right. In other words, the choice of non-standard time function  can be interpreted instead as the introduction of an effective spatial warping function,
$$\tau = \arctan(t) + \tfrac{\pi}{2} \; \; \longrightarrow \; \; \tilde{f}(\tau) = \sin^2(\tau)$$
and note that this warping function pinches down to zero at both ends;
$$\lim_{\tau \, \to \,0^+} \tilde{f}(\tau) \; = \; 0 \; = \; \lim_{\tau \,\to \,\pi^-} \tilde{f}(\tau)$$
The picture is shown schematically in Figure \ref{arctan_fauxbang_fauxcrunch}, (visualized with compactified spatial topology). In the eyes of the null distance function, the point $\overline{p}_0 \in \overline{M}_N$ might be viewed as a `big bang point'. But this effect is artificial, since the spacetime here is Minkowski space. Hence, we might call $\overline{p}_0$ a `faux bang point'. Similarly, we might interpret $\overline{q}_0$ as a `faux crunch point'. Letting $(\overline{M}_R, \overline{d}_R)$ be the metric completion of $(M, d^R_\tau)$, then as noted above, we know $(M, d^R_\tau) = (\field{R}^{n+1}, d_{\field{E}^{n+1}})$, which is complete, so the completion $(\overline{M}_R, \overline{d}_R) \approx (M, d^R_\tau)$ adds no new points, and the unique extension $\overline{\tau}_R : \overline{M}_R \to [0, \pi]$ has $(\overline{\tau}_R)^{-1}(0) = \emptyset$. In particular, note the disagreement:
$$\emptyset = (\overline{\tau}_R)^{-1}(0)  \not \approx (\overline{\tau}_N)^{-1}(0) = \{\overline{p}_0\}$$

\vspace{1pc}
To put this situation in the context or GRW faithfulness, fix any two points $p = (t_p, p_S)$ and $(t_q,q_S)$, with $p_S = q_S$ and $t_p < t_q$. Then, as in Lemma \ref{Mink_MVT_lemma}, there is a $t^* \in (t_p, t_q)$, such that:
$$\hat{d}_\tau(p,q) \;  = \; \phi'(t^*) \cdot (t_q - t_p)$$
Since $\phi'(t^*) = 1/[(t^*)^2 + 1] \to 0$ as either $t_p \to \infty$, or as $t_q \to - \infty$, we see that $\hat{d}_\tau$ fails the GRW faithfulness property (F1), (and thus also (F6)). The corresponds to the `artificial' null distance pinching as $t \to \pm \infty$. More precisely, as $t \to - \infty$, the effective TSW time warping function $f_T = 1/\phi' \to \infty$, and the effective GRW spatial warping function $\tilde{f} \to 0$, and we get the `faux bang point', $\overline{p}_0$. As $t \to \infty$, $f_T = 1/\phi' \to \infty$, $\widetilde{f} \to 0$, and we get the `faux crunch point', $\overline{q}_0$. 
\begin{flushright} $\Diamond$\end{flushright}
\end{exm}

\vspace{1pc}

\vspace{2pc}
\begin{exm} [$\tau = t^3$ on Future Half-Minkowski] \label{ex_tcubed_futMink} Consider again the future half-Minkowski space, $M = \field{M}^{n+1}_+ = \{t > 0\} \subset \field{M}^{n+1}$. But now consider the time function $\tau : M \to (0,\infty)$ given by
$$\tau(t, {\bf x}) = t^3 = \phi(t)$$
Again, by Corollary \ref{timelikegrad} and Theorem \ref{null_dist_properties_thm}, the induced null distance function $\hat{d}_\tau$ is definite, and induces the manifold topology. 

\vspace{1pc}
Let $(\overline{M}_N, \overline{d}_N)$ be the metric completion of $(M, \hat{d}_\tau)$, and let $\overline{\tau}_N : \overline{M}_N \to [0, \infty)$ be the unique continuous extension. Since both $\phi(t) \to 0$ and $\phi'(t) \to 0$, then as in Lemma \ref{null_pinching}, we again have $\hat{d}_\tau(p,q) \le \tau(p) + \tau(q)$, for all points $p, q \in M$, which shrinks to zero as both points $p, q$ approach the $\{t = 0\}$ slice, and $(\overline{\tau}_N)^{-1}(0) = \{\overline{p}_0\}$ is a single element. Indeed, in the context of Proposition \ref{null_time_vs_space}, we have that:
$$\hat{d} \, \bigg(\; t^3 \; , \; -dt^2 + d{\bf x}^2 \; \bigg) \; = \; \hat{d} \, \bigg(\; \tau \; , \; -d\tau^2 + 9\tau^{4/3}d{\bf x}^2 \; \bigg)$$
In other words, the nonstandard time function $\tau = t^3$ can be swapped for the effective spatial warping function $\tilde{f}(\tau) = 3\tau^{2/3}$. 

\vspace{1pc}
Letting $(\overline{M}_R, \overline{d}_R)$ be the metric completion of $(M, d^R_\tau$), and $\overline{\tau}_R : \overline{M}_R \to [0, \infty)$ the unique continuous extension of $\tau : (0, \infty) \to M$, then, as in Example \ref{exfuturehalfMinkowski} above, we have $(\overline{\tau}_R)^{-1}(0) \approx \{0\} \times \field{R}^n \approx \field{R}^n$, as sets, and in fact as metric spaces. On the other hand, as we saw above, $(\overline{\tau}_N)^{-1}(0) = \{\overline{p}_0\}$ is a single point. So again we have dramatic disagreement;
$$\field{R}^n = (\overline{\tau}_R)^{-1}(0) \not \approx (\overline{\tau}_N)^{-1}(0) = \{\overline{p}_0\}$$

\vspace{1pc}
Again, to put this situation in the context or GRW faithfulness, fix any two points $p = (t_p, p_S)$ and $(t_q,q_S)$, with $p_S = q_S$ and $t_p < t_q$. As in Lemma \ref{Mink_MVT_lemma}, there is a $t^* \in (t_p, t_q)$, such that:
$$\hat{d}_\tau(p,q) \;  = \; 3(t^*)^2 \cdot (t_q - t_p)$$
Since $3(t^*)^2 \to 0$ as $t_q \to 0^+$, we see that here again $\hat{d}_\tau$ fails the GRW faithfulness property (F1), and that this  corresponds to the `artificial' null distance pinching as $t \to 0^+$, where the effective TSW time warping function $f_T = 1/\phi' \to \infty$, and the effective GRW spatial warping function $\tilde{f} \to 0$, giving the `faux bang point', $\overline{p}_0$. 
\begin{flushright} $\Diamond$\end{flushright}
\end{exm}

\vspace{1pc}
\begin{figure}[h]
\begin{center}
\includegraphics[width=8cm]{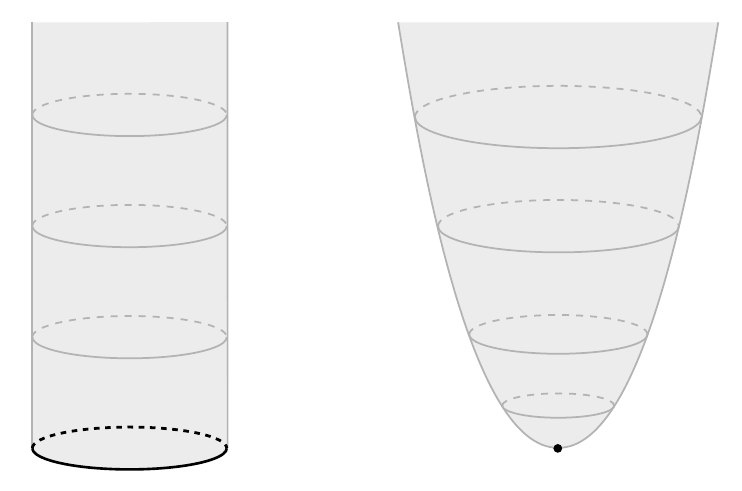}
\end{center}
\caption[]{If we take $\tau = t^3$ on the future half-Minkowski space, $M = \field{M}^{n+1}_+$, as in Example \ref{ex_tcubed_futMink}, the Riemannianized completion $\overline{M}_R$ is the closed half-Euclidean space, $[0, \infty) \times \field{R}^n$, and is shown on the left. The null distance completion, $\overline{M}_N$, is shown on the right, with its `faux bang point' $\overline{p}_0$. (Note: The pictures are again visualized schematically as being spatially compactified.)}
\label{t_cubed_fauxbang}
\end{figure}

\begin{figure}[h]
\begin{center}
\includegraphics[width=8cm]{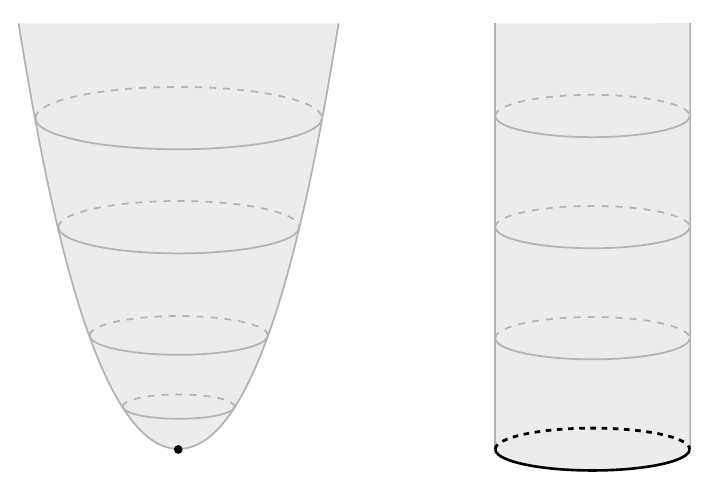}
\end{center}
\caption[]{If we take $\tau = \sqrt{t}$ on $g = -dt^2 + 4td\theta^2$, as in Example \ref{sqrt_blowup}, the Riemannianized completion $\overline{M}_R$ is shown on the left, with its big bang point. The null distance completion, $\overline{M}_N$, shown on the right effectively `blows up' the big bang point.}
\label{sqrt_blowupfig}
\end{figure}

\vspace{2pc}
\begin{exm} [$\tau = \sqrt{t}$ on $g = -dt^2 + 4t d\theta^2$] \label{sqrt_blowup} Consider now the 2-dimensional GRW spacetime given by
$$(M,g) = ( \, (0, \infty) \times \field{S}^1 \, , \, -dt^2 + 4t d\theta^2 \, )$$
with spatial warping function $f(t) = 2\sqrt{t}$, together with the time function $\tau(t,\theta) = \sqrt{t} = \phi(t)$. For the null distance function, as in Proposition \ref{null_time_vs_space}, we have:
$$\hat{d} \, \bigg(\; \sqrt{t} \; , \; -dt^2 + 4t d\theta^2 \; \bigg) \; = \; \hat{d} \, \bigg(\; \tau \; , \; -d\tau^2 + d\theta^2 \; \bigg)$$
where $0 < t < \infty$ on the left, and also $0 < \tau < \infty$ on the right. That is, we can swap the nonstandard time function $\tau = \sqrt{t}$ for the new spatial warping function $\tilde{f}(\tau) = 1$, i.e., no warping at all. This of course means, in particular, that $\tilde{f}(\tau) \to 1$ as $\tau \to 0^+$. The situation here is thus essentially opposite to that in Examples \ref{ex_arctan_Mink} and \ref{ex_tcubed_futMink}, in that the `genuine big bang point' in our original spacetime is `blown up' by the effective warping of the nonstandard time function $\tau = \sqrt{t}$. See Figure \ref{sqrt_blowupfig}. To place this situation in the context or GRW faithfulness, fix any two points $p = (t_p, p_S)$ and $(t_q,q_S)$, with $p_S = q_S$ and $t_p < t_q$. By Lemma \ref{Mink_MVT_lemma}, there is a $t^* \in (t_p, t_q)$, such that we have:
$$\hat{d}_\tau(p,q) \;  = \; \dfrac{1}{2\sqrt{t^*}} \cdot (t_q - t_p)$$
Since $\phi'(t^*) = 1/(2\sqrt{t^*}) \to \infty$ as $t_q \to 0^+$, we see that, in this case, $\hat{d}_\tau$ fails the GRW faithfulness property (F2), and that this corresponds to an `artificial null expansion' as $t \to 0^+$, where the effective TSW time warping function $f_T = 1/\phi' \to 0$, the effective GRW spatial warping function $\tilde{f} = 1$ is bounded away from zero, and we do not detect the `genuine big bang point'. 
\begin{flushright} $\Diamond$\end{flushright}
\end{exm}

\vspace{1pc}

\begin{figure}[h]
\begin{center}
\includegraphics[width=7.5cm]{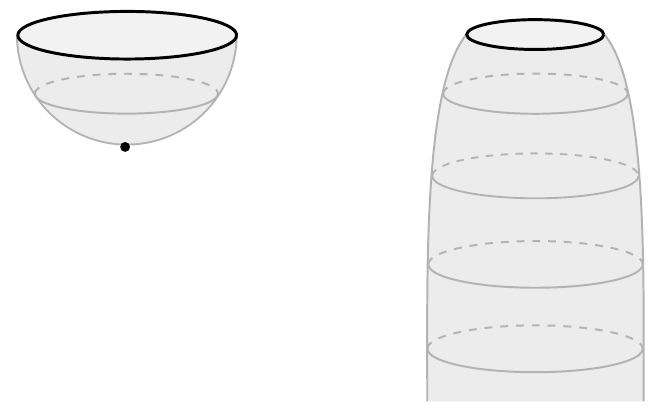}
\end{center}
\vspace{-1pc}
\caption[] {The `hemispherical spacetime' $g = -dt^2 + \sin^2(t)d\theta^2$ considered in Example \ref{hemisphere}. The Riemannianized completion, with respect to $\tau = \ln t$ is the standard closed round hemisphere, shown on the left. The null distance completion induced by $\tau = \ln t$ is shown in the right, and effectively `blows up' the big bang point, though in a much different way than in Example \ref{sqrt_blowup}.}
\label{blowupbang}
\end{figure}

\vspace{2pc}
\begin{exm} [$\tau = \ln t$ on $g = -dt^2 + \sin^2(t) d\theta^2$] \label{hemisphere} We now consider the 2-dimensional `hemispherical' GRW spacetime
$$(M,g) = ( \, (0, \tfrac{\pi}{2}) \times \field{S}^1 \, , \, -dt^2 + \sin^2(t) d\theta^2 \, )$$
together with the time function $\tau(t,\theta) = \ln t = \phi(t)$. By Corollary \ref{corGRWRiemannize}, the Riemannianization is the standard round open hemisphere, with the pole removed. To compute the null distance function, as in Proposition \ref{null_time_vs_space}, we have:
$$\hat{d} \, \bigg(\; \ln t \; , \; -dt^2 + \sin^2(t) d\theta^2 \; \bigg) \; = \; \hat{d} \, \bigg(\; \tau \; , \; -d\tau^2 + \bigg(\dfrac{\sin(e^\tau)}{e^\tau}\bigg)^2 d\theta^2 \; \bigg)$$
where $0 < t < \pi/2$ on the left, and $- \infty < \tau < \ln(\pi/2)$ on the right. That is, we can swap the nonstandard time function $\tau = \ln t$ for the new spatial warping function $\tilde{f}(\tau) = \sin(e^\tau)/e^\tau$. And note that $\tilde{f}(\tau) \to 1$ as $\tau \to - \infty$. Note that this is again opposite to the situation in Examples \ref{ex_arctan_Mink} and \ref{ex_tcubed_futMink}, in that the `genuine big bang point' in our original spacetime is obscured by the effective warping of the nonstandard time function $\tau = \ln t$. Alternatively, the null distance induced by $\tau = \ln t$ `blows up' the big bang point, though here in a much different way than in Example \ref{sqrt_blowup}. See Figure \ref{blowupbang}. In terms of GRW faithfulness, fixing any two points $p = (t_p, p_S)$ and $(t_q,q_S)$, with $p_S = q_S$ and $t_p < t_q$, then by Lemma \ref{Mink_MVT_lemma}, there is a $t^* \in (t_p, t_q)$, such that we have:
$$\hat{d}_\tau(p,q) \;  = \; \dfrac{1}{t^*} \cdot (t_q - t_p)$$
Since $\phi'(t^*) = 1/t^* \to \infty$ as $t_q \to 0^+$, we see again that $\hat{d}_\tau$ fails the GRW faithfulness property (F2), and that this corresponds to the `artificial null expansion' as $t \to 0^+$, where the effective TSW time warping function $f_T = 1/\phi' \to 0$, the effective GRW spatial warping function $\tilde{f} \to 1$, and we do not detect the `genuine big bang point'. 
\begin{flushright} $\Diamond$\end{flushright}
\end{exm}

\vspace{1pc}

\vspace{1pc}
\subsection{GRW Singularities}

\vspace{1pc}
In this section, we focus on past finite GRW spacetimes of the form,
$$(M,g) = ((0,b) \times S, -dt^2 + f(t)^2h)$$
where $0 < b \le \infty$. Let $d_S$ denote the Riemannian distance function of $(S,h)$, and let $\mathrm{diam}(S)$ denote the diameter of $S$ with respect to $d_S$. We will assume that $S$ has positive dimension, (i.e., is not a single point), and thus that $\mathrm{diam}(S) > 0$. However, we allow for the case $\mathrm{diam}(S) = \infty$. In this setting, we will again consider abstract distance functions $d$ which satisfy some of the GRW faithfulness conditions, and more concretely, the standard Riemannianized, null, elevator distances, $d^R = d^R_t$, $d^N = d^N_t$, $d^E = d^E_t$, induced by the standard time function, $\tau(t,x) = t$. As above, we shall index these distances by $d^Z \in \{d^R, d^N, d^E\}$, where we symbolically take $Z \in \{R,N,E\}$. Fixing any choice of $Z \in \{R,N,E\}$, we then let $(\overline{M}_Z, \overline{d}_Z)$ be the metric completion of $(M,d^Z)$. Moreover, since $\nabla \tau$ is everywhere unit timelike, standard time $\tau(t,x) = t$ is, in particular, locally anti-Lipschitz, and by Corollaries \ref{extending_tau_wrt_nulldist}, \ref{extending_tau_wrt_Riem}, \ref{elevator_Lipschitz}, $\tau(t,x) = t$ admits a unique continuous extension, which we denote by $\overline{\tau}_Z : \overline{M}_Z \to \field{R}$.

\vspace{1pc}
We note that in the past finite GRW setting, for $M = (0, b) \times_f S$, the standard time function $\tau(t,x) = t$ is precisely the `cosmological time function' of $M$, as in \cite{AGHcosmo}. Indeed, this forms part of the basis for the separate, ongoing work on more `generalized big bangs', mentioned above, and described in \cite{SormaniOberwolfach}. Here, we focus first on the `classical big bangs', in the GRW setting, as in the standard Friedmann models. When completing $M = (0,b) \times_f S$ with respect to a distance function, and extending $\tau(t,x) = t$ to $\overline{\tau} : \overline{M} \to \field{R}$, we shall refer to $(\overline{\tau})^{-1}(0)$ as the `initial singularity', though we emphasize that the term is employed somewhat informally, and is not meant to indicate, e.g., curvature blowup, etc.

\vspace{2pc}
We first note the following:

\begin{cor} [GRW Faithfulness (F2) and Initial Singularities] \label{GRW_initial_sing_nonempty} Consider a GRW spacetime $(M,g) = ((0,b) \times S, -dt^2 + f(t)^2h)$, where $0 < b \le \infty$. Fix any distance function $d$ on $M$, and let $(\overline{M}, \overline{d})$ be the metric completion of $(M,d)$. If $d$ satisfies the GRW faithfulness condition (F2), and if the standard time function $\tau(t,x) = t$ admits a continuous extension $\overline{\tau} : \overline{M} \to \field{R}$, then the `initial singularity' $(\overline{\tau})^{-1}(0)$ is a nonempty subset of $\overline{M} - M$. In particular, this holds for $(\overline{\tau}_Z)^{-1}(0)$, for any choice $d = d^Z \in \{d^R, d^N, d^E\}$.
\end{cor}

\begin{proof} Fix any spatial location $p_S \in S$, and any sequence $t_k \to 0^+$. Then by Lemma \ref{F2_vertical_Cauchy}, the sequence $p_k = (t_k,p_S)$ is Cauchy in $(M,d)$, and as in Proposition \ref{extendfunctionprop}, we have $[\{p_k\}] \in (\overline{\tau})^{-1}(0)$. In particular, since any choice $d^Z \in \{d^R, d^N, d^E\}$ satisfies all the GRW faithfulness properties, and we have such an extension $\overline{\tau}_Z$, this also applies to $(\overline{\tau}_Z)^{-1}(0)$.
\end{proof}

\vspace{2pc}
The nature of the `singularity' guaranteed in Corollary \ref{GRW_initial_sing_nonempty} will depend on the geometry of $M$, with two basic cases, which we now begin to address. We first address the case of `bounded spatial warping' and in particular, no spatial pinching. The following is simply a special case of Lemma \ref{faith_completed_slices}:

\begin{cor} [GRW Faithful Initial Singularities: Bounded Warping] \label{GRW_sing_bounded_warping} Consider a GRW spacetime, $M = ((0,b) \times S , -dt^2 + f^2(t)h)$, where $0 < b \le \infty$. Suppose that $d$ is a distance function on $M$ which satisfies the GRW faithfulness conditions (F4) and (F5), with constants $C_4$ and $C_5$. Let $(\overline{M}, \overline{d})$ be the metric completion of $(M, d)$. Let $(\overline{S}, \overline{d}_S)$ be the metric completion of $(S,d_S)$. Let $(\overline{I}, |\cdot|)$ be the metric completion of $(I, |\cdot|)$. We will also identify $\overline{I}$ with the closure of $I$ in $\field{R}$. Suppose that the standard time function $\tau(t,x) = t$ admits a continuous extension $\overline{\tau} : \overline{M} \to \overline{I}$. Let $f_{\inf} = \inf \{f(t) : 0 < t < b\}$ and $f_{\sup} = \sup \{f(t) : 0 < t < b\}$, and suppose that $0 < f_{\inf} \le f_{\sup} < \infty$. 

\ben
\item [(1)] Then the initial singularity is given by:
$$(\overline{\tau})^{-1}(0) \; = \; \{ \, [\{(t_k, x_k)\}]_M : \, [\{t_k\}]_I = 0 \in \overline{I} \, , \, [\{x_k\}]_S \in \overline{S} \, \} \; \approx \; \{0\} \times \overline{S}$$
\item [(2)] If the initial singularity $(\overline{\tau})^{-1}(0)$ is minimal in the sense that the limit $f_0 := \lim_{t \, \to \, 0^+} f(t)$ exists, and $f_0 = f_{\inf}$, then for any two elements $[\{p_k\}], [\{q_k\}] \in (\overline{\tau})^{-1}(0)$, with representative sequences $p_k = (t_k, x_k)$ and $q_k = (u_k, y_k)$, we have:
$$C_5 \cdot f_0 \cdot \overline{d}_S([\{x_k\}], [\{y_k\}]) \; \le \; \overline{d} ([\{p_k\}], [\{q_k\}]) \; \le \; C_4 \cdot f_0 \cdot \overline{d}_S([\{x_k\}], [\{y_k\}])$$ 
\item [(3)] Note that all of the above applies in the case of the standard Riemannianized, elevator, null distance functions, $d^Z \in \{d^R, d^E, d^N\}$, with $C_4 = C_5 = 1$. Hence, in this setting, for any such choice $d^Z$, we have $(\overline{\tau}_Z)^{-1}(0) \approx \{0\} \times \overline{S}$, and if $(\overline{\tau}_Z)^{-1}(0)$ is minimal as in (2), then the metric space structure of the initial singularity is precisely:
$$(\, (\overline{\tau}_Z)^{-1}(0) \, , \, \overline{d}_Z \, ) \; \approx \; (\, \{0\} \times \overline{S} \, , \, f_0 \cdot \overline{d}_S\, )$$
\een
\end{cor}

\vspace{2pc}
We now move on to the case of `spatial pinching'. We first note the following simple fact:

\begin{lem} [Diameter of Sublevel Set Estimate] \label{level_set_bound} Let $d$ be a distance function on a set $M$, and let $f : M \to (0, \infty)$ be a positive function. For each $t > 0$, consider the (strict) sublevel set $M_t := \{p \in M : f(p) < t\}$. If there is a constant $C > 0$ such that, for all $p,q \in M$, we have $d(p,q) \le C(f(p) + f(q))$, then $\mathrm{diam}(M_t) \le 2 \, C\, t$.
\end{lem}

\vspace{1pc}
We showed in Lemma \ref{GRW_faithful_pinching} that spatial pinching is reflected faithfully by any distance function satisfying the GRW faithfulness condition (F4). The following now shows the significance of the more complicated sixth condition (F6). 

\begin{lem} [GRW Faithful Sequential Pinching and Early Universe Diameter] \label{GRW_faithful_seq_bang_lem} Consider a GRW spacetime $(M,g) = ((0,b) \times S, -dt^2 + f(t)^2h)$, where $0 < b \le \infty$. For $t \in (0, b]$, set $M_t := (0,t) \times S$, so that $M_t$ is the `universe before time $t$'. Fix any distance function $d$ on $M$ which satisfies the GRW faithfulness conditions (F4) and (F6). Then the following conditions (S1) and (S2) are equivalent:

\vspace{.2pc}
\ben
\item [(S1)] $\, \Lim{k \, \to \, \infty} f(t_k) = 0$, for some sequence $t_k \to 0^+$
\vspace{.2pc}
\item [(S2)] $\Lim{\, t \, \to \, 0^+} \mathrm{diam}(M_t) = 0$
\een
Moreover, let $(\overline{M}, \overline{d})$ be the metric completion of $(M,d)$. If the standard time function $\tau(t,x) = t$ admits a continuous extension $\overline{\tau} : \overline{M} \to \field{R}$, then conditions (S1) and (S2) are also equivalent to:
\ben
\item [(S3)] $(\overline{\tau})^{-1}(0) = \{\overline{p}_0\}$ is a single big bang point.
\vspace{.5pc}
\een

\end{lem}

\begin{proof} That (S1) implies (S2) follows from Lemmas \ref{GRW_faithful_pinching} and \ref{level_set_bound}. For (S2) $\Longrightarrow$ (S1), we prove the contrapositive. Suppose then that there are positive constants $0 < \delta < b$ and $A > 0$, such that $f(u) \ge A$ for all $0 < u < \delta$. Setting $A_{(0,\delta)} = \inf \{f(u) : 0 < u < \delta \}$, note that $A_{(0,\delta)} \ge A$. Fix $t \in (0, \delta)$. Fix any two points $p, q \in M_t$. Applying property (F6) to the interval $(t_1, t_2) = (0,\delta)$, we have:
\begin{align*}
d(p,q) & \; \ge \; C_6 \cdot \min \{ \, E_{(0,\delta)}(t_p,t_q) \, , \, A_{(0,\delta)} \cdot d_S(p_S,q_S) \, \}\\[1pc]
& \; = \; C_6 \cdot \min \{ \, 2\delta - t_p - t_q \, , \, A_{(0,\delta)} \cdot d_S(p_S,q_S) \, \}\\[1pc]
& \; \ge \; C_6 \cdot \min \{ \, 2\delta - 2t \, , \, A \cdot d_S(p_S,q_S) \, \}\\[1pc]
\Longrightarrow \hspace{2pc} \mathrm{diam}(M_t) & \; \ge \; C_6 \cdot \min \{ \, 2\delta - 2t \, , \, A \cdot d_S(p_S,q_S) \, \}
\end{align*} 

\vspace{.5pc}
\noindent
Since this last estimate holds for arbitrary $p_S, q_S \in S$, it follows that:
$$\mathrm{diam}(M_t) \; \ge \; C_6 \cdot \min \{ \, 2\delta - 2t \, , \, A \cdot \mathrm{diam}(S) \, \}$$
where $0 < \mathrm{diam}(S) \le \infty$. This estimate then implies that, for all $0 < t < \delta/2$, we have $\mathrm{diam}(M_t) \ge C_6 \cdot \min \{ \delta , A \cdot \mathrm{diam}(S)\}$. In particular, if the limit exists,
$$\Lim{\, t \, \to \, 0^+} \mathrm{diam}(M_t) \ne 0$$

\vspace{1pc}
Now let $(\overline{M}, \overline{d})$ be the metric completion of $(M,d)$ and suppose $\tau(t,x) = t$ admits a continuous extension $\overline{\tau} : \overline{M} \to \field{R}$. We will show (S2) $\Longleftrightarrow$ (S3). First suppose that (S2) holds. Since $d$ satisfies (F4) it satisfies (F2), by Lemma \ref{lem_F2_F3_F4}. Hence, by Corollary \ref{GRW_initial_sing_nonempty}, $(\overline{\tau})^{-1}(0)$ is nonempty. Let $[\{p_k\}], [\{q_k\}] \in (\overline{\tau})^{-1}(0)$. This means $p_k = (t_k,x_k)$ and $q_k = (u_k,y_k)$ are Cauchy sequences in $(M, d)$, with $\tau(p_k) = t_k \to 0$ and $\tau(q_k) = u_k \to 0$. Fix $\e > 0$. By the condition on the diameters in (S2), there is a $\delta > 0$ such that $\mathrm{diam}(M_t) < \e$, for all $0 < t \le \delta$. Since $t_k, u_k \to 0$, there is a $k_0 \in \field{N}$, such that, for all $k \ge k_0$, we have $p_k, q_k \in M_\delta$, and thus $d(p_k,q_k) < \mathrm{diam}(M_\delta) <  \e$. This then gives that:
$$\overline{d}([\{p_k\}], [\{q_k\}]) \; = \; \lim_{k \, \to \, \infty} \, d(p_k,q_k) \; < \; \e$$
Since $\e > 0$ was arbitrary, we have $[\{p_k\}] = [\{q_k\}] =: \overline{p}_0$ in $\overline{M}$, and thus the `initial singularity' $(\overline{\tau})^{-1}(0) = \{\overline{p}_0\}$ consists of a single `big bang point'. Now suppose conversely that (S3) holds, that is, that $(\overline{\tau})^{-1}(0) = \{\overline{p}_0\}$ consists of a single point. Take any two points $p, q \in M$, $p = (t,x)$ and $q = (u,y)$. Consider the sequences $p_k := (t/k, x)$ and $q_k := (u/k, y)$. Recall again that since $d$ satisfies (F4) it satisfies (F2). Hence, by Lemma \ref{F2_vertical_Cauchy}, $\{p_k\}$ and $\{q_k\}$ are Cauchy sequences in $(M,d)$, and thus correspond to elements $[\{p_k\}], [\{q_k\}] \in \overline{M}$. But since $\tau(p_k) = t_k \to 0$, and $\tau(q_k) = u/k \to 0$, we have $[\{p_k\}], [\{q_k\}] \in (\overline{\tau})^{-1}(0) = \{\overline{p}_0\}$, and thus $[\{p_k\}] = [\{q_k\}] = \overline{p}_0$. Identifying $w \in M$ with $[w] \in \overline{M}$ as usual, and using the triangle inequality in the metric completion $(\overline{M}, \overline{d})$, and condition (F2), we get:
\begin{align*}
d(p, q)  \; = \; \overline{d}(p,q) & \; \le \; \overline{d}(p,\overline{p}_0) \; + \; \overline{d}(q, \overline{p}_0) \\[1pc] 
& \; = \; \lim_{k \, \to \, \infty} d(p,p_k) \; + \; \lim_{k \, \to \, \infty} d(q,q_k) \\[1pc] 
& \; \le \; \lim_{k \, \to \, \infty} C_2 \cdot (t - t/k) \; + \; \lim_{k \, \to \, \infty} C_2 \cdot (u - u/k) \; = \;  C_2 \cdot (t + u)
\end{align*}
Thus, by Lemma \ref{level_set_bound}, we have $\mathrm{diam}(M_\delta) \le 2 \, C_2  \, \delta$, which implies (S2).
\end{proof}

\begin{figure}[h]
\begin{center}
\includegraphics[width=6cm]{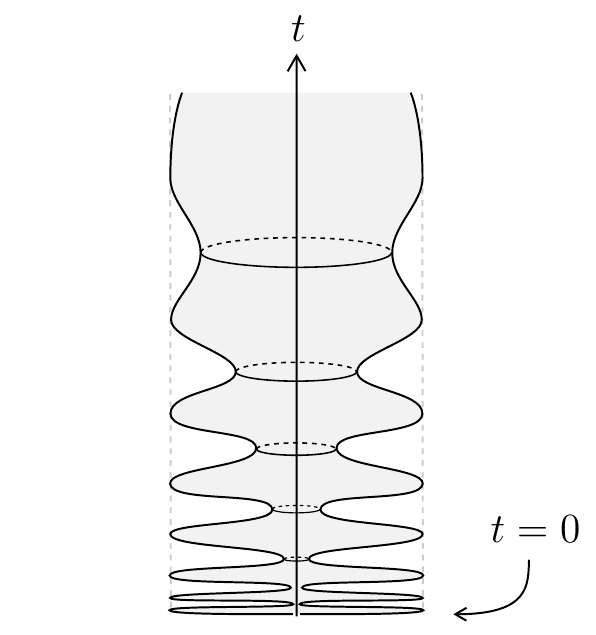}
\vspace{.1pc}
\caption[]{A spacetime illustrating Lemma \ref{GRW_faithful_seq_bang_lem}} 
\label{oscibang}
\end{center}
\end{figure}

\vspace{2pc}
As suggested by Lemma \ref{GRW_faithful_seq_bang_lem} and Figure \ref{oscibang}, we need to use \emph{intrinsic} diameters to be able to encode the basic GRW condition `$f(t) \to 0$' in terms of distance:

\begin{lem} [Spatially Finite, GRW Faithful Uniform Pinching] \label{GRW_faithful_uniform_bang_lem} Consider a GRW spacetime $(M,g) = ((0,b) \times S, -dt^2 + f(t)^2h)$, with $0 < b \le \infty$, and $\mathrm{diam}(S) < \infty$, (for example, if $S$ is compact). For any choice of $u, v \in \field{R}$, with $0 \le u < v \le b$, let $M_{(u,v)} := (u,v) \times_f S$ denote `the universe between times $t = u$ and $t = v$', which is an open subspacetime of $M = M_{(0,b)}$. Suppose that $d$ is a distance function on $M$, which is a length metric. Let $d_{(u,v)}$ denote the corresponding intrinsic distance function of $M_{(u,v)}$, and let $\mathrm{diam}_\mathrm{int}(M_{(u,v)})$ denote the intrinsic diameter of $M_{(u,v)}$. Suppose that each $d_{(u,v)}$ satisfies the GRW faithfulness properties (F4) and (F6) on $M_{(u,v)}$, with uniform constants $C_4$ and $C_6$, i.e., with these constants being independent of $u$ and $v$. Then the following are equivalent:

\vspace{.2pc}
\ben
\item [(U1)] $\Lim{\, t \, \to \, 0^+} f(t) = 0$

\vspace{.2pc}
\item [(U2)] $\Lim{\, t \, \to \, 0^+} \bigg[\, \Sup{0 \,\le \, u \, < \, v \, \le \, t} \bigg ( \mathrm{diam}_{\mathrm{int}}(M_{(u,v)})  \bigg) \bigg] = 0$
\vspace{.5pc}
\een
\end{lem}

\begin{proof} To prove (U1) $\implies$ (U2), suppose that $f(t) \to 0$ as $t \to 0^+$. Fix $\e > 0$. Let $\delta > 0$ such that $0 < f(t) < \e$ for all $0 < t \le \delta$. Since we are free to shrink $\delta$ further at this point, we can assume that $\delta < \e \cdot \mathrm{diam}(S)$. Fix $0 \le u < v \le \delta$. Fix $p, q \in M_{(u,v)}$. Since $d_{(u,v)}$ on $M_{(u,v)}$ satisfies (F4), with uniform constant $C_4$, we may apply Lemma \ref{lem_F2_F3_F4} on $M_{(u,v)}$, to get:
\begin{align*}
d_{(u,v)}(p,q) & \; \le \; C_4 \cdot \bigg( \, |t_q - t_p| \, + \, A_{[t_p,t_q]} \, d_S(p_S,q_S)\, \bigg)\\[1pc]
& \; \le \; C_4 \cdot (\, \delta \, + \, \e \cdot d_S(p_S,q_S) \,)\\[1pc]
& \; \le \; C_4 \cdot (\, \delta \, + \, \e \cdot \mathrm{diam}(S) \,) \; \le \; 2 C_4 \cdot \mathrm{diam}(S) \cdot \e
\end{align*}

\vspace{1pc}
\noindent
Taking the supremum of the left over $p,q \in M_{(u,v)}$ then gives:
$$\mathrm{diam}_{\mathrm{int}}(M_{(u,v)}) \; \le \; 2 C_4 \cdot \mathrm{diam}(S) \cdot \e$$
Since this holds for all choices of $u$ and $v$ with $0 \le u < v \le \delta$, we have:
$$\Sup{0 \,\le \, u \, < \, v \, \le \, \delta} \bigg ( \mathrm{diam}_{\mathrm{int}}(M_{(u,v)})  \bigg) \; \le \; 2 C_4 \cdot \mathrm{diam}(S) \cdot \e$$
Since $\mathrm{diam}(S) < \infty$, this completes the proof of (U1) $\Longrightarrow$ (U2). For (U2) $\Longrightarrow$ (U1), we prove the contrapositive. Suppose then that there is a positive number $A > 0$ and a sequence $t_k \to 0^+$, such that $f(t_k) \ge A$, for all $k \in \field{N}$. Fix any $\delta > 0$. Then there is a $k_0 \in \field{N}$ such that $0 < t_{k_0} < \delta$. Hence, $f(t_{k_0}) \ge A$, and by continuity, we can find a sufficiently small $\delta_0 > 0$ such that, letting $u_0 := t_{k_0} - \delta_0$ and $v_0 := t_{k_0} + \delta_0$, we have $0 < u_0 < v_0 < \delta$, and $A_{(u_0,v_0)} = \inf\{f(t) : u_0 < t < v_0\} \ge A/2$. Fix any two points $p, q \in M_{(u_0,v_0)}$, in the $t = t_{k_0}$ time slice, $p = (t_{k_0},p_S)$, $q = (t_{k_0}, q_S)$. Since $d_{(u_0,v_0)}$ satisfies (F6) on the spacetime $M_{(u_0,v_0)}$, we have $E_{(u_0,v_0)}(t_{k_0}, t_{k_0}) = 2\delta_0$, and 
\begin{align*}
 C_6 \cdot \min \{ \, 2\delta_0 \, , \, A_{(u_0,v_0)} \cdot d_S(p_S,q_S)  \,\} & \; \le \; d_{(u_0,v_0)}(p,q) \\[1pc]
 \Longrightarrow \hspace{4pc} C_6 \cdot \min \{ \, 2\delta_0 \, , \, \tfrac{A}{2}  \cdot d_S(p_S,q_S) \, \} & \; \le \; d_{(u_0,v_0)}(p,q)\\[1pc]
\Longrightarrow \hspace{4pc} C_6 \cdot \min \{ \, 2\delta_0 \, , \, \tfrac{A}{2}  \cdot d_S(p_S,q_S) \, \} & \; \le \; \mathrm{diam}_{\mathrm{int}}(M_{(u_0,v_0)})
\end{align*}

\vspace{.5pc}
\noindent
Since this holds for arbitrary $p_S, q_S \in S$, it follows that:
$$C_6 \cdot \min \{ \, 2\delta_0 \, , \, \tfrac{A}{2}  \cdot   \mathrm{diam}(S) \, \} \; \le \; \mathrm{diam}_{\mathrm{int}}(M_{(u_0,v_0)})$$
Finally, since $C_6$ is uniform, taking a supremum on the right gives:
$$C_6 \cdot \min \{ \, 2\delta_0 \, , \, \tfrac{A}{2}  \cdot   \mathrm{diam}(S) \, \}  \; \le \; \Sup{0 \,\le \, u \, < \, v \, \le \, \delta} \bigg ( \mathrm{diam}_{\mathrm{int}}(M_{(u,v)})  \bigg)$$
This completes the proof of $\cancel{(U1)} \; \Longrightarrow \; \cancel{(U2)}$, i.e., (U2) $\Longrightarrow$ (U1).
\end{proof}

\vspace{2pc}
\begin{figure}[h]
\begin{center}
\includegraphics[width=13cm]{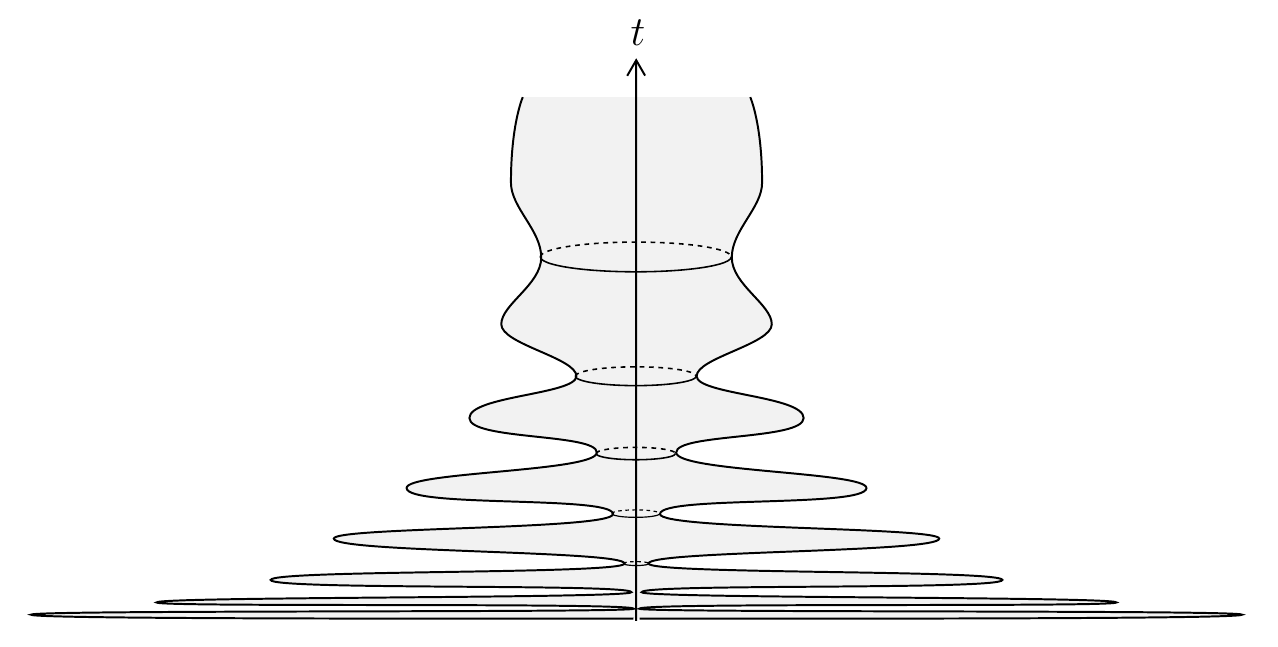}
\vspace{0pc}
\caption[]{Another spacetime illustrating Lemma \ref{GRW_faithful_seq_bang_lem} and Theorem \ref{GRW_faith_bang_thm}. } 
\label{oscibang2}
\end{center}
\end{figure}

\vspace{3pc}
Since the standard Riemannianized, null, and elevator distances, $d^R$, $d^N$, $d^E$, are all length metrics, which by Theorem \ref{RNE_GRW_faithful} satisfy all the GRW faithfulness properties, with uniform constants $C_i = 1$, for all $1 \le i \le 6$, Corollaries  \ref{extending_tau_wrt_nulldist}, \ref{extending_tau_wrt_Riem}, \ref{elevator_Lipschitz}, \ref{GRW_initial_sing_nonempty} and Lemmas \ref{GRW_faithful_seq_bang_lem}, \ref{GRW_faithful_uniform_bang_lem} give the following:

\begin{thm} [GRW Riemannianized, Null, and Elevator Big Bang Points] \label{GRW_faith_bang_thm} Consider a GRW spacetime $(M,g) = ((0,b) \times S, -dt^2 + f(t)^2h)$, where $0 < b \le \infty$. For $0 < t \le b$, let $M_t := (0,t) \times S$ denote the `universe before time $t$'. For $0 \le u < v \le b$, let $M_{(u,v)} := (u,v) \times_f S$ denote `the universe between times $t = u$ and $t = v$'. Let $d^R$, $d^N$, $d^E$ be the standard Riemannianized, null,  elevator distances, and index these by $d^Z \in \{d^R, d^N, d^E\}$. Let $(\overline{M}_Z, \overline{d}_Z)$ be the metric completion of $(M,d^Z)$ and let $\overline{\tau}_Z : \overline{M}_Z \to \field{R}$ be the unique continuous extension of $\tau(t,x) = t$. Then, for any choice $d^Z \in \{d^R, d^N, d^E\}$, we have the following:

\ben
\item [(1)] For arbitrary spatial diameter $0 < \mathrm{diam}(S) \le \infty$, the following conditions are equivalent: 
\ben
\item [(S1)]  $f(t_k) \to 0$, for some sequence $t_k \to 0^+$.

\vspace{.5pc}
\item [(S2)] The diameters $\mathrm{diam}^Z(M_t)$ shrink to zero as $t \to 0^+$.

\vspace{.5pc}
\item [(S3)] $(\overline{\tau}_Z)^{-1}(0) = \{\overline{p}_0\}$ is a single big bang point.
\een

\vspace{.5pc}
\item [(2)] If $0 < \mathrm{diam}(S) < \infty$, the following conditions are equivalent: 
\ben
\item [(U1)] $f(t) \to 0$ as $t \to 0^+$.

\vspace{.5pc}
\item [(U2)] $\sup \, \{\, \mathrm{diam}^Z_{\mathrm{int}}(M_{(u,v)}) : 0 \,\le \, u \, < \, v \, \le \, t\}$ shrinks to zero as $t \to 0^+$.
\een
\een
\end{thm}

\vspace{2pc}
\begin{rmk} [Spatially Infinite GRW Models] Lemma \ref{GRW_faithful_seq_bang_lem} holds regardless of spatial topology, including for spatially open models with infinite spatial diameter, $\mathrm{diam}(S) = \infty$. On the other hand, Lemma \ref{GRW_faithful_uniform_bang_lem} fails miserably without the spatially finite assumption, $\mathrm{diam}(S) < \infty$. Evidently, in the case $\mathrm{diam}(S) = \infty$, we have not yet produced a way to encode the distinction between the weaker condition of `sequential pinching', that $f(t_k) \to 0$ along some sequence $t_k \to 0^+$, and the stronger condition of `uniform pinching', that $f(t) \to 0$ as $t \to 0^+$, purely in terms of, say, the standard Riemannianized, null, or elevator distances, or other GRW faithful distance.
\end{rmk}

\pagebreak

\vspace{2pc}
\begin{exm} [Power-Warped Future Half-Planes] For any power $\a \in \field{R}$, consider the spatial warping function $f(t) = t^\a$, and the two-dimensional GRW spacetime $M = (0,\infty) \times_f \field{R}$, with spacetime metric tensor
$$g = -dt^2 + t^{2\a}dx^2$$

\vspace{1pc}
\begin{figure}[h]
\begin{center}
\vspace{1pc}
\includegraphics[width=13cm]{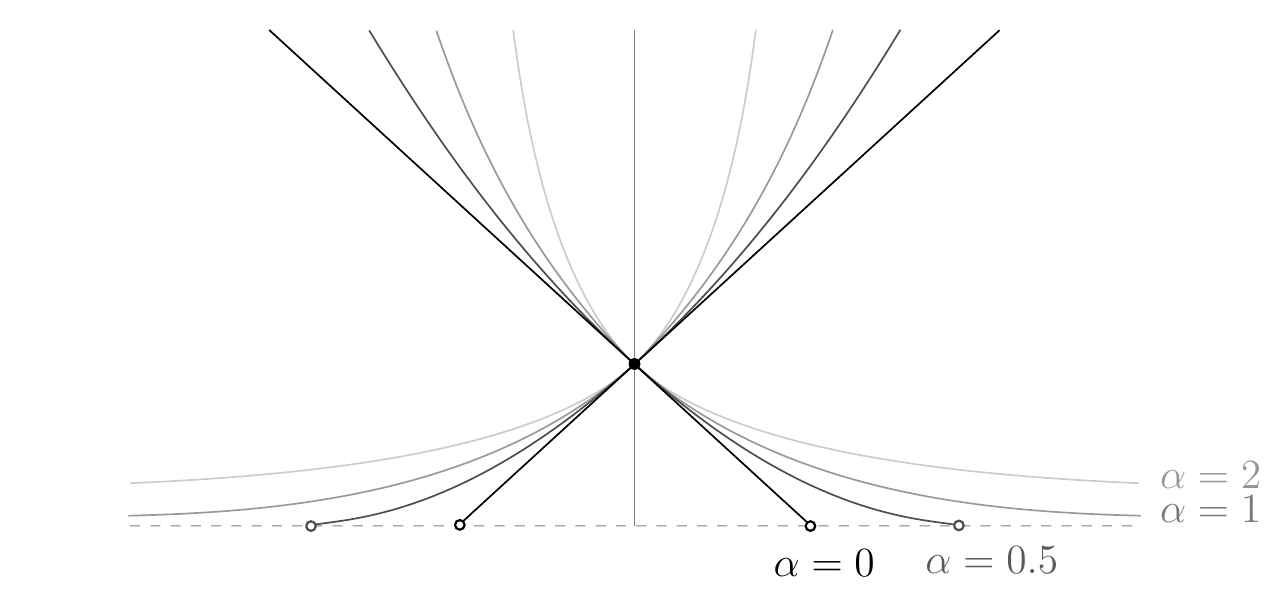}
\caption[]{The past and future causal cones from a point in the power-warped future half-plane, $((0, \infty) \times \field{R}, -dt^2 + t^{2\a}dx^2$), as a function of the parameter $\a$.} 
\label{power_warped_cones}
\end{center}
\end{figure}
\end{exm}

\noindent
Note that for $\a = 0$ this is simply the future half-Minkowski plane, $\field{M}^{1+1}_+$. For general power $\a \in \field{R}$, a curve $\eta(t) = (t, \s(t))$ is a null iff $\s'(t) =\pm \, t^{-\a}$. To determine the past and future of a point $(t_0,x_0) \in M$, we take $\s(t_0) = x_0$. Then, for example, taking the plus sign, $\s'(t) = t^{-\a}$, we have:

$$\s(t) - \s(t_0) = \int_{t_0}^t u^{-\a}du = \left\{
     \begin{array}{lr}
       \dfrac{t^{1-\a} - t_0^{1-\a}}{1-\a} \hspace{1pc} &  \a \ne 1\\
       \\
       \ln t - \ln t_0 &  \a =1
     \end{array}
   \right.$$

\vspace{1pc}
\noindent
This gives the past and future of the point $(t_0,x_0)$ as a function of $\a$. As $\a$ increases, the pasts flare out, while the futures focus in. For $-\infty < \a < 1$, the point $(t_0,x_0)$ has a finite view of past infinity, and the past causal boundary is $\field{R}$, as a set. When $1 \le \a < \infty$, $(t_0,x_0)$ can see all of past infinity, and the past causal boundary is a single point. The cases $\a = 0, 0.5, 1, 2$ are illustrated in Figure \ref{power_warped_cones}. Note, furthermore, that for $\a > 0$, the warping function $f(t) = t^\a$ satisfies $f(t) \to 0$ as $t \to 0^+$. Then, for example, as in Example \ref{exfuturehalfMinkowski}, and Theorem \ref{GRW_faith_bang_thm}, we have the following:

\begin{prop} [Power-Warped Future Half-Planes] \label{power_warped_half_planes} For $\a \ge 0$, consider the 2-dimensional GRW spacetime 
$$(M,g) = ((0, \8) \times \field{R} , -dt^2 + t^{2\a}dx^2)$$
\ben
\item [(1)] Metric Completions: Consider the standard Riemannianized, elevator, and null distance functions $d^R$, $d^E$, $d^N$, induced by $\tau(t,x) = t$. Let $d^Z \in \{d^R,d^E, d^E\}$. Then, as sets, we have the following:
\ben
\item [(a)] For $\a = 0$, $\overline{M}_Z = [0,\8) \times \field{R}$ and $(\overline{\tau}_Z)^{-1}(0) = \{0\} \times \field{R} \approx \field{R}$.

\vspace{.5pc}
\item [(b)] For $\a > 0$, $\overline{M}_Z = \{\overline{p}_0\} \cup M$ and $(\overline{\tau}_Z)^{-1}(0) = \{\overline{p}_0\}$.
\een

\vspace{.5pc}
\item [(2)] Past Causal Completion: Let $\d^-(M)$ and $M^-$  be the past causal boundary and completion of $M$, respectively. Then, as sets, we have:
\ben
\item [(a)] For $0 \le \a < 1$, $M^- = [0, \infty) \times \field{R}$ and $\d^-(M) = \{0\} \times \field{R} \approx \field{R}$

\vspace{.5pc}
\item [(b)] For $\a \ge 1$, $M^- = \{\overline{p}_0\} \cup M$ and $\d^-(M) = \{\overline{p}_0\}$
\een
\een
\end{prop}

\vspace{1pc}
\begin{rmk} [Metric Space Initial Singularities vs Past Causal Boundary] Proposition \ref{power_warped_half_planes} shows that, even in the past finite case, and even when the corresponding `boundaries are nice', there can be a big disagreement between the `initial singularity' coming from a choice of distance function and its metric completion, compared to the past causal boundary. 
\end{rmk}

\pagebreak
\section{Appendix} \label{sec_Appendix}

\vspace{1pc}
\subsection{Extending Causality to a Metric Completion}

\vspace{1pc}
In studying metric completions of spacetimes, a natural question is whether we can extend causality to such completions. Consider, for example, the following tentative definition:

\begin{Def} [Extending Causality to a Metric Completion] \label{Def_causality_completion} Let $(M,g)$ be a spacetime. Fix a distance function $d$ on $M$, and let $(\overline{M}, \overline{d})$ be the metric completion of $(M,d)$. For two elements $\overline{p}, \overline{q} \in \overline{M}$, let us say that 
$$\overline{p}\;  \le \; \overline{q}$$
if there are representative Cauchy sequences $\{p_k\}, \{q_k\}$ in $(M,d)$, with $[\{p_k\}] = \overline{p}$ and $[\{q_k\}] = \overline{q}$, for which $p_k \le q_k$, for all $k \in \field{N}$.
\end{Def}

\vspace{1pc}
This notion of causality in the completion gives, for example, the following:

\begin{lem} [Anti-Lipschitz Extensions to Metric Completion]\label{antiLip_on_completion} Let $(M,g)$ be a spacetime. Fix a distance function $d$ on $M$, and let $(\overline{M}, \overline{d})$ be the metric completion of $(M,d)$. Suppose that $\tau : M \to \field{R}$ is continuous and $\l$-anti-Lipschitz with respect to $d$. If $\tau$ admits a continuous extension $\overline{\tau} : \overline{M} \to \field{R}$, then $\overline{\tau}$ is also $\l$-anti-Lipschitz with respect to $\overline{d}$, that is, for all $\overline{p}, \overline{q} \in \overline{M}$, we have:
$$\overline{p}\;  \le \; \overline{q} \; \; \Longrightarrow \; \; \l \cdot \overline{d} \, (\, \overline{p} \, , \,\overline{q} \, ) \; \le \; \overline{\tau}(\overline{q}) - \overline{\tau}(\overline{p})$$
\end{lem}

\begin{proof} Fix $\overline{p}, \overline{q} \in \overline{M}$, with $\overline{p} \le \overline{q}$. Let $\{p_k\}$ and $\{q_k\}$ as in Definition \ref{Def_causality_completion}. Then, applying the definitions, the $\l$-anti-Lipschitz property for $\tau$, and Proposition \ref{extendfunctionprop}, we have:
\begin{align*}
\l \cdot \overline{d} \, (\, \overline{p} \, , \,\overline{q} \, ) & \; = \; \lim_{k \to \infty} \, [\, \l \cdot d(p_k, q_k) \,]\\[1pc]
 & \; \le \; \lim_{k \to \infty} \, [\, \tau(q_k) - \tau(p_k) \, ] \; = \; \overline{\tau}(\overline{q}) - \overline{\tau}(\overline{p})
\end{align*}
\end{proof}

\vspace{1pc}
In the case of null distance, Definition \ref{Def_causality_completion} gives the following:

\begin{lem} [Null Distance on Causal Pairs in Completion]\label{antiLip_on_completion} Let $(M,g)$ be a spacetime. Suppose that $\tau : M \to \field{R}$ is a locally anti-Lipschitz (continuous) time function. Let $(\overline{M}_N, \overline{d}_N)$ be the metric completion of $(M, \hat{d}_\tau)$, and let $\overline{\tau}_N : \overline{M}_M \to \field{R}$ be the unique continuous extension of $\tau$. Then for all $\overline{p}, \overline{q} \in \overline{M}_N$, we have:
$$\overline{p}\;  \le \; \overline{q} \; \; \Longrightarrow \; \;  \overline{d}_N  (\, \overline{p} \, , \,\overline{q} \, ) \; = \; \overline{\tau}_N(\overline{q}) - \overline{\tau}_N(\overline{p})$$
\end{lem}

\begin{proof} Fix $\overline{p}, \overline{q} \in \overline{M}$, with $\overline{p} \le \overline{q}$. Let $\{p_k\}$ and $\{q_k\}$ as in Definition \ref{Def_causality_completion}. Then, applying the definitions, and Propositions \ref{null_dist_causal_pairs} and \ref{extendfunctionprop}, we have:
\begin{align*}
 \overline{d}_N (\, \overline{p} \, , \,\overline{q} \, ) & \; = \; \lim_{k \to \infty} \, [\, \hat{d}_\tau(p_k, q_k) \,] \; = \; \lim_{k \to \infty} \, [\, \tau(q_k) - \tau(p_k) \, ] \; = \; \overline{\tau}_N(\overline{q}) - \overline{\tau}_N(\overline{p})
\end{align*}
\end{proof}

\vspace{2pc}
We plan, however, to study the question of extending causality to metric completions more systematically in separate work, and do not purse this further here.

\vspace{1pc}
\subsection{A Nontrivial Example of Riemannianization}

\vspace{1pc}
\begin{exm} \label{Riem_example2} Consider the future half-Minkowski plane, 
$$\field{M}^{1+1}_+ = ( \, (0, \infty) \times \field{R} \, , \, -dt^2 + dx^2 \, )$$
Consider the timelike vector field
$$T = (\tfrac{1}{t^2} + 1)^{1/2}\d_t + \tfrac{1}{t}\d_x$$
The corresponding Riemannianized metric tensor is given by
$$g_T^R = (\tfrac{2}{t^2} + 1 )dt^2 - \tfrac{4}{t}(\tfrac{1}{t^2} + 1)^{1/2}dtdx + (\tfrac{2}{t^2} + 1 )dx^2$$
Note that, for large $t$, we have $T \approx \d_t$ and $g_T^R \approx dt^2 + dx^2$. On the other hand, as $t \to 0^+$, $T$ tips over to the null direction and the components of $g_T^R$ blow up in size. Nonetheless, $g_T^R$ induces the standard Euclidean measure on the entire half plane, since:
$$\mathrm{det}(g_T^R) = (\tfrac{2}{t^2} + 1)^2 - \tfrac{4}{t^2}(\tfrac{1}{t^2} + 1) = \tfrac{4}{t^4} + \tfrac{4}{t^2} + 1 - \tfrac{4}{t^4} - \tfrac{4}{t^2} = 1$$
Letting $M = (0, \infty) \times \field{R}$, then the Gaussian curvature of the resulting 2-dimensional Riemannian manifold $(M,g^R_T)$ is given at each point $(t,x) \in M$ by
$$K(t,x) \; = \; - \dfrac{6}{t^4}$$
\end{exm}

\pagebreak


\providecommand{\bysame}{\leavevmode\hbox to3em{\hrulefill}\thinspace}
\providecommand{\MR}{\relax\ifhmode\unskip\space\fi MR }
\providecommand{\MRhref}[2]{%
  \href{http://www.ams.org/mathscinet-getitem?mr=#1}{#2}
}
\providecommand{\href}[2]{#2}

\end{document}